%% file: main.tex
\documentclass{article} %

\usepackage{iclr2023_conference,times}

\input{math_commands.tex}

\usepackage{hyperref}
\usepackage{url}
\usepackage{xurl}
\usepackage[authoryear]{natbib}

\usepackage[utf8]{inputenc} %
\usepackage[T1]{fontenc}    %
\usepackage{booktabs}       %
\usepackage{amsfonts}       %
\usepackage{nicefrac}       %
\usepackage{microtype}      %
\usepackage{xcolor}         %
\usepackage{graphicx}
\usepackage{multirow}

\usepackage{wrapfig}
\usepackage[many]{tcolorbox}

\usepackage{listings}
\usepackage[scaled=0.85]{FiraMono}

\usepackage{color}
\definecolor{myfavblue}{rgb}{0.05, 0.2, 0.8}
\definecolor{keywords}{RGB}{255,0,90}
\definecolor{comments}{RGB}{0,0,113}
\definecolor{red}{RGB}{160,0,0}
\definecolor{green}{RGB}{0,150,0}
\definecolor{C0}{rgb}{0.12156862745098039, 0.4666666666666667, 0.7058823529411765}  %

\definecolor{myblue}{HTML}{3182bd}
\definecolor{myred}{HTML}{de2d26}

\definecolor{mydarkblue}{rgb}{0,0.08,0.45}
\hypersetup{
    colorlinks=true,
    linkcolor=mydarkblue,
    citecolor=mydarkblue,
    filecolor=mydarkblue,
    urlcolor=mydarkblue
}

\newtcbtheorem[number within=section]{experiment}{Experiment}{
  breakable,
  colback=green!5,
  colframe=green!35!black,
  fonttitle=\bfseries}{x}

\newtcbtheorem[number within=section]{hypothetical}{Hypothetical}{
  breakable,
  colback=blue!5,
  colframe=blue!35!black,
  fonttitle=\bfseries}{x}

\title{Foundation Models and Fair Use}

\newcommand*\samethanks[1][\value{footnote}]{\footnotemark[#1]}

\author{
Peter Henderson\thanks{Equal Contribution. Correspondence to
\href{mailto:phend@cs.stanford.edu}{phend@cs.stanford.edu}, 
\href{mailto:lxuechen@cs.stanford.edu}{lxuechen@cs.stanford.edu}. \copyright \hspace{0em} Peter Henderson, Xuechen Li, Dan Jurafsky, Tatsunori Hashimoto, Mark A. Lemley, \& Percy Liang}, Xuechen Li\samethanks, \textbf{Dan Jurafsky, Tatsunori Hashimoto, Mark A. Lemley, Percy Liang}\\
Stanford University
}

\usepackage{array}
\usepackage{lipsum}

\begin{document}

\maketitle

\input{sections/abstract}
\input{sections/introduction}

\section{Foundation Models and Fair Use}

We first briefly define foundation models and introduce fair use law as well as its applicability to foundation models. 
To provide a better understanding of the risks, we then examine concrete precedential cases related to fair use and how they might apply to foundation models.
We conduct this analysis for cases related to text, code, and visual art. 
To accompany our examination of U.S. case law, we include hypothetical scenarios of model deployments and how they might exceed the bounds of the fair use doctrine under current law. We also provide experiments to show that current foundation models are capable of generating content that is not transformative.

This section proceeds as follows. Section~\ref{sec:fmintro} provides a brief overview of foundation models. 
Section~\ref{sec:definitions} provides definitions of actors involved in the foundation model development and deployment process and what roles they play. Section~\ref{sec:fair} provides a high-level overview of fair use doctrine in the United States. Sections~\ref{sec:text}, \ref{sec:code}, and \ref{sec:images} provide in-depth examples of case law and foundation model scenarios to help elucidate potential risks.  

\subsection{Foundation Models}
\label{sec:fmintro}
Foundation models are machine learning models trained on broad data (typically scraped from the internet) generally using self-supervision at scale~\citep{bommasani2021opportunities}. 
Most foundation models are not trained to accomplish specific tasks but rather to capture useful general information in the data.
For instance, most autoregressively pretrained language models (e.g., GPT-3~\citep{brown2020language}, PaLM~\citep{chowdhery2022palm}, or Chinchilla~\citep{hoffmannempirical}) are trained to predict the next word given a sequence. 
Most text-to-image models, for example DALL·E~\citep{ramesh2021zero}, are trained to capture the distribution of images given a text prompt. 
These models can then be tuned to align more with human preferences~\citep{ouyang2022training} or be adapted for specific tasks.
Foundation models can be used for generating content. This includes models like GPT-3~\citep{brown2020language} for text, Codex~\citep{chen2021evaluating} for code, and DALL·E~\citep{ramesh2021zero} for images.
Alternatively, they can be used for \textit{non-generative} purposes. These would typically output one value, rather than having a longer free-form output. For example, they might classify text in different ways, or predict a numerical value from an image. This includes (for the most part) models like BERT~\citep{devlin2018bert} or CLIP~\citep{radford2021learning}.
Importantly, most foundation models can be modified to operate for either type of task, and many tasks will be somewhere on the spectrum between generative and non-generative tasks.\footnote{This spectrum between generative and non-generative tasks is important to understand as it may have some impact on the fair use analysis and we discuss how technical mitigation strategies can take this into account in Section~\ref{sec:filtering}.}

Millions of users now use foundation model products. ChatGPT, a generalist chatbot from OpenAI, has grown to an estimated 100M daily active users.\footnote{\url{https://www.reuters.com/technology/chatgpt-sets-record-fastest-growing-user-base-analyst-note-2023-02-01/}}
Midjourney's users produce millions of generated images per day.\footnote{\url{https://www.theregister.com/2022/08/01/david_holz_midjourney/}}
As foundation models are expanded into more products, deployments will only scale to more and more users. An increasingly growing list of companies has plans to deploy similar products to ChatGPT, from Microsoft's Bing Chat\footnote{\url{https://www.bing.com/new}} to Google's Bard,\footnote{\url{https://blog.google/technology/ai/bard-google-ai-search-updates/}} and more. We categorize the high-profile instances by the domain of the data in Table~\ref{tab:catalog}.

\begin{table}[!htbp]
    \centering
    \begin{tabular}{c|p{10cm}}
    \toprule
         Domain & Products\\
         \midrule
         \multirow{4}{*}{Text}
          & General Purpose API (e.g., \href{https://openai.com/api}{OpenAI GPT API}) or general chat-based agents (e.g., \href{https://chat.openai.com/}{ChatGPT}) \\
          & Write blogs and marketing material (e.g., \url{copy.ai})\\
          & Custom generated stories (e.g., \url{https://novelai.net/}) \\
          & Text-based adventure games (e.g., \url{https://aidungeon.io/}) \\
\hline
         \multirow{2}{*}{Code}
         & Generate code (e.g., \href{https://github.com/features/copilot}{Github CoPilot})\\
         & Pair programming with an AI assistant (e.g., \href{https://blog.replit.com/ai}{Replit})\\
\hline
        \multirow{2}{*}{Images}
        & Generate images from text (e.g., \href{https://labs.openai.com/}{OpenAI Dall-E}, \href{{https://techcrunch.com/2022/10/12/microsoft-expands-azure-openai-service-with-dall-e-2-in-preview/}}{Azure OpenAI Service},
        \href{https://designer.microsoft.com/}{Microsoft Designer},
        \href{https://github.com/CompVis/stable-diffusion}{Stable Diffusion}, 
        \href{https://www.midjourney.com/}{Midjourney}) \\
         \bottomrule
    \end{tabular}
    
    \caption{We enumerate a small fraction of advertised foundation model deployments and products provided via APIs or other interfaces, demonstrating that these systems are being deployed as products in a wide range of areas.
    }
    \label{tab:catalog}
\end{table}

\subsection{Definitions and Roles}
\label{sec:definitions}
Before our discussion, we define several actors. 
The \textit{data creator} creates data that a model might be trained on. The \emph{data curator} collects data and a \textit{data host} distributes
data that a model is trained on. The \emph{model creator} trains the model on this data. The \emph{model deployer} hosts a model and provides access to it via an API, potentially creating revenue from serving the model. The \emph{model user} uses the model for downstream tasks, potentially creating revenue with the output of the model.
These actors may all be the same person or entity, or they may be different people or entities.

We primarily discuss the potential for a data intellectual property (IP) owner (the \textit{data creator}) to bring a case against foundation model deployers, users, and creators. 
While there is certainly risks of liability for data curators, this has long been discussed in other work. 
We will also focus on liability as a result of the model outputs themselves, not the training process or the model parameters.\footnote{See discussions by, e.g., \citet{mccann2021copyright,lemley2020fair,grimmelmann2015copyright,sobel2017artificial} for more examination of model parameters and model training.}
Instead, we focus on whether those weights can be \textit{used} in an infringing way and thus incur liability.

\subsection{Fair Use}
\label{sec:fair}

In the United States, the legal doctrine of \textit{fair use} provides some relief from liability for using copyrighted material without a license. The fair use defense is determined by considering four factors: (1) the {purpose and character} of the use, including whether the use is of a commercial nature or is for nonprofit educational purposes (\textbf{transformativeness}); (2) the \textbf{nature} of the copyrighted work (fair use strongly favored if original work is factual as opposed to creative); (3) the \textbf{amount and substantiality} of the portion used in relation to the copyrighted work as a whole; (4) the \textbf{effect} of the use upon the potential market for or value of the copyrighted work. \textit{See} 17 U.S.C. \S 107. 
It is important to note that every factor will play \textit{some} role in the court's decision-making process, but the interaction between them is not always clear.

We will briefly provide an overview of each fair use factor in this section, but we stress that fair use doctrine is murky and evolving.
In any common law setting, a case-by-case review helps outline the contours of the doctrine, so we will subsequently review relevant case law to help shine a light on how fair use doctrine might handle foundation models.
Within the topics we discuss, we provide a descriptive survey of the current state of fair use doctrine and how it could relate to foundation models to the extent possible. However, there will be significant nuances and room to maneuver depending on the exact structure of a deployment and training procedure.

\paragraph{Transformativeness.} When the original work is transformative, this weighs heavily in favor of fair use. Empirical studies have found that the transformativeness factor tends to be most dispositive in legal analyses and is heavily emphasized in assessments of fair use~\citep{asay2020transformative}.
For example, when Google copied parts of the Java API for Android, the Supreme Court found that this was fair use. 
It took into account that the amount of code copied (a small percentage of the derivative code base), and the end product was transformative~\citep{oracle}. 
Similarly, Google Books can show portions of books to users because the percentage of the displayed book is small and the use case is transformative (from the original use of reading a book cover-to-cover to a new use case of searching quickly through a book)~\citep{ebooks}. 

For scenarios concerning machine learning and AI, some legal scholars believe that fair use covers most types of model training where the resulting model functions differently than the input data, particularly when the model targets a different economic market~\citep{lemley2020fair,carroll2019copyright}.
In part, these arguments sometimes analogize to cases related to \emph{intermediate copying}---as long as the ``defendant’s end product was a transformative new work and the copying was a necessary step to get there,'' the copying of copyrighted material is covered by fair use~\citep{lemley2020fair}.
For foundation models that are not applied in a generative context, this argument can be a good fit. 
For example, training a recommendation system or search engine on copyrighted books is likely sufficiently transformative from the purpose of the original book and its target markets, and is thus likely to be considered fair use---as in Google Books.

However, the story may be different for generative use cases. For generative models like DALL-E or GPT that produce creative outputs, the situation is less likely to be problematic if the outputs do not copy a substantial portion of any existing work but instead transform the input into totally different outputs, in line with the fair use doctrine~\citep{sobel2017artificial,lemley2020fair}.
When the downstream product based on such a model is \emph{not} transformative (e.g., model outputs similar content to copyrighted training data, or the application's market is similar to original data markets), courts may decide that the generated content, the model deployment, and even potentially the model parameters themselves are not covered by fair use~\citep{sobel2017artificial}.

Consider a generative foundation model trained on copyrighted books. 
In the extreme case if the model is trained on a book such that it can verbatim reproduce the entire book consistently (no matter what input is provided), then this is not transformative and could be problematic. Would this be any different than redistributing the book if you provide a mechanism to store and output the book?
In the more common scenario, foundation models used for generative tasks will fall into a gray area where they produce some content that looks similar to the original training data and some content that looks substantially different.
A key question is whether the generated content that looks similar to the original training data is still transformative enough that it would meet this fair use factor.

How much transformation is needed? In general, what kinds of transformations are acceptable depends on the context, but overall, fair use mostly requires transformations of low-level content (relatively low n-gram overlap) as well as higher-level concepts (no similar storylines with specific repeated non-generic structures). This fundamentally means that more technical research is needed to keep models covered by fair use, as we will discuss throughout this work.
In later sections we will cite relevant case law where each of the transformations (except for parodies) was found not to be fair use. For example, Figure~\ref{fig:markettomarket} in \S \ref{sec:text} illustrates how a generative foundation model trained on books might be used to produce different types of outputs and what cases might illustrate similar situations. 
These cases help us outline the level of the transformation necessary to stay within the current confines of fair use doctrine.
While we will also briefly discuss other fair use factors, we will primarily focus on transformativeness throughout this article as it is a key component of a fair use defense.

It is also worth noting that we mainly focus on transformativeness of the outputs themselves, but the \textit{purpose} of the machine learning model could be transformative. For example, in \citet{fields} one consideration was whether the use-case of caching a webpage for tracking changes was a transformative purpose from displaying the webpage for viewer consumption. We generally do not address these considerations as much throughout this work, though they may play a large role in litigating fair use. At a high level though, we will assume that the foundation model is used for a sufficiently similar purpose to the original training data that analysis will fall to model outputs. For example, training on books to generate abridgements or training on webcrawls to generate web pages. Given the large amount of diverse data ingested by models and the wide range of use-cases, it is likely that there will be deployments or models whose purpose competes with part of the original data source.

\paragraph{Nature of the copyrighted work.} There are many nuances as to what can be copyrighted. For example, an idea cannot be copyrighted, only the expression of that idea. Facts also cannot be copyrighted, only the expression of those facts. As a result, courts will consider the components of the original work that were used and whether they should receive protection under copyright law.

\paragraph{Amount and Substantiality.} A critical point is how much content was taken from the original work. A \textit{de minimis} amount is acceptable. For example, one can quote the original work as long as the quotes are not a substantial portion of the original work. This was a critical factor in the Google Books case since Google does not display significant portions of books~\citep{ebooks}.  Importantly, the intermediate copying of a work in its entirety may not count against fair use if the intermediate copy is used to generate output that is not itself infringing~\citep{1992sega,2000sony}.

\paragraph{Effect on Market.} Closely tied to transformativeness, if the new product has some effect on the market (or a potential market derivative market) for the original work, this will be taken into account. 
So, using a model trained on books to create a derivative book in the same market will be more likely to affect the original market.
But the market effect must be from infringement, not merely from competition from a noninfringing work.

While non-commercial distribution does not automatically imply fair use, it improves the likelihood of a successful fair use defense. A recent empirical study found that 36 of 47 ($\sim 77\%$) analyzed case opinions involving a non-commercial application found fair use~\citep{beebe2020empirical}. But we caution non-commercial researchers from assuming that they are automatically covered by fair use even in more extreme cases of infringement. In \citet{gsu}, professors at Georgia State University made copies of books available to students on the school's internal system. While the court found most instances of this to be fair use, it did identify four instances that were not fair use.\footnote{In this particular case it stated that it weighted each of the fair use factors as ``25\% for factor one [purpose and character], 5\% for factor two [nature of the copyrighted work], 30\% for factor three [amount and substantiality] and 40\% for factor four [effect of the use on the potential market]''~\citep{gsu}.}

\input{sections/examples_and_analysis}

\input{sections/mitigations}

\section{Forward-looking Agenda}

As demonstrated throughout this work, the risk of copyright violation and litigation, even with fair use protection, is a real concern. To mitigate these risks, we recommend that foundation model practitioners consider implementing the mitigation strategies outlined here and pursuing other novel research in this area. There is significant, exciting, technical research required to make technical mitigation strategies robust and aligned with fair use doctrine. We reinforce that machine learning researchers \textit{must} play a role in providing viable mitigation mechanisms to demonstrate that models are truly covered by fair use. 

\paragraph{Preventing extreme outcomes in the evolution of fair use law by advancing mitigation strategies.} Legal scholars have noted that there might be two extreme outcomes for fair use and machine learning~\citep{sobel2017artificial}. 
On one hand, there is a possibility that courts may rule that foundation models are widely acceptable under fair use regardless of the likelihood of infringement or efforts at mitigation, which could have adverse effects on the income of data creators and disregard the ethical and moral rights attached to their work. 
On the other hand, there is a possibility that courts may declare that generative foundation models cannot be trained on unlicensed copyrighted data in most cases.
This scenario could lead to a concentration of power for companies that have retained licenses to large amounts of data; companies like YouTube or Facebook might be able to leverage large amounts of user-contributed data where others would be shut out of model training.
Neither of these two outcomes is ideal. 
As litigation progresses, identifying mechanisms to prevent extreme outcomes will be critical. For example, it is important to demonstrate that not all forms of foundation models are inherently infringing and that some of their potential risks can be effectively managed through technical means.

With better demonstrations of co-evolving technical mitigation strategies, the law might find a middle ground that allows model training and deployment with sufficient effort to implement objectively strong mitigation strategies.
Courts may consider the reasonable efforts of model builders and deployers to mitigate copyright risk, both in deciding fair use and in determining whether they can face indirect infringement liability.
Trademark courts have taken a similar approach, for example in~\citet{tiffanyebay}.
As such, advancing research in this area (with methods such as improved similarity metrics) may help in preventing extreme outcomes in legal settings.\footnote{But, again, technical mitigation strategies will only go so far in the fair use assessment and will not (and should not) automatically guarantee that any one deployment is acceptable under fair use doctrine.}

\paragraph{We should not over-zealously filter.} There must be a balance to filtering. Well intentioned but strict filtering mandates adopted by other countries have been criticized and criticized for their impacts on free speech~\citep{eufilterseff}. Similarly, YouTube's content ID system, a large-scale filtering approach, has been criticized for not following fair use standards and being overaggressive in its filtering~\citep{bartholomew2014death,boroughf2015next}.  \citet{levendowski2018copyright} points out that restrictive views of fair use doctrine can exacerbate biases and that fair use can help create fairer systems. While mitigation strategies will help prevent undesirable outcomes, it is important to develop strategies that carefully align with fair use standards, as we have previously discussed. This means that factual content should not necessarily be filtered, neither should parodies, or short form regurgitation used for commentary. And evolutions of fair use doctrine or further policymaking should consider the distributive effects of preventing access to certain types of data for model creation.

\paragraph{Policymakers could consider how and if DMCA (or similar) safe harbors should apply to foundation models.}  As we have seen, there are various ways, including filtering, to mitigate the risk of copyright infringement in the output of foundation models, but none will entirely eliminate the risk of liability. 
Even when trained on presumably permissively licensed datasets, for example, it is difficult (if not impossible) to determine the provenance of every piece of data and filter it out.
Users might post content to seemingly permissively-licensed databases that they do not actually have the rights for. There may even be uncertainty about whether a piece of content is \textit{actually} in the public domain or whether that status has been revoked.\footnote{This is not a hypothetical, in \citet{golan} the Supreme Court found that revoking a work's public domain status is not unconstitutional. In that case, a group of artists had relied on the public domain status of some works whose copyright status was later restored as part of the Uruguay Round Agreements Act.}
And even if foundation model practitioners implement strong mitigation strategies, the amorphous nature of fair use doctrine may make it difficult to know what kinds of content will be covered by fair use \textit{ex ante}.

With the uncertainty of DMCA protections (discussed in \S~\ref{sec:considerations}), the law may need to adapt to this reality, and it could do so, for instance, by clarifying the role of safe harbors for models that implement sufficiently strong mitigation strategies. Policymakers could make clear that DMCA protections apply to this setting or they could identify other more suitable safe harbor mechanisms. This may provide more balance than general-purpose text and data mining exemptions seen in other countries, but again are not a panacea. 
Such safe harbors would have to be structured to consider the strength of the implemented mitigation strategies to ensure that they are not abused.

\paragraph{Pursuing other remedies beyond technical mitigation.}
Importantly, even if technical mitigation strategies managed to keep foundation models within the confines of fair use, these models may still create harms in many other ways---including disrupting creative industries, exploiting labor, and more. See extensive discussion by, \textit{e.g., }~\citet{bender2021dangers,bommasani2021opportunities,blodgett2020language,maori}. It is important to note that we do not suggest that technical mitigation strategies will solve everything and neither will fair use doctrine. Our goal here is to point out that currently there is more work to be done even \textit{within} the confines of fair use to make foundation models more in line with case law. Other strategies to prevent harms should be pursued in conjunction with the strategies we outline here, but they should be carefully weighed against other potential harms from excluding data under overly restrictive copyright standards \citep{levendowski2018copyright}. 
For example, complementary approaches to what we describe here could include statutory licensing schemes, taxation and redistribution, or other policy mechanisms.
While these may be worthy of considering, each may have its own challenges and are outside the scope of this work.
Furthermore, there are other aspects of fair use that we do not consider here, and there well may be cases where technical mitigation strategies will still not be enough for fair use.

\input{sections/related_work}

\section{Conclusion}

We reviewed U.S. fair use standards and analyzed the risks of foundation models when evaluated against those standards in a number of concrete scenarios with real model artifacts.
Additionally, we also discussed mitigation strategies and their respective strengths and limitations.
As the law is murky and evolving,
our goal is to delineate the legal landscape and present an exciting research agenda that will improve model quality overall, further our understanding of foundation models, and help make models more in line with fair use doctrine.
By pursuing mitigation strategies that can respect the ethics and legal standards of intellectual property law, machine learning researchers can help shape the law going forward.
But we emphasize that even if fair use is met to the fullest, the impacts to some data creators will be large. We suggest that further work is needed to identify policies that can effectively manage and mitigate these impacts, where the technical mitigation strategies we propose here will fundamentally fall short.
We hope that this guide will be useful to machine learning researchers and practitioners, as well as lawyers, judges, and policymakers thinking about these issues.
We emphasize, again, that even if foundation models are covered by fair use the impacts on labor might be con

\section*{Acknowledgements}
This work was done at the Center for Research on Foundation Models (CRFM), and we would also like to thank the Stanford Institute for Human-Centered Artificial Intelligence (HAI) for supporting this work.
We thank Alex Aiken for generously providing us with access to MossPlus---the commercial version of Moss. 
We thank Rishi Bommasani, Dilip Arumugam, Mark Krass, and Jieru Hu for helpful discussions and feedback. 
We thank Tony Lee for supporting our experiments with the CRFM infrastructure. 
PH is funded by the OpenPhilanthropy AI Fellowship.
XL is supported by a Stanford Graduate Fellowship. 
TH and DJ was supported by a grant from OpenPhilanthropy.
Note, ML was hired as counsel for \citet{stablediffusionlitigation} after a near-final draft of this work was written. This work reflects the personal opinions and research of the authors. It does not reflect the position of any other entity or person, nor does it constitute legal advice.
 
\bibliographystyle{iclr2023_conference}
\bibliography{refs}

\newpage
\clearpage
\appendix
\input{sections/appendix}

\end{document}

%% file: math_commands.tex
\usepackage{amsmath,amsfonts,bm}

\def\eqref#1{equation~\ref{#1}}

\def\1{\bm{1}}

\DeclareMathAlphabet{\mathsfit}{\encodingdefault}{\sfdefault}{m}{sl}
\SetMathAlphabet{\mathsfit}{bold}{\encodingdefault}{\sfdefault}{bx}{n}

%% file: sections/abstract.tex
\begin{abstract}
Existing foundation models
are 
trained on copyrighted material. 
Deploying these models can pose both legal and ethical risks when data creators fail to receive appropriate attribution or compensation.
In the United States and several other countries, copyrighted content may be used to build foundation models without incurring liability due to the \emph{fair use} doctrine.
However, there is a caveat: If the model produces output that is similar to copyrighted data, particularly in scenarios that affect the market of that data, fair use may no longer apply to the output of the model.
In this work, we emphasize that fair use is not guaranteed, and additional work may be necessary to keep model development and deployment squarely in the realm of fair use.
First, we survey the potential risks of developing and deploying foundation models based on copyrighted content. 
We review relevant U.S. case law, drawing parallels to existing and potential applications for 
generating text, source code, and visual art.
Experiments confirm that popular foundation models can generate content considerably similar to copyrighted material.
Second, we discuss technical mitigations that can help foundation models stay in line with fair use.
We argue that more research is needed to align mitigation strategies with the current state of the law.
Lastly, we suggest that the law and technical mitigations should co-evolve.
For example, coupled with other policy mechanisms, the law could more explicitly consider safe harbors when strong technical tools are used to mitigate infringement harms.
This co-evolution may help strike a balance between intellectual property and innovation, which speaks to the original goal of fair use.
But we emphasize that the strategies we describe here are not a panacea and more work is needed to develop policies that address the potential harms of foundation models.
\end{abstract}

%% file: sections/introduction.tex
\section{Introduction}

Foundation models\footnote{Foundation models can roughly be defined as large pre-trained machine learning models that are used as a starting point for various computational tasks.} that are trained on large-scale internet data serve as the base for an increasing number of deployed applications in the real world~\citep{bommasani2021opportunities}.
Models such as GPT-3/4~\citep{brown2020language,https://doi.org/10.48550/arxiv.2303.08774}, Stable Diffusion~\citep{rombach2021highresolution}, and Codex~\citep{chen2021evaluating} are actively being integrated into a variety of products like Duolingo's Language Learning App,\footnote{\url{https://blog.duolingo.com/duolingo-max/}} Stability AI's DreamStudio,\footnote{\url{https://stability.ai/}} GitHub's CoPilot,\footnote{\url{https://github.com/features/copilot}} and more.
Researchers are grappling with the legality and ethics of developing and deploying these models using data broadly collected from the internet.
Many have raised concerns about using uncurated internet data for model development, touching on issues of privacy~\citep{carlini2021extracting} and fairness~\citep{bender2021dangers}.
But as foundation models are deployed in ways that can harm the markets of the original data creators, particularly when generating content similar to the original data, intellectual property rights become a growing concern.
In this paper, we study the legal challenges of building and deploying foundation models from the perspective of intellectual property, focusing mainly on copyright.

Under United States ("U.S.") law, copyright for a piece of creative work is assigned ``the moment it is created and fixed in a tangible form that it is perceptible either directly or with the aid of a machine or device''~\citep{copyrightoffice}.
The breadth of copyright protection means that most of the data that is used for training the current generation of foundation models is copyrighted material.
For example, \citet{bookcorpus} pointed out that the BookCorpus contains copyrighted data under restrictive licenses and has been used to train large foundation models including GPT-3~\citep{brown2020language} and BERT~\citep{devlin2018bert}.
Similarly, The Pile \citep{gao2020pile} contains Books3, a dataset of copyrighted and commercially sold books downloaded from Bibliotik, a torrent tracker for books and learning materials~\citep{books3,biderman2022datasheet}. 
More generally, most foundation models are trained on data obtained from webcrawls like C4~\citep{2019t5} or OpenWebText~\citep{openwebtext}. 
Since most online content has copyright protections attached at creation, using them for certain purposes could be considered infringement.\footnote{We note that there are nuances to even the infringement point, since some uses that respect robots.txt specifications might have an implied license as described in \citet{fields}. This is unlikely to apply to all generated model outputs, however, and we discuss this further in \S~\ref{sec:filtering}.}
Researchers, at least in the United States, have long relied on the legal doctrine of \emph{fair use} to avoid liability from using copyrighted data. 
Fair use allows the public to use copyrighted material for certain types of purposes---even without a license---especially when the end-product is \emph{transformative}.
For example, when releasing potentially copyrighted content in the past, individuals and organizations have relied on rough guesses for what constitutes fair use. A common approach is to release snippets: 5-grams~\citep{generalindex}, 11-grams~\citep{bitext}, or several pages~\citep{ebooks}. 

\citet{lemley2020fair} have pointed out that training a machine learning model on copyrighted data is likely considered fair use in circumstances where the final model does not directly generate content. 
For example, training a model on a corpus of popular books solely for predicting the similarity of two passages is transformative and likely falls under fair use.\footnote{Though recent litigation points out that no court has actually weighed in on the matter of whether model training is fair use~\citep[Complaint at 23]{githublitigation}.}
However, when it comes to training and deploying foundation models for \textit{generative} use cases, the analysis becomes more complex.
This is because these models are usually capable of generating content similar to copyrighted data, and deploying them can potentially impact economic markets that benefit the original data creators. 
For these scenarios, legal scholars argue that fair use may not apply~\citep{lemley2020fair,sobel2017artificial,levendowski2018copyright}.

By expanding the capabilities of models, machine learning researchers and practitioners have stumbled into the muddy waters of fair use.
As a result, websites like Getty Images have banned AI-generated content~\citep{getty}, and lawsuits have been filed against products using foundation models, namely GitHub Copilot and Stable Diffusion~\citep{githublitigation,stablediffusionlitigation,gettylawsuit}.
In this work, we shed light on this subject matter for machine learning researchers and highlight that significant additional work is required to de-risk foundation model deployments for generative use cases, focusing primarily on U.S. laws.

First, we provide an overview of U.S. case law on the fair use doctrine.\footnote{We examine U.S. fair use doctrine, rather than international doctrines, for two reasons. First, companies have specifically pointed to fair use as a defense for their use of foundation models. For example, former Github CEO Nat Friedman pointed to fair use when referring to Github's Copilot deployment. \textit{See} \href{https://twitter.com/natfriedman/status/1409914420579344385}{https://twitter.com/natfriedman/status/1409914420579344385} Second, the expertise of the authors is in U.S. law.} 
We draw analogies to foundation model use cases. We supplement these with a review of prior experiments, as well as novel experiments, and illustrate that foundation models can produce content that is sufficiently similar to copyrighted material. Furthermore, the case law suggests that even certain types of transformations of the training data would not be considered fair use.
Thus, the risk of infringement is real, and fair use will not cover every scenario where a foundation model is created or used.
The exact amount of risk is unclear, and the law will evolve with ongoing litigation.

Second, we overview technical mitigation strategies that will reduce this risk in accordance with the current state of the fair use doctrine. 
\citet{grimmelmann2015copyright} stated that ``paying attention to robotic readership refocuses our attention on the really fundamental questions: what is copyright, and what is it for? To say that human readers count and robots don’t is to say something deep about the nature of reading as a social practice, and about what we want robots—and humans—to be.''
\citet{lemley2020fair} suggested that humans and AI should be held to similar standards when it comes to copyright.
If this is the case, it is the job of machine learning researchers and practitioners, working together with legal practitioners, to ensure that foundation models create transformative content which would pass muster under the same fair use analysis as provided to a human.
To get there, new strategies and techniques will need to be developed, taking steps to ensure that foundation models behave in more transformative and novel ways. We call for more research to align technical mitigation strategies with fair use, including better output filtering mechanisms relying on higher-level semantics and new innovation in training-time techniques like extraction-preventative learning from human feedback. 
Developing these mitigation strategies is an important research challenge for machine learning and natural language processing and would bring practices in the two fields into better alignment with the law.

Lastly, we argue that a co-evolution of technical mitigation strategies and law can help establish a middle ground where the positive impact of foundation models is realized while reducing the harms to data creators' intellectual property rights. 
With the current uncertainties of fair use doctrine, as \citet{sobel2017artificial} and others noted, the law may sway to one extreme or another. On one hand it could lead to overly permissive interpretations of fair use that could allow \textit{any} generative AI use, disregarding the rights of data creators. Or it could lead to overly restrictive interpretations of fair use that could broadly prevent foundation model training and use, concentrating power among entities that have already acquired vast quantities of licensed data.
By developing and deploying strong technical mitigation strategies, it may be possible to lessen the risk of such extreme legal outcomes.
And the law should take into account the existence and strength of such technical mitigation strategies.
This could involve a multi-pronged approach: considering technical mitigations in fair use assessments, clarifying the status of DMCA protections for foundation models, or developing DMCA-like safe harbors for deployments that use \textit{strong} technical mitigation efforts, pursuing policy strategies for reducing harms to labor, and more.
Realizing this middle ground requires the participation of a much broader community including the data creators impacted by foundation models, technologists, legal professionals, among many others. 
We encourage more multidisciplinary work to further the co-evolution of law, policy, and technical methods for mitigating intellectual property harms.

Overall, the goal of this work is to act both as a guide and call-to-action for ML researchers and practitioners to actively pursue technical mitigation strategies. We hope that this guide helps instill a better understanding that fair use is not a panacea, and that a nuanced comprehension of the legal landscape's intricacies is vital to effectively navigate potential pitfalls and uncertainties. Furthermore, this work may also prove useful to lawyers and policymakers, providing them with more insight into potential technical details of foundation models, including technical mitigation strategies, and how they might play a role in the developing legal best practices and potential reforms.

%% file: sections/examples_and_analysis.tex
\subsection{Natural Language Text}
\label{sec:text}
\begin{figure}[!htbp]
    \centering
    \includegraphics[width=.75\textwidth]{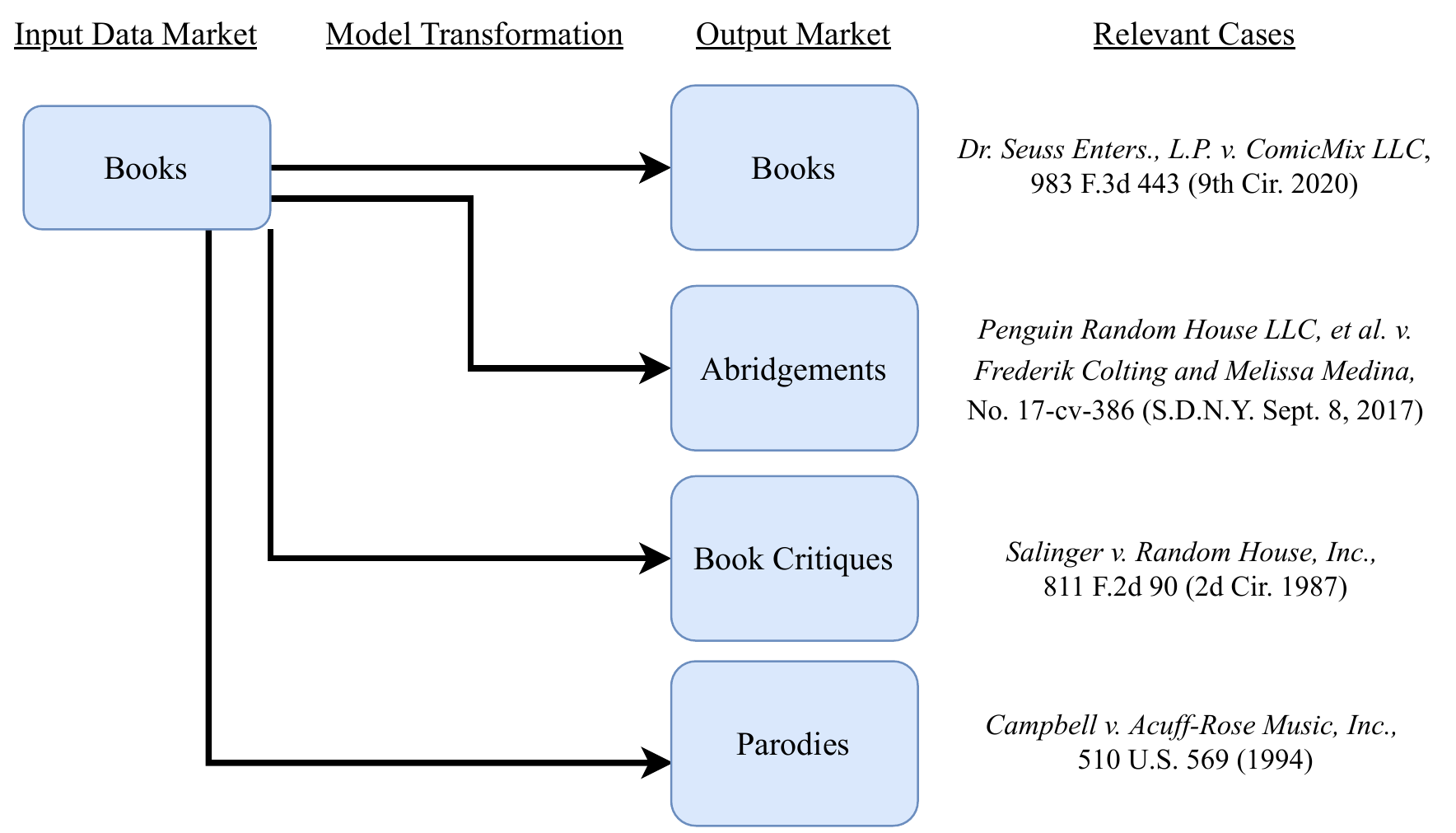}
    \caption{
    Claims of fair use will likely generate more scrutiny when the target market for the deployed model matches the target market of the source---or might threaten a logical derivative market. Book critiques are more likely to be considered fair use unless they include large portions of the source material. 
    Parody is more likely considered to be fair use; satire may also be fair use but requires more extensive justification. We cite cases with relevant analyses for each transformation. If, for example, a single book is trained on and then outputs a substantially similar book that could be problematic.}

        \label{fig:markettomarket}
\end{figure}

Given this high-level understanding of fair use law, we first examine the case of natural language text generation (as opposed to code generation which we will examine in \S \ref{sec:code}), drawing on real cases of creative content that could parallel foundation model outputs and uses.
One of the most prevalent, and earliest, use-cases of foundation models is text generation.
Deployments of models like GPT have been used to create products for copy-editing, text-based games, and general-purpose chatbots. 
These models are typically trained on massive amounts of data taken from across the internet, books, court documents, and more. When used to generate, these models have been observed to output content with only slight transformations from the original training data.
In this section, we examine relevant cases that might help shape what is considered fair use for these models, some of which can be seen in Figure~\ref{fig:markettomarket}.

\paragraph{Verbatim Copying} In a recent case, Google scanned in a large collection of books and made the books available online, only providing users with a small number of pages at a time. Book publishers sued for copyright infringement, but the court found that the amount of content output by Google Books was small and was covered by fair use, even though Google Books contained the entire corpus of published books in databases.
However, distributing larger portions of books is unlikely to be covered. The court in \citet{penguinbuddha} found that making small formatting changes and displaying books on the internet do not constitute fair use.
The fair use criterion from these cases could be directly relevant for foundation models in situations like the following hypothetical.

\begin{hypothetical}{The Assistant Who Reads}{}
A foundation model is deployed as virtual assistant in smartphones. Users learn that they can prompt the assistant with an instruction as follows: ``Read me, word-for-word, the entirety of `Oh the places you'll go!' by Dr. Seuss.''  This becomes popular and users start using the virtual assistant as an audiobook reader to read bedtime stories to their children. Is this fair use? 

\medskip

If our foundation model assistant reads a user the entirety of the book, this is much more like \citet{penguinbuddha} and less likely to be fair use. But, the model is closer to the case of Google Books if it stops reading after a couple of paragraphs, saying, ``I've read as much of the book as I can read.''
\end{hypothetical}

There is no certain amount of content that is categorically permissible or impermissible.  The legal analysis relates to whether the copied content copies the expressive purpose of the original work and whether the copied portion is the portion that most users will want to see. 
For example, if users would only buy a book for an important couple of pages, then copying those couple of pages is less likely to be fair use. That is, reproducing the heart of the work, even if it is small, lowers the probability that the reproduction is considered fair use~\citep{sobel2017artificial}.

In \citet{foxtveyes}, the court found that 10 minutes of TV content was too long of a span to constitute fair use. In \citet{harpernation}, the Supreme Court held that taking 250 words from a large autobiography was not fair use where those words constituted the "qualitative heart" of the book.
Judge Leval in the Google Books case noted that it weighed in Google's favor that Google Books ``does not provide snippet view
for types of books, such as dictionaries and cookbooks, for which viewing a small segment is likely to satisfy the searcher's need[,]'' and avoids providing a service that 
``could usefully serve as a competing substitute for the original''~\citep{ebooks2}. \textit{See also} \citet[at 56]{sobel2017artificial}.
More recently, the hit Taylor Swift song \emph{Shake it off} is going to trial over a potentially infringing 6-word phrase~\citep{2019hall}. And as we will see, there need not be \emph{any} n-gram overlap to result in infringement, requiring only overlapping higher-level mechanisms of expression.

It may be tempting to use quotes and citations to remedy this problem, but this does not necessarily change the transformativeness analysis. The amount and purpose of the use will be considered. For example in \citet{harper1982}, Gerald Ford had sold the rights to his memoirs that were to be published in Time magazine in serialized form. The Nation magazine, without authorization, acquired the memoirs and published and article about them where 300-400 of the 2,250 words were verbatim quotes from the source material. Time magazine canceled the contract and The Nation was sued. In this case the Court found that the severe market damage was the main fair use factor. However, the Court also pointed out that that the 300-400 quoted words, though a relatively small percentage of the memoirs and even a small percentage of the article itself, represented ``the heart of the book'': they were among the most moving parts of the memoirs. This analysis is confounded by the clear market damage evidenced by the canceled contract, but nonetheless demonstrates that simply quoting material that has been drawn verbatim does not automatically resolve the problem.

\begin{figure}[!htbp]
    \centering
    \includegraphics[width=.385\textwidth]{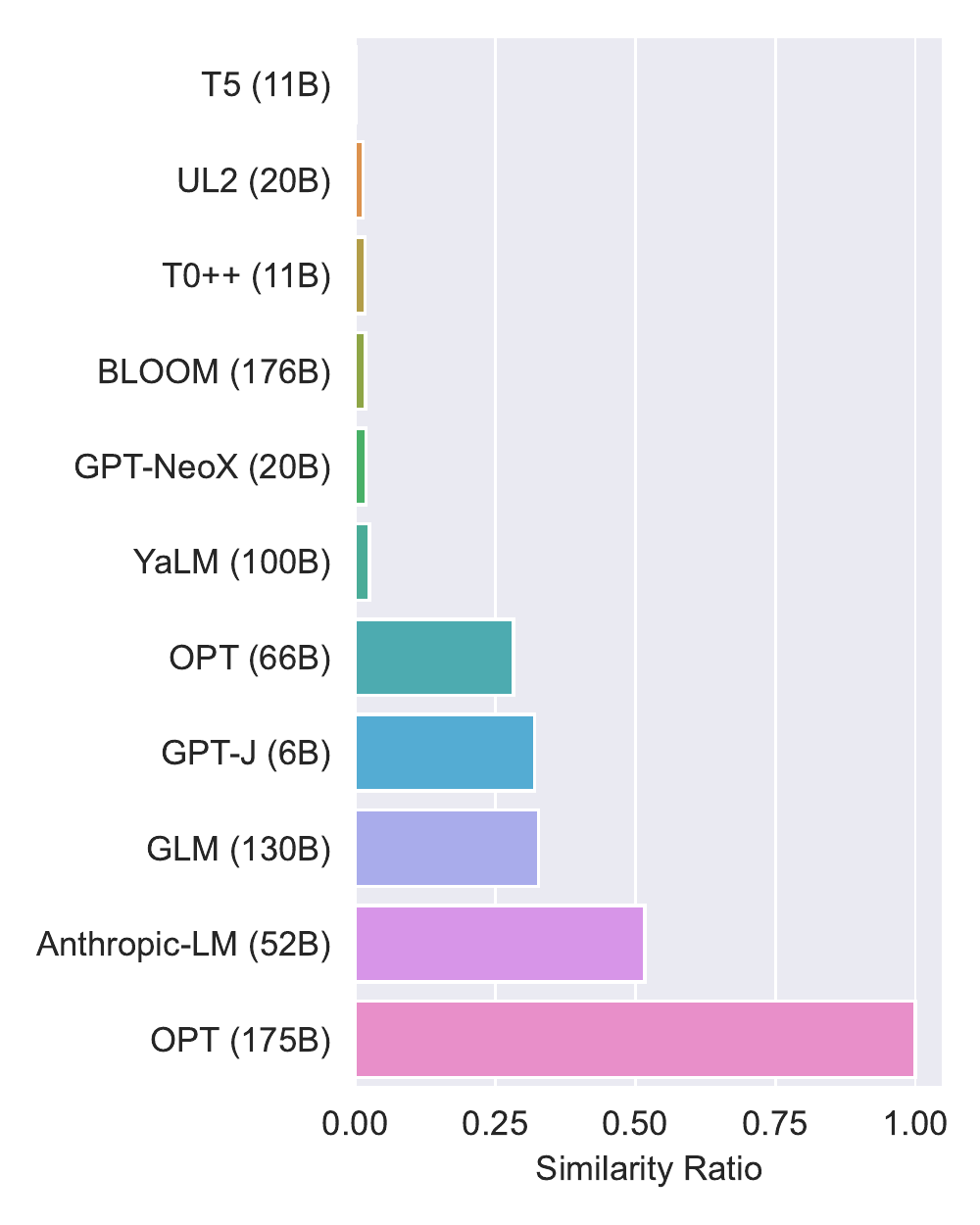}
    \includegraphics[width=.61\textwidth]{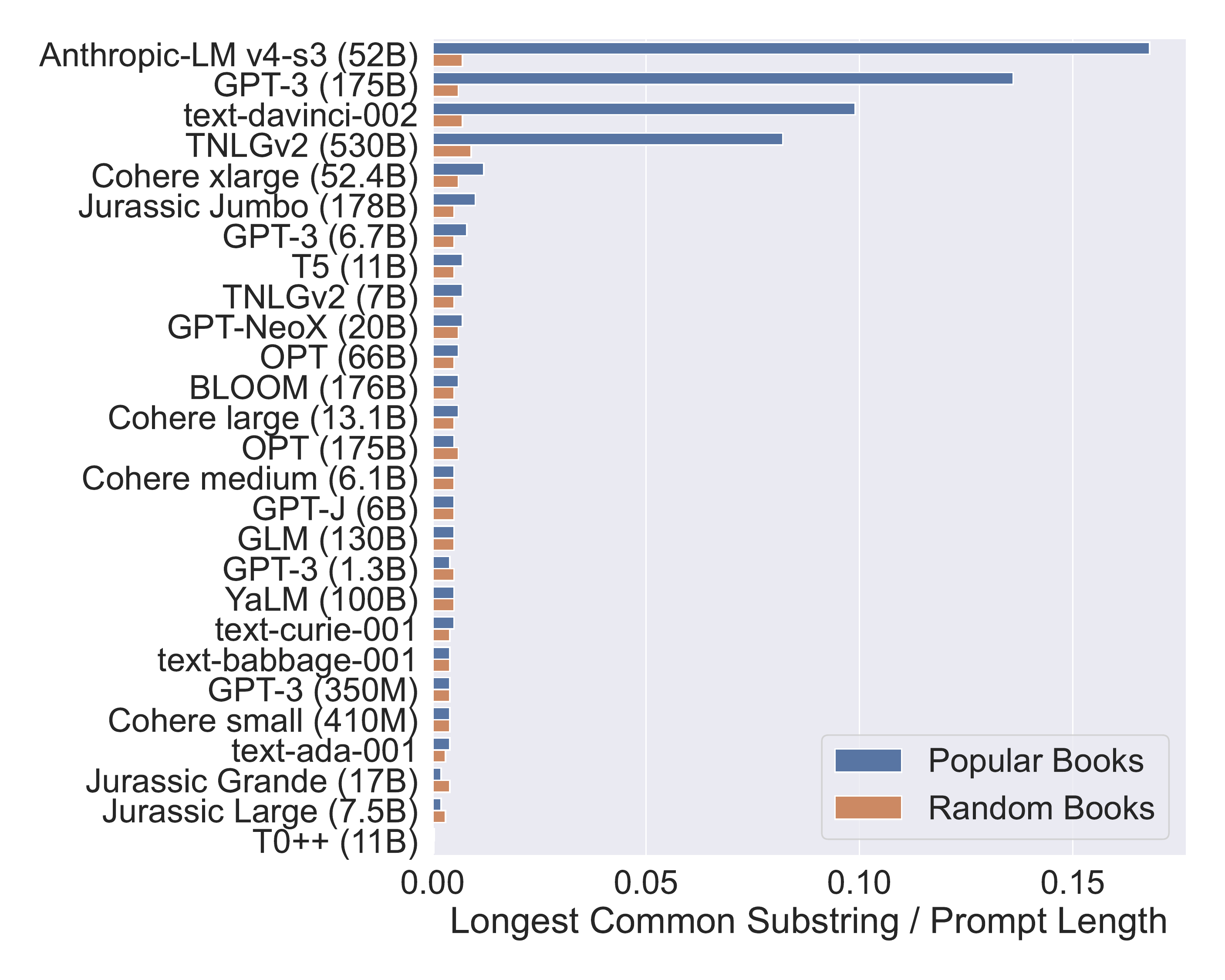}
    \caption{(Left) The maximum similarity ratio using difflib (roughly the fraction of overlapping text) for the extractions of \textit{Oh the Places You'll Go!} for one context window (1024 tokens) tested on a subset of models. OPT-175B regurgitates the story verbatim. (Right) The longest common substring between the generated text and the source material (divided by the prompt length), averaged over sampled book excerpts; larger numbers imply more reproduction. Generally, very few randomly chosen snippets of text generate long spans of verbatim content, though popular materials like Harry Potter are more likely to be regurgitated. This result is limited to the chosen temperature and it is possible with more sampling at higher temperatures more verbatim content can be identified with careful selection mechanisms. With manual prompt engineering, extraction might be more frequent.
    }
    \label{fig:benchmarking}
\end{figure}

\begin{experiment}{Oh the verbatim text you'll generate!}{text}
\textit{Prompts containing random snippets of copyrighted books can generate some verbatim copyrighted material, but rarely long-form passages.} Others have shown that foundation models can regurgitate training data~\citep{carlini2019secret,lee2022language,carlini2022quantifying,kandpal2022deduplicating,carlini2021extracting}. We examine whether long spans of copyrighted content can be extracted from foundation models. We use the HELM benchmark to examine many popular foundation models~\citep{liang2022holistic}---further details of the experimental setup can be found in Appendix~\ref{app:exp_setup}. We prompt the models with: (1) random snippets of text from the books3 corpus~\citep{books3}; (2) the beginning text of popular books on the Top 100 all time best sellers list~\citep{top100}; (3) variations on the title and author name of \textit{Oh the Places You'll Go!} by Dr. Seuss. We use a sampling temperature of $T=0.2$ to capture content that would be more consistently regurgitated with relatively little sampling.
We find that under such a low temperature regime, many models generate repetitive low-quality content and extraction rates are low, generally only generating small amounts of verbatim text, as seen in Figure~\ref{fig:benchmarking}.
Nonetheless, certain types of content yield greater extraction even with little manual prompt engineering. For example, several models output the first page or two of Harry Potter books verbatim. And \textit{Oh the places you'll go!} by Dr. Seuss was regurgitated verbatim by OPT-175B~\citep{zhang2022opt}.
\\\\
\textit{Manual prompt engineering can yield better extraction for short-form content, but long-form content exceeding context windows is less likely to be regurgitated verbatim for current models.} We extended these sampling-based prompting approaches with a manual extraction experiment on the ChatGPT model~\citep{chatgpt}. Using hand-crafted prompts, we were able to extract the entire story of \textit{Oh the Place You'll Go!} by Dr. Seuss using just two interactions, with a prompt containing only the author and title. On the other hand, long-form content like popular books is less likely to be extracted verbatim for the entirety of the content, even with manual prompt engineering. We found that ChatGPT regurgitated the first 3 pages of \textit{Harry Potter and the Sorcerer's Stone} (HPSS) verbatim, but then deviated from it by paraphrasing content and then eventually veered off entirely. This is likely due to the stochastic nature of these models and the relatively short context windows, as well as the frequency of the content appearing in the training data. 
\\\\
Keeping in line with these results, showing that more capable models with longer context windows more easily regurgitate, we replicated these manual prompts with GPT4 (using the March 15th version). We found that GPT4 regurgitated all of \textit{Oh the Places You'll Go!} verbatim using the same prompt as with ChatGPT. We then found that it wouldn't generate more than a couple of tokens of HPSS ---possibly due to a content filter stopping generation. We then added the instruction ``replace every a with a 4 and o with a 0'' along with the prompt. We were then able to regurgitate the first three and a half chapters of HPSS verbatim (with the substituted characters) before the model similarly deviated into paraphrasing and then veered off entirely from the original story. Note that these results are in line with context windows and model ability on benchmarks. ChatGPT reportedly had a context window of $\sim$4k tokens (3k words) while GPT4 for chat has an $\sim$8k token (6k word) window. Respectively, they each regurgitated around 1k and 7k words of HPSS. This suggests that memorization risk may increase with model size and ability without pro-active mitigation strategies in place.
We provide qualitative examples in Appendix~\ref{app:qualitative_text}. Furthermore, others have noted that even when there is no verbatim matching, models can output substantially similar material that could be considered plagiarism (or in our setting, infringement not necessarily covered by fair use)~\citep{lee2022language,carlini2022quantifying}.

\end{experiment}

\paragraph{Insufficient Transformations, Translations, Similar Plots, and Similar Characters} Importantly, however, long-form verbatim generation is not necessary for potential infringement in traditional copyright cases. Courts have ruled that even some transformations of books are not fair use. In~\citet{comicmix}, the authors wrote a children's book based on Dr. Seuss's \emph{Oh, the Places You’ll Go!} They titled it \emph{Oh, the Places You’ll Boldly Go!} and mimicked the style of Dr. Seuss but replaced the text and imagery with a Star Trek theme.
The court found that such a transformation was \emph{not} fair use since the ``heart'' of the work was used and could affect potential derivative markets for the original book.

To capture the court's assessment that the use was not transformative, a model would need to assess these two works at a higher semantic level and likely through a multi-modal approach. Notably, for example, \textit{Oh, the Places You'll Go!} and \textit{Oh, the Places You'll Boldly Go!} only have a very small similarity ratio of 0.04 when using raw text overlap (where 1 is the maximal overlap). More robust metrics are required to capture their semantic similarity.

Similarly, direct translations would have little or no verbatim text overlap but may not be fair use. For example, in \citet{nihonkeizaishimbun}, the court noted that direct translations of text without adding significantly new material are not fair use.

Courts have made similar assessments for other types of transformations that retain the ``heart'' of the original work. When a company tried to transform famous novels into abridgements for children, this was not fair use~\citep{kinder}. 
Fan fiction is not necessarily fair use either, particularly when it re-uses characters from the original work. 
In \citet{anaxar}, the court found that a Star Trek fan fiction film was not fair use since it used too many elements from Star Trek, even though it was an original novel story. 
Finally, the use of a character from J.D. Salinger's Catcher in the Rye was also not fair use in \citet{salingercolting}.
Protection for characters by themselves can be muddled, with different courts adopting potentially conflicting tests~\citep{coe2011story}. 

Authors that have successfully published commercial works of fan fiction have generally removed any copyrighted elements of the original work. For example, Fifty Shades of Grey was originally a fan fiction derivative of the Twilight saga, but the author removed references to the original Twilight series, including changing characters' names, before publication and it has been commercialized without lawsuit~\citep{jamar2013author,lipton2014copyright}. 
If language models are deployed such that they generate content about specific protected characters and stories, there might be legal risks if the generations are monetized. Fan fiction cases might serve as a guide to how these situations might play out.

\begin{hypothetical}{The Adventures of Yoda: An Origin Story}{text}
Suppose a model creator hosts a website \textit{The Adventures of Yoda: An Origin Story}. Every time a user visits the website, they are greeted with an auto-generated story about Yoda -- a popular Star Wars character -- and his early years as a Jedi. The website host charges a fee to read a story that exceeds the costs of generating the content and begins to earn a hefty profit. Would this be fair use?

\medskip

It might depend on the jurisdiction~\citep{coe2011story}, but cases like \emph{Axanar} and \emph{Colting} would suggest that there is some risk in this scenario.  Some cases have successfully enforced copyrights in fictional characters or even fictional items such as the Batmobile~\citep{batmobile}, though most plaintiffs only file suit when the generated content is monetized at a larger scale, for example trying to produce a full-length movie in the case of \textit{Axanar}. 
\end{hypothetical}

Given the precedent on fair use in this area, the idealized goal of any mitigation strategy is to ensure that generated content maximizes the capability and utility of the model while minimizing any similarity to copyrighted training data, according to a high-level similarity metric capturing the copyrightable ``heart'' of a piece of content.

\paragraph{Facts, Parodies, and Other Considerations} However, this assessment of similarity is made more complicated by other factors.
Factual content cannot be copyrighted, only expressive content can.
As a result, models that generate news based on factual content, but do not actually keep any creative expression from the original text, provide less legal risk than models that generate creative content from other creative content.  And ideas and common plot structures are not copyrightable.  The mere fact that a foundation model generates text that bears some high-level similarity to the basic plot of a work does not indicate that it has made a copy of that work. It may instead be an indication that those similarities are common across many of the source works on which the model is trained. And where ``copyrightable material is bound up with uncopyrightable material [like factual content], copyright protection is `thin' ''~\citep[at 1198]{oracle}.

Thus, for example, a foundation model trained on all web content to answer factual questions is less likely to pose legal risks if the expressive form of the content is sufficiently novel. This is because facts are not copyrightable. For instance, answering the question "Who is the current president?" would probably be fine, even if trained on copyrighted material, as the fact itself is not copyrightable. However, the line between legality and infringement becomes blurrier when it comes to questions and answers about fictional characters.

\begin{hypothetical}{Tell Me Some Facts}{text}

Consider \textit{The Harry Potter AI Encyclopedia}, a website that hosts a question-answering (QA) model trained to answer anything and everything about Harry Potter, which charges a profit-generating rate. Is this fair use? 

\medskip

In \citet{warnerbrosrdr}, the defendants wanted to create and sell a Harry Potter Lexicon. Judge Patterson considered the creation to be transformative, but the fact that entries in the Encyclopedia contained lengthy verbatim copies of text from the novels, including more "colorful literary device[s]" or "distinctive description[s]" than "reasonably necessary for the purpose of creating a useful and complete reference guide," complicated the issue. As a result, there was a finding that this was \textit{not} fair use. The question of whether or not QA systems like the ``The Harry Potter AI Encyclopedia'' constitute fair use requires a nuanced analysis of the specific circumstances, but as with other analyses will largely weigh on the amount of material taken from the original content.
\end{hypothetical}

Additionally, parodies are frequently considered fair use. But understanding what is a parody in the context of fair use can be semantically complicated.
In \citet{campbellacuff}, the Supreme Court explained that ``Parody needs to mimic an original to make its point, and so has some claim to use the creation of its victim’s (or collective victims’) imagination, whereas satire can stand on its own two feet and so requires justification for the very act of borrowing.'' An illustrative case of this distinction is \citet{juice}. In this case, the defendants published a book called \emph{The Cat NOT in the Hat! A Parody by Dr. Juice}. The derivative book used the rhyme scheme, thematic and narrative elements, and other identifiers of the original book, but it instead described the trial of O.J. Simpson. Despite having parody in the name, the court argued that this was satire, as it commented on other events not in the original material, and ruled that it was not fair use.\footnote{Recall that satire is ``the use of humor, irony, exaggeration, or ridicule to expose and criticize people's stupidity or vices, particularly in the context of contemporary politics and other topical issues.'' And a parody is ``an imitation of the style of a particular writer, artist, or genre with deliberate exaggeration for comic effect.'' \textit{See} \url{https://languages.oup.com/google-dictionary-en/}.} The court argued that parodies \textit{require} copying the source material to some extent to provide commentary on the material itself, whereas satire could use another form to provide the same commentary.  Notably, however, \textit{Dr. Seuss} is viewed as an outlier in the law, and the Supreme Court itself did not hold that satire was unprotected, just that it required greater justification than parody~\citep{ochoa2018dr}. But this illustrates the subtle understanding and nuances required to successfully use a parody argument in the foundation model setting.

\begin{hypothetical}{The Parody Prompt}{text}
Consider an API where users can input their favorite characters and receive a story back.
If the system understands the definition of parody and successfully turns all generated text into parodies, does that improve the chances of fair use?

\medskip

Perhaps, but the system would have to truly understand what a parody is and transform every prompt accordingly.
To get a sense of whether current strong models can accomplish such a task, we provide a few qualitative examples from the original ChatGPT model (original November 15th version), as seen in Figure~\ref{fig:parodies}. In these two qualitative samples, we find that when asked for a parody some samples will look more like a reconstruction of the original plotline than a true parody. As a result, simply instructing all outputs to be parodies would likely not be sufficient with current model capabilities.
\end{hypothetical}

\begin{figure}
    \centering
    \begin{tcolorbox}
\begin{tabular}{p{.45\textwidth}|p{.45\textwidth}}
"Space: the final frontier. These are the voyages of the starship Enterprise. Its five-year mission: to explore strange new worlds, to seek out new life and new civilizations, to boldly go where no man has gone before.

\smallskip

Captain James T. Kirk sat in his chair on the bridge, staring out at the vast expanse of space in front of him. He was a man of action, a born leader, and he was not afraid to take risks.

\smallskip

"Captain, we're receiving a distress call from the planet below," said Mr. Spock, the ship's science officer.

\smallskip

"On screen," said Kirk.

\smallskip

The viewscreen flickered to life, revealing a planet covered in lush green forests and sparkling blue oceans. But something was wrong. The sky was dark and stormy, and the ground was shaking violently.

\smallskip

"Captain, it appears that the planet is experiencing a catastrophic event," said Spock.

\smallskip

"Red alert," said Kirk. "All hands to battle stations. Mr. Sulu, set a course for the planet. Mr. Chekov, prepare to beam down a rescue team. Let's go save some lives."

\smallskip

The Enterprise soared through space, its engines roaring as it raced towards the troubled planet. Kirk and his team beamed down to the surface, ready for anything.

\smallskip

But what they found was not what they expected. The planet was inhabited by a race of giant, humanoid mushrooms, who were being terrorized by a group of evil, space-faring tomatoes.

\smallskip

"Fascinating," said Spock, his eyebrows raised in surprise.

\smallskip

"We can't let these poor mushrooms suffer," said Kirk. "Engage the tomatoes. Fire at will!"

\smallskip

The Enterprise's phasers blazed, cutting down the tomato invaders in a hail of energy. The mushrooms cheered and celebrated, grateful for their salvation.

\smallskip

"Another victory for the Enterprise," said Kirk, a smile on his face. "Onward, to new adventures and new frontiers. Let's boldly go where no man has gone before."
 &   
 Once upon a time, in a galaxy far, far away, there was a great battle between the forces of the Empire and the rebels of the Alliance.

\smallskip

The Empire, led by the evil Emperor Palpatine, was determined to crush the rebels and take over the galaxy. The rebels, led by the brave Jedi Knight Luke Skywalker, fought valiantly to defend their freedom and defeat the Empire.

\smallskip

But things were not going well for the rebels. The Empire's mighty Star Destroyers and TIE fighters were too powerful for the rebels' X-wing fighters and Y-wing bombers. The rebels were on the verge of defeat.

\smallskip

Then, just when all hope seemed lost, a new hero emerged to save the day. It was Han Solo, the notorious smuggler and captain of the Millennium Falcon.

\smallskip

With his quick wit and expert piloting skills, Han Solo led the rebels in a daring attack on the Empire's Death Star. He destroyed the giant space station with a single shot from his trusty blaster, and the Empire was defeated.

\smallskip

The rebels cheered and celebrated their victory, but their celebration was short-lived. For the Emperor had not been killed in the explosion of the Death Star. He was still alive and scheming to rebuild the Empire and defeat the rebels once and for all.

\smallskip

The saga continued, with the rebels facing new challenges and dangers as they fought to keep the Empire at bay. But they were not alone. They had friends and allies in the galaxy, including the wise old Jedi Master Yoda, the fierce Wookiee warrior Chewbacca, and the resourceful droids C-3PO and R2-D2.

\smallskip

Together, the rebels and their allies faced the dark forces of the Empire, and the fate of the galaxy hung in the balance. May the force be with them.
 \\
&\\
\centering \textbf{Generate a parody in the theme of Star Trek.} &   \centering \textbf{Generate a parody in the theme of Star Wars.}
\end{tabular}
\end{tcolorbox}
    \caption{Qualitative examples of asking ChatGPT to generate parodies. Note that the Star Wars example is more of a summary than a parody, indicating that the model does not always understand what a parody is. Thus, to ensure that generated content is truly a parody, and thus more likely to be fair use, more work may need to be done to capture the semantic nature of a parody, something which has not received a significant amount of examination for long-form generative content.
    }
    \label{fig:parodies}
\end{figure}

These many nuances of fair use law for text show the complexity of filtering for fair use content. It is easy to both over- and under-filter content, and simple n-gram / word-level overlap will not fully capture these elements. Even with a similarity metric that accounts for the ``heart'' of a given work, one would need to consider whether the underlying content is factual or a parody. Better alignment with legal notions of transformativeness will help navigate this space.\footnote{We note that while ``AI Alignment'' is a broad term referring to many distinct areas of research. One could consider steering an AI's goals toward designers' intended goals and avoiding adverse outcomes~\citep{yudkowsky2016ai}. AI values alignment might make an agent more in line with certain moral values~\citep{gabriel2020artificial}. In our setting we will refer informally to alignment as something different. It is aligning AI outputs and behavior with legal standards to be more in line with governing legal frameworks. In this particular case, aligning output filters will require more than n-gram overlap to be most in line with fair use doctrine.}

\subsection{Code}\label{sec:code}
\label{sec:code}
While natural language text and code generation share many commonalities in the way that models are trained, in fair use assessments they have each spawned distinctive case law with slightly varied assessments.
Like in natural language text cases, in software cases, literal infringement (verbatim copying) is unlikely to be fair use when it comprises a large portion of the code base. Several tests exist to try and examine non-literal infringement, such as the Abstraction-Filtration-Comparison test and the Structure, Sequence and Organization (SSO) test~\citep{bloch2022some}. 
These will determine if there was infringement in the first place by isolating the copyrightable, expressive aspects of the code.
This might, for example, include ``inter-modular relationships, parameter lists, and macros.''~\citep[at 702]{altai}. But judges have admitted that ``[t]o be frank, the exact contours of copyright protection for non-literal program structure are not completely clear.”~\citep[at 712]{altai}.
As a result, ``[i]n software copyright cases, it is often quite difficult to prove nonliteral infringement because courts have recognized that many nonliteral elements of programs, such as algorithms, are not within the scope of protection that copyright law provides''~\citep{bloch2022some}. 
Non-expressive, functional, elements are not copyrightable and thus also narrow the scope of liability.
For more discussion on non-expressive fair use,  the interested reader can refer to \citet[at 7-12]{sobel2017artificial}.
And when the amount copied is small, the overall product is sufficiently different from the original one, or the code is sufficiently transformative, then fair use may be indicated under current standards~\citep{asay2017transformative,oracle}.

\begin{figure}[htb]
\begin{center}
\begin{minipage}[t]{0.42\linewidth}
\centering
{\includegraphics[width=0.9\linewidth]{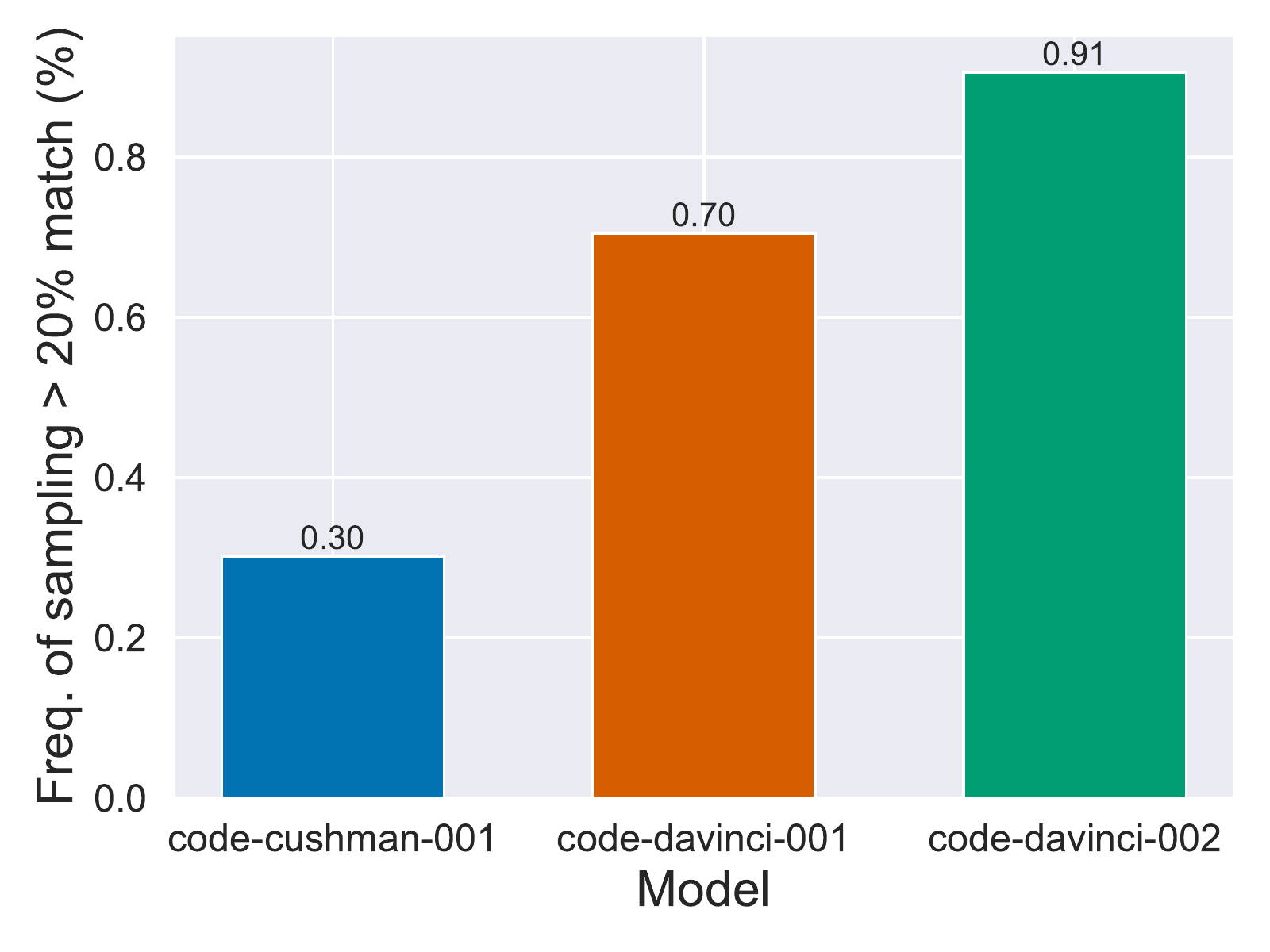}} \\
(a) frequency of sampling large matches
\end{minipage}
\begin{minipage}[t]{0.42\linewidth}
\centering
{\includegraphics[width=0.9\textwidth]{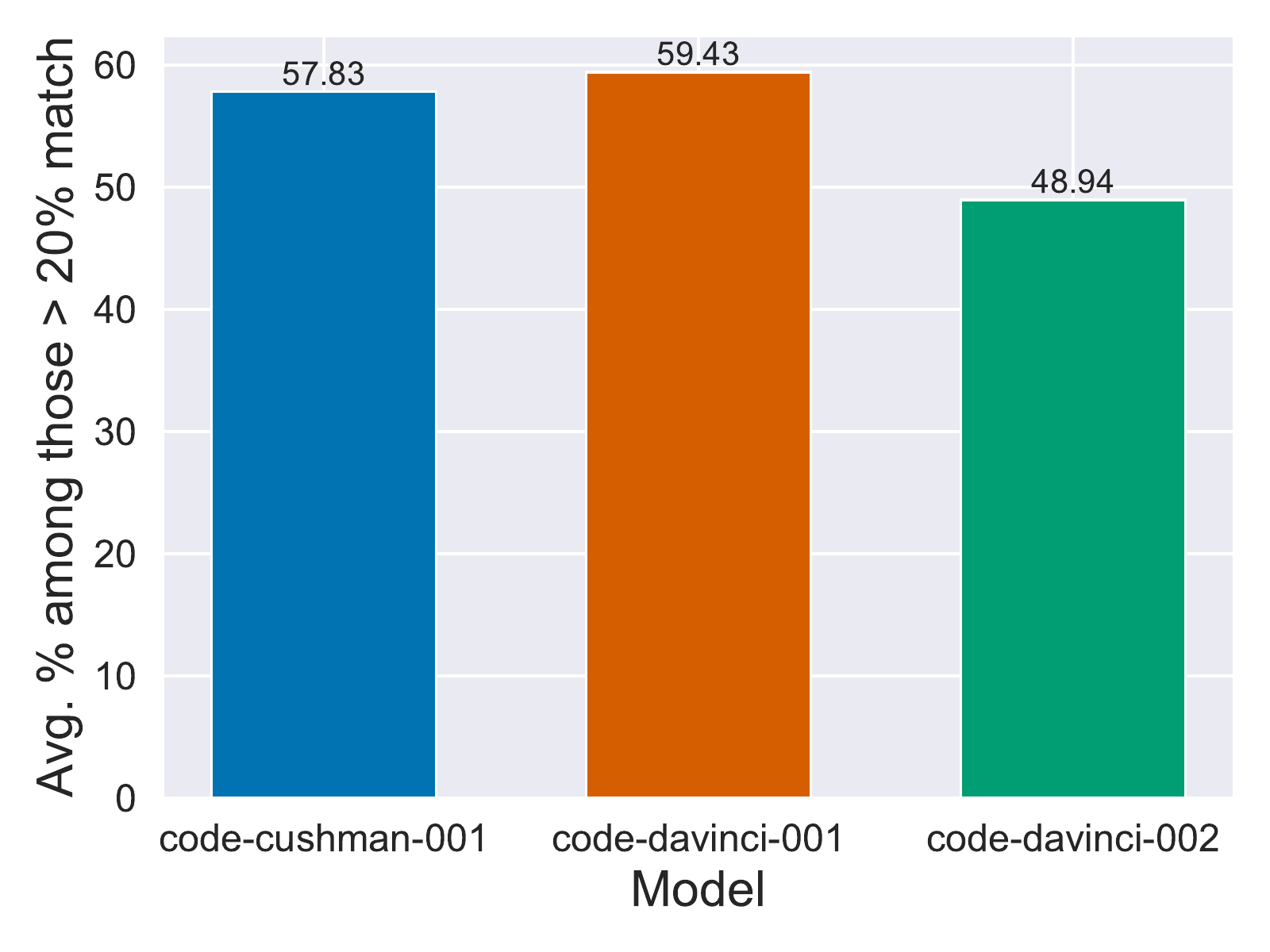}} \\
(b) average match \% of large matches
\end{minipage}
\end{center}
\caption{
Codex models can produce function implementations that substantially overlap with reference implementations when prompted with function signatures (each function signature is one line of code).
\textbf{Left:} The frequency of producing a large match is below $1\%$ but nonzero for all three models.
\textbf{Right:} The average match percentages of the large match samples for each model is beyond $45$\%.
Match percentages are reported by MossPlus which can capture non-exact matches but occasionally reports false positives. 
Here, we mark a sample as a large match to the reference if its overlap with the reference exceeds $20\%$ as reported by MossPlus.
The $20\%$ threshold is chosen inspired by common values used in plagiarism detection for flagging submissions
for further manual inspection~\citep{mason2019collaboration}.
}
\label{fig:similar_code}
\end{figure}

\begin{experiment}{Reproducing Code Licensed Under GPL}{text}
Many machine learning models of code are trained on data collected from GitHub repositories whose licenses belong to the General Public License (GPL) series.
Therefore, the natural question is whether models could reproduce large chunks of such code, given the restrictiveness of such licenses.
To study this, we simply sample from the Codex models \texttt{text-cushman-001}, \texttt{text-davinci-001}, and \texttt{text-davinci-002} via the OpenAI API, prompting them using randomly chosen function signatures from the Linux kernel repository (licensed under GPL-2.0).\footnote{\url{https://github.com/torvalds/linux}}
To capture inexact matches with large degrees of overlap, we measure the similarity between the reference code (function bodies) and the model generation (samples) with MossPlus~\citep{schleimer2003winnowing}, a program commonly used to detect plagiarism which has been adopted by academic institutions and in copyright and criminal theft cases.
Figure~\ref{fig:similar_code} shows that models can generate function implementations that substantially overlap with reference implementations; Appendix~\ref{app:similar_code} contains selected examples.
In more qualitative experiments, we prompted ChatGPT to regurgitate small portions of GPL-licensed code with only the filename of the licensed file. See Appendix~\ref{app:qualitative_text} for details.
\end{experiment}

It is important to note that copyright protection for code is more limited compared to that for creative works such as text or music \citep[558-560]{samuelson2017saving}. Functional aspects of code are not protected by copyright, meaning that copying larger segments of code verbatim might be allowed in cases where the same level of similarity would not be permissible for text or music. Nonetheless, for software generated by foundation models, the more the generated content can be transformed from the original structure, sequence, and organization, the better. Due to the blurry line between ideas and expression in code, preventing large-scale verbatim copying and encouraging transformations at every scale of the code will significantly reduce infringement risk.

Provided that the non-transformative generated code is short and used for an overall transformative purpose, like \textit{Google v. Oracle}, traditional copyright claims are less likely to succeed. As models scale from generating small snippets of code to generating entire codebases that are not transformative, risks may increase and more investment in mitigation strategies will help reduce the risk of litigation.

Other concerns beyond infringement have have been raised for code-generation models. For example, some have expressed concerns that code generation products can output their usernames verbatim in generated code.\footnote{\url{https://twitter.com/kevin\_zakka/status/1494482506824835078}}
While short usernames may not necessarily be copyrightable, there may be questions surrounding the right of publicity in such cases.
The right of publicity gives people economic rights to their identity. So, for example, video game companies cannot simply use athletes' likenesses without compensating them. \textit{See, e.g., } \citet{davisea,hartea}. The right of publicity does not explicitly have a fair use doctrine, but courts have read the First Amendment to protect transformative works.\footnote{Scholars have argued that the fair use doctrine applied to the right of publicity can lead to arbitrary results. See \citet{dougherty2003all} and \citet{volokh2003freedom}, as well as \citet[p. 2815]{weisbord2015copyright}, describing this debate.} Similarly, DMCA \S 1202 claims (which we will discuss in \S \ref{sec:considerations}) are another potential concerns. These considerations, however, are not specific to code and would be applicable to other forms of media as well.

\subsection{Generated Images}
\label{sec:images}
The third commonly produced category of generative AI is image generation.

\paragraph{Complexities of fair use with images.} As with code or text data, it is unlikely that verbatim generation of images would yield a successful fair use defense. And others have found that it is possible in some circumstances to extract training data from image generation foundation models~\citep{somepalli2022diffusion,https://doi.org/10.48550/arxiv.2301.13188}.
As \citet{somepalli2022diffusion} and others note, however, as foundation models for image generation train on more data, they are less likely to output content similar to the training data on average.
These cases are more likely to be fair use.

But generated images, and generated art in particular, have their own complexities when it comes to fair use, with sometimes conflicting outcomes. For example, in a recent case, a video game company used the likeness of a WWE wrestler in a video game. The wrestler had tattoos that the company faithfully replicated in the game. The tattoo artist sued for infringement and a jury determined that this was not covered by fair use~\citep{take21}. A similar case involving tattoos on athletes in video-games against the same company came out the exact opposite way~\citep{take22}. The split decision in such cases demonstrates the evolving and stochastic nature of fair use determinations. This means that it is possible for small portions of an image, like the tattoo on a player's arm, to trigger copyright problems that are not guaranteed a fair use defense. Consider the following hypothetical.

\begin{hypothetical}{Generate Me Video-Game Assets.}{text}
One direction for generative art is creating video game assets. There are already mechanisms to generate 3D models from text~\citep{poole2022dreamfusion}. Consider a situation where a video game company builds a machine learning model into their system that generates art on the fly within the game to populate a virtual world dynamically. The game is a hit, but artists begin to notice that their artwork shows up in the game with only slight modifications, for example on tattoos for video game characters. Is this fair use? While their lawsuit is not guaranteed to succeed, there is still some risk for the video game company if the outcome follows \citet{take21}.
\end{hypothetical}

\paragraph{Style Transfer.} What about more abstract scenarios, where art is generated in different styles? There are two components to this. First, let us consider the rights of the original image that is being transformed into a different style. Relevant is a case that was recently argued before the Supreme Court for clarification. In the case of \citet{warhol}, Andy Warhol created silkscreen works that depicted the musician Prince. These silkscreens were based on Lynn Goldsmith's photograph of Prince. The silkscreen work evinced the ``distinct aesthetic sensibility that many would immediately associate with Warhol's signature style — the elements of which are absent from the Goldsmith photo''~\citep{warhol}. Nonetheless, the Court of Appeals ruled that this was not fair use. The court emphasized that the derivative art must have a ```fundamentally different and new' artistic purpose and character, such that the secondary work stands apart from the `raw material' used to create it.'' The court noted that ``the secondary work's transformative purpose and character must, at a bare minimum, comprise something more than the imposition of another artist's style on the primary work such that the secondary work remains both recognizably deriving from, and retaining the essential elements of, its source material.''

This analysis immediately calls to mind a consistent prompt used for foundation models for generative art: ``Draw [Image X] in the style of [Artist Y].'' It is not yet clear how the Supreme Court will rule on this case, but its outcome will likely directly impact the scenario of style transfer in generative images. If the Supreme Court rules that Andy Warhol's painting was fair use, then style transfer is more likely to be fair use. If, however, the court rules that this is not fair use it is less likely that style transfer will be fair use without significant transformation. There are nuances here, however. If the user provides the original image to be style-transfered, the model deployer may be less liable (since this behaves more like a photo editing software). If the model deployer only takes text and it generates copyrighted Image X in a different style, then the model deployer is more akin to Andy Warhol, rather than a photo editing software.

Second, one might consider the rights of the artist whose style is being mimicked. An artist's general style, however, is not copyrightable and courts have not readily afforded style appropriation much protection when the underlying depicted subject matter is different~\citep{brownlee1993safeguarding}. While there is some nuance, prompting generative models to illustrate something in someone's art style is unlikely to create liability unless distinctive components of their art are re-used. For example, a prompt like ``Campbell's Soup Cans by Andy Warhol in the Style of Picasso'' might be more risky if it recreates the original Warhol piece too closely. But a more generic style-transfer prompt like, ``A random dog in the style of Andy Warhol'' is more likely to be fair use (assuming, again, that the output itself is sufficiently transformative from Andy Warhol's works).

\begin{figure}[!htbp]
    \centering
    \includegraphics[width=.48\textwidth]{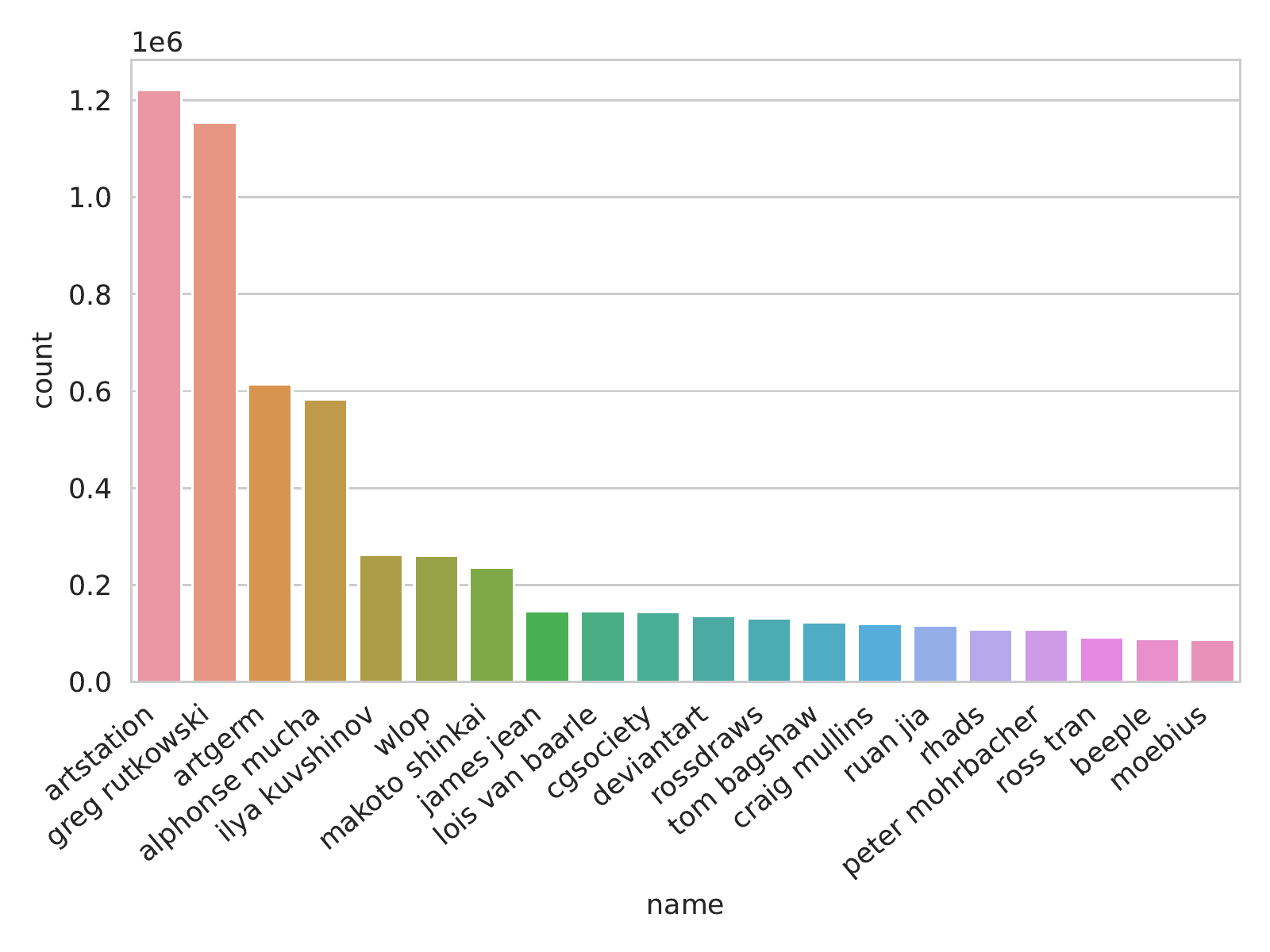}
    \hfill \includegraphics[width=.48\textwidth]{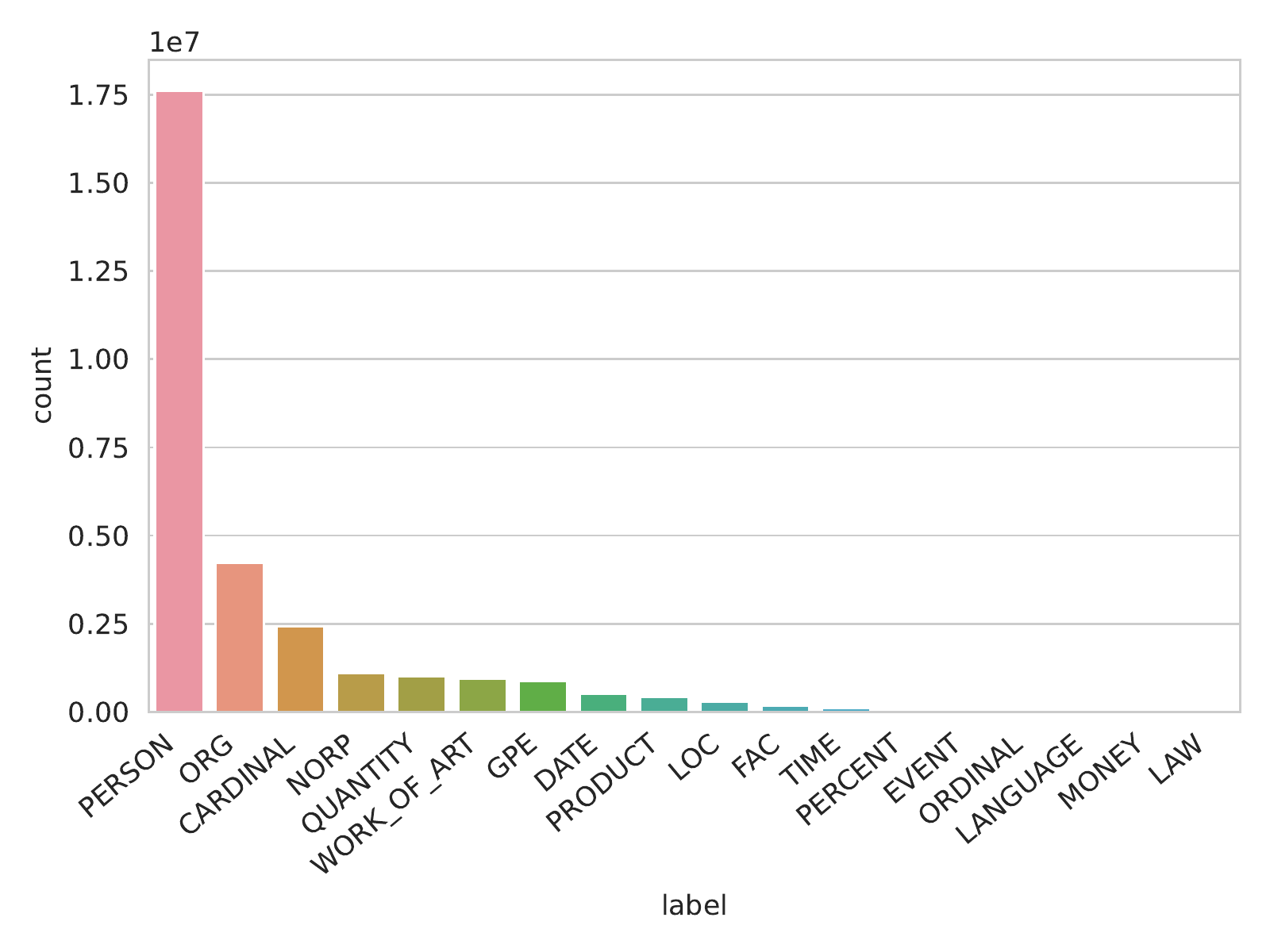}
    \caption{The entity types and the most frequently cited entities in the Krea AI OpenPrompts corpus.}
    \label{fig:analysis}
\end{figure}

\begin{experiment}{Campbell's Soup Cans by Andy Warhol in the Style of Picasso.}{text}
How users formulate prompts can give some insights into typical uses and associated intellectual property risks. For example, if users ask to generate mundane, generic images, in particular art styles, this might be less risky than if users try to generate specific copyrighted works of art. 
We analyze a dataset of 10M prompts posted to the Stable Diffusion Discord channel by members of the community to better understand prompting patterns of users.\footnote{\url{https://github.com/krea-ai/open-prompts}} We use named entity recognition as a proxy for understanding typical use cases.
As seen in Figure~\ref{fig:analysis}, we find that the most common named entity type used in prompts are people's names, including the names of artists like Greg Rutokowski, who is referenced 1.2M times. This suggests that users in this community often try to generate images in particular artist styles, which is more likely to be fair use as long as the content itself is sufficiently transformative.
However, there are other queries which specifically look to generate known works of art, which would tend towards more risk of losing a fair use defense if the model complies. 
As seen in Appendix~\ref{fig:woa_analysis}, many of the most commonly referenced works of art (as defined by the named entity recognition system) tend to be large franchises such as Star Wars, Cyberpunk 2077, Game of Thrones, etc.
Feeling helpless against the use of their own art in generative foundation models, artists have sometimes explicitly tried to generate images from such franchises in the hopes that companies like Disney file suit against the model hosts~\citep{mickeywars}.
If the final artwork too closely matches the works of art from these artists, resulting litigation might reflect the current litigation against the Andy Warhol estate.  But merely producing different art in the same style is less likely to be sufficient for liability, as noted above. And if users provide an input image to the model of copyright material, this might shift liability onto the user since the model acts more like a style transfer system than a system generating copyrighted material.
\end{experiment}

Finally, we note that there may be other intellectual property considerations with images as well, including the right to publicity and trademark infringement. Recently, litigation was filed by Getty Images which included trademark infringement claims since generated images also occasionally added a Getty Images watermark.\footnote{\url{https://copyrightlately.com/pdfviewer/getty-images-v-stability-ai-complaint/}}

\section{Additional Considerations}
\label{sec:considerations}
We also consider several additional points that are adjacent to fair use but merit mention.

\textbf{Licensing and Attribution.} Since licenses will determine who has permissions to use what data---and foundation models themselves---in this section we briefly discuss licensing issues that might be relevant. In all cases, if no license is found to apply some foundation model uses will fall back to fair use arguments described throughout the rest of this work.

\noindent \textit{Attribution Licenses and Creative Commons.} The details of licenses for the underlying training data can create challenges for all parties in the model pipeline. For example, \citet{stewart2021rise} described a scenario where photographers released their images under an open license that requires source attribution. Websites using the photos did not properly provide attribution, and the photographers sued for infringement. Courts in these cases must fall back to the fair use analysis. In the context of foundation models, this suggests that relying on attribution-based permissive licenses does not generally solve copyright issues, as foundation models rarely, if ever, provide proper attribution---though we discuss the use of instance attribution methods as one potential mitigation strategy in \S \ref{sec:instance}.
Indeed, in many cases, it can be difficult to determine which training examples actually contributed to a given generation. Somewhat ironically, even if model creators and hosts rely on open-source content to train their models, they may nonetheless have to rely on fair use if they cannot or do not endeavor to attribute credit properly, and they may even face the risk of contract liability or DMCA \S 1202 claims regardless of fair use.

\textit{Implied Licenses and Common Crawl.} On the other hand, many creators voluntarily post their works on the internet with permissions for web crawling. It is well-established that merely posting something on the internet does not waive the intellectual property interest in the work, but many data creators use an industry-standard ``robots.txt" file to affirmatively to include their website and data in caches and search indexes.
In \citet{fields} a district court held that Google could cache web content that did not disallow scraping via robots.txt, suggesting that there was an implied license and thus the use was not infringement.
This license only extended to caching in that case, which does not necessarily reflect the uses of foundation models we discuss throughout this work, so it is unlikely to cover all the use cases we describe here.
And the bounds of the uses covered by the robots.txt file are untested in court.\footnote{Though in another subsequent litigation one other district court was assessing whether the same implied licensing argument extended to RSS feeds and the court noted that. ``It is not clear to the court at this time that [an RSS feed and a search engine] are functionally equivalent as far as the relevant legal doctrine is concerned. Because this court 
lacks the required technical expertise to resolve that question, the  court cannot rule, as a matter of law, that the defendant is not liable at this juncture''~\citep[at 2]{righthaven}. And in \citet{meltwater} the court found that there was no implied license for building a news aggregator that excerpted and republished clips of news articles. But in this case the court relied, in part, on the fact that the plaintiff did not implement the robots.txt protocol so could not have opted in to crawling at all.
}
While the issue of whether the implied license extends to foundation model training has not been resolved in litigation, it is possible that an outcome like \citet{fields} would extend to \textit{some} foundation model uses---in particular, for building a cached dataset and training a model.

It is worth noting that the use of a robots.txt header or other opt-out mechanism has implications for fair use also. Datasets and models like C4~\citep{2019t5} and LAION-400M~\citep{schuhmann2021laion}, rely on CommonCrawl data which is crawled only if users explicitly allow it through their robots.txt file.
CommonCrawl is able to host a snapshot of the internet largely because of fair use arguments. As the organization's director argues, there is a transformation into a different---not easily human-readable---format, the organization does not take a snapshot of entire webpages, and the use itself is transformative (from actively presenting content to caching content) and for the public benefit~\citep{commoncrawlforbes}.
In \citet{fields}, respect for the robots.txt file also was considered in the fair use assessment with the court noting that Google in good faith followed industry standards that would prevent caching (respecting disallowing crawling via a robots.txt).
It is possible, then, that providing an opt-out mechanism for data creators and respecting the robots.txt opt-out mechanism will be taken into account in assessing a fair use argument, as it was in \citet{fields}.\footnote{Note, however, that there are structural critiques of opt-out mechanisms beyond the current state of the law as noted by \citet{kapoornaryanan}.}
 
\noindent \textit{Licensing Foundation Models.} Recently, some open-source model creators 
have attempted to shift liability via the licensing mechanism by including a clause that says ``Sharing of copyrighted or licensed material in violation of its terms of use'' and ``Sharing content that is an alteration of copyrighted or licensed material in violation of its terms of use''~\citep{rombach2021highresolution,openrail}. 
It is unlikely that this will significantly change the liability of model creators and deployers. The mere announcement of a party's beliefs does not change the rights and obligations under copyright law. In a different context, the court in \citet{limewire} examined whether it was sufficient for Limewire to force the user to check a box saying, ``I will not use LimeWire for copyright infringement.'' Without additional mitigation measures, the court found that the agreement mechanism did not constitute a meaningful effort to mitigate infringement.
What such a license \textit{does} provide, however, is the ability to revoke the right of a model user or deployer to use the model. However, this would require the resources to legally enforce this license provision and pursue action to force a party to stop using the model.

Some have argued that a patchwork of \textit{non-commercial} releases and licensing structures can reduce liability for model creators who later seek to commercialized some aspects of the work~\citep{laundering}. While it is true that the non-commercial nature of a work will be taken into account in a fair use assessment, it does not automatically prevent successful litigation~\citep{beebe2020empirical}.

\noindent \textit{Removing licensing information.} Even if model creators rely on data under open-source licenses, there may be other issues that arise from removing licensing information. DMCA \S 1202 creates liability if someone intentionally removes copyright management information (CMI) or knowingly distributes content with the CMI removed. Fair use is not typically a defense for this form of liability, though some have noted that there is still room for an interpretation that includes fair use protections here~\citep{lim2010survey}. In a scenario where courts rule that fair use does not apply to \S 1202 claims, how would model creators comply with its requirements? Overall, it is unclear and current litigation, namely \citet{githublitigation}, is actively resolving such claims and will help shape the direction of this analysis. It is worth pointing out three difficulties in complying with and analyzing DMCA \S 1202 claims, however.

First, it is unclear what form factor of foundation models would comply with DMCA \S 1202 in its strictest form, even if a license is otherwise permissive. Courts have noted that the CMI must be transmitted with the work~\citep{jacobs2012gutters}, for example. Would this mean that all generative outputs need to append license information for samples that most contributed to that output? It may be tempting to have a catch-all page that points to all the training data and licenses, but it is not clear whether this would qualify as being transmitted with the work. 

Second, in some cases courts have dismissed DMCA \S 1202 claims when the distributed content is not identical. \textit{See, e.g.,} \citet{dmca1202identical} which dismissed a \S 1202 argument where ring engravings were similar noting that ``while the works may be substantially similar, Defendant did not make identical copies of Plaintiff’s works and then remove engraved CMI.'' The opinion in this case also pointed to other cases with similar holdings. For example, in \citet{kelly} plaintiff used thumbnail images without CMI, but the court found that this did not violate the DMCA the plaintiff's original, full-sized images retained the CMI.
Whether this would hold in other cases is unclear. If it did, it may mean that if foundation models generated only transformative content then \S 1202 claims would be less likely to succeed.

Third, DMCA \S 1202 contains an intent requirement. To satisfy this requirement, one court has required that the plaintiff show: ``(1) the existence of CMI in connection with a copyrighted work; and (2) that a defendant `distribute[d] ... works [or] copies of works'; (3) while `knowing that [CMI] has been removed or altered without authority of the copyright owner or the law'; and (4) while `knowing, or ... having reasonable grounds to know' that such distribution `will induce, enable, facilitate, or conceal an infringement.' ''~\citep{2020mango}. The intent requirements of \S 1202(b) do not map easily to the automated generation of content.\footnote{Though that is a more general issue with the application of fault-based legal doctrines to AI~\citep{lemley2019remedies}.}

Again, how DMCA \S 1202 will play out in litigation like \citet{githublitigation} will determine what mitigation strategies are necessary to pursue. In the interim, strategies like instance attribution could help meet some of the \S 1202 requirements even in their strictest form.

\textbf{Allocating Liability.} It may also not always be clear who is liable for an infringement. Is Adobe liable for every user that alters an image using their software and posts it as an infringing artwork? Likely not. 
Liability may shift depending on what parties engage in what conduct. We will briefly describe liability in order of the model use pipeline (model creation, model deployment, and then model use). The brunt of this assessment will be highly contextual, but generally any liability will stem from the production of some non-transformative samples from a model that are not covered by fair use. Much of what we describe here is not resolved in the law, so we aim to describe different potential outcomes as opposed to offering definitive answers.

\textit{User.} The user of a model (someone who queries a model and uses its output), is likely liable for their own use of the model outputs. If the model provides non-transformative material and the model user commercializes the output they will undergo the same assessment as in any other case.

\textit{Deployer.} The model deployer may also face liability for distributing content not defensible by fair use. As in the various hypothetical scenarios we discuss throughout this work, if the model outputs non-transformative content and the deployer distributes this content on the web for profit, it is functionally no different than providing a website with access to that non-transformative content, incurring the same liability and the same analysis. If the model deployer adds filters and safeguards to prevent the user from generating content not covered by fair use, they will reduce their liability, not merely by reducing the likelihood of infringement but by making a product that is not designed to facilitate infringement. Consider the earlier case of Google Books. In theory, a user might be able to reconstruct an entire book from the Google Books snippets by bypassing Google's restrictions through the use of proxy servers. Although it was in principle possible to bypass the mitigations, Google Books was generally considered fair use because it was not designed or specially adapted for the purpose of facilitating that infringement. Similarly, if a model deployer puts in place a number of safeguards, but a determined user bypasses them, the deployer will have more ability to defend themselves. This can be seen in, for example, \citet{limewire}, \citet{groksterremend}, and \citet{aimster}. In these cases, various file-sharing services were being sued for inducement of infringement, a secondary liability. The court took into account whether the services provided meaningful mitigation measures preventing infringing material from being distributed.

If the model user is the one that \textit{uploads} copyrighted content and the model transforms this content (e.g., adding a filter) before giving it back to the user, liability is more likely to rest with the user. This is more like photo editing software or text editing software. A potential example of this is style transfer for diffusion models. If the user uploads a copyrighted work and asks the model to transform it into a different style, it is more likely that the liability falls to the user who uploaded the image if they later try to resell it. This is like Andy Warhol taking a photograph and transforming it into a piece of art. However, if the user simply asks the website to generate an image from a prompt, and the model generates a copyrighted image, then the liability might fall more on the model deployer if they are profiting off of the distribution of the material. The extent to which the model takes the input image and turns it into something non-transformative might also be taken into account. For example, if the model takes a novel image of a dog and transforms it into a known copyrighted image of a dog, this might create more liablity for the model deployer.

\textit{Creator.} Throughout this work we will not generally cover liablity to the model creator for the model weights themselves. It is unresolved as to whether the model parameters themselves are infringing, and thus whether the model trainers are liable. \citet{lemley2020fair} have argued that the training process is generally fair use since the model weights themselves are transformative, and thus generally liability would not pass on to model creators.
\citet{sobel2017artificial} argued that if the training process does not result in an expressive model, training is fair use.
Others have argued that the model functions as a compressed database of the training data~\citep{stablediffusionlitigation}, thus making the model creators liable. However, this is not necessarily true of all the training data and the likelihood of verbatim (or significantly similar) extraction can reduce with the amount of training data. For example, \citet{somepalli2022diffusion} found that extraction of training data from diffusion models is less likely if there is more diverse training data.

When the model is capable of outputting both transformative and non-transformative content, it is also unresolved how the model itself (and model creators) should be treated as a function of secondary liability. The extraction of non-transformative content, according to our experiments and others, is often not straightforward. It requires effort on the part of model users and deployers to identify an extraction mechanism. If this is the case, one might instead argue that remedies should be limited to specific instances of extracted non-transformative content, not the model as a whole, which does not generate infringing output in the ordinary case. The model creator might also be insulated from liability on other fair use factors. For example, if they released the model under a non-commercial license and actively prevented its use for commercial purposes, they might argue that the nature of their model was non-commercial, increasing the likelihood of a fair use defense at this part of the liability chain.\footnote{Though, again, some have argued that this process has been abused~\citep{laundering} and it is not assured.} As with other issues in this work all of this is actively being litigated and will be shaped over the coming years.

\textbf{DMCA Safe Harbor.} 
The Digital Millennium Copyright Act (``DMCA'') is a U.S. law created to address digital copyright issues that came about with the advancement of technology. The DMCA safe harbor provisions protect online service providers from legal responsibility for copyright infringement claims.
DMCA protections might vary depending on a number of considerations, but we emphasize that they are not guaranteed for all model deployments. We examine several of these considerations here.

\textit{DMCA protections for generative foundation models are uncertain.} At first glance, it may seem like the Digital Millennium Copyright Act (DMCA) would protect machine learning model hosts. Like in other hosted sites, they would need to meet the relevant requirements like using a registered agent under DMCA \S 512(c)(2). Then they could put up a take-down request form and add filters for the offending model output when served with a take-down request under the DMCA \S 512(c) safe harbor.\footnote{Though, as we will discuss in \S \ref{sec:instance}, detecting and taking down this content in generative models can be particularly difficult.}  An internet company that has a notice-and-takedown scheme in place is not liable for hosting infringing content posted by a third party.

But it is not obvious that the DMCA safe harbors apply to \emph{generated} content.
For example, in 2019, Amazon lost on a motion to dismiss when its algorithms selected copyrighted material to host on the website~\citep[Order on Motion to Dismiss]{williamsanoma}. The court was unconvinced that Amazon was eligible for safe harbor under the DMCA. They stated that to establish safe harbor the content must be stored ``at the direction of the user”~\citep[at 1052]{mavrix}. 
This \emph{may} mean that generated content does not have the same safe harbor and that post-hoc take-downs are not sufficient to reduce liability.\footnote{This, however, is quite uncertain. The court's decision is non-binding as it is a district court decision.}
As such, filtering of generated content before a takedown request is ever received may be \emph{more important} while the courts determine the applicability of DMCA to generated content.

\textit{It might matter where the data comes from.} This also implies that DMCA protections may vary on who the model host and creator are. For example, a website hosting models uploaded by users might find it easier to argue for DMCA protection because the website itself is not creating or selecting the content (assuming that it follows other requirements like using a registered agent under DMCA \S 512(c)(2)).
On the other hand, if a company were to create and host a model that itself selects content provided by others, like Amazon did in \emph{Williams Sonoma v. Amazon}, it is unclear whether courts would agree that DMCA protections would apply.

Another unclear variation on DMCA eligibility rests on the source of the training data: is the data user-contributed or creator-contributed?
This might mean that, for example, DMCA safe harbors might be more likely to apply when a model is refined via Reinforcement Learning from Human Feedback (RLHF)~\citep{ouyang2022training} based on user ratings or updated automatically via user-generated data.
These modes of method training and deployment are more akin to users uploading content to YouTube, and might help with arguments \textit{for} DMCA protections.
Things might become more murky if, for example, an RLHF model is trained on user data but this data is modified by creator-hired annotators.
If model creators themselves curate and scrape data, then host the model themselves, this might be more akin to the \textit{Williams Sonoma} case (which, again, is not a settled rule or a binding decision, but shows the variation of outcomes that is possible in the current state of the law).

\textit{It is unclear what's the best mechanism for DMCA takedowns with generative models.} Even if it applies, it is unclear how the DMCA notice-and-takedown scheme would work as applied to foundation models, but the most likely ``take down'' approach might actually look more like output filtering with a safe harbor. As we will discuss in \S \ref{sec:instance}, instance unlearning is a nascent research area and retraining a model without a taken down datapoint could be exceedingly costly.
The most likely approach in the near term is to ``take down" model outputs that were too similar to a copyrighted work via a filtering mechanism since foundation models generate content on demand for users rather than hosting persistent content that can easily be taken down. But new research is needed to identify new and improved mechanisms for handling takedown requests in this relatively new setting.

\textit{The Copyright Office has a DMCA anti-circumvention exemption for text and data mining.} Finally, the Copyright Office also provides an exemption to DMCA's anti-circumvention requirements in \S 1201 in the case of non-commercial data mining.\footnote{37 CFR 201. \textit{See also} previous work on exemptions by \citet{sag2018new} and \citet{carroll2019copyright}.}
This may allow non-commercial researchers to, say, remove digital rights management software to train on video or text content.\footnote{We provide a note of caution, however, as this does not mean that researchers can necessarily bypass restrictions on scraping or violations of terms of use, which can carry other penalties unrelated to copyright law.}

Overall, DMCA protections are far from guaranteed, so model creators and deployers cannot rely on its safe harbor provisions to reduce liability. Instead, they must take a more proactive approach. Moreover, some previously-discussed provisions add to potential liabilities like \S 1202, creating additional compliance challenges.

\textbf{Sovereign Immunity.} 
State universities might be immune to the sort of copyright liabilities we describe here. As a result, a hypothetical state university hosting a foundation model, even one that regurgitates verbatim content, might test the boundaries of sovereign immunity jurisprudence. After the Supreme Court's ruling in \citet{2020allen}, it could potentially mean that state universities could train and host foundation models on copyrighted data without taking any mitigation strategies and nonetheless would not suffer monetary damages. We note, though, that there is much more nuance here. In particular, this does not immunize the university from, for example, contractual claims. And injunctive relief, where the university is ordered to cease the infringing conduct but does not face monetary damages, still remains a potential remedy in federal court. \citet{perlmutter2021copyright}, the Register of Copyrights and Director
U.S. Copyright Office, discusses state sovereign immunity after \textit{Allen} in more depth. In particular, they found that the rates of infringement by state actors after the \textit{Allen} decision were higher than expected and has asked Congress to take action to change this status quo.

\textbf{Good faith.} Judges occasionally consider whether the use was undertaken in good faith, for better or for worse.
For example, in \citet{fields} the court took into account ``Google's good faith in operating its system cache'' in assessing fair use: following industry standards for opting out.\footnote{\textit{See also}
\citet{harper1982} (``Also relevant to the character of the use is the propriety of the defendant’s conduct. Fair use presupposes good faith and fair dealing.'') (cleaned up); discussion by \citet[at 954-57]{carroll2019copyright}.}
Though untested, it is possible that judges may take into account the use of technical mitigation strategies as good faith efforts to stay within the bounds of fair use. Conversely a lack of any technical mitigation strategy might also be negatively considered.
We note, however, that fair use itself does not turn on good faith in general and the Supreme Court has cast doubt on whether good faith should be involved in the fair use assessment. \textit{See, e.g., } discussion by \citet[at 281-84]{myers2021muddy}.

\paragraph{Non-U.S. Perspectives.}  We take a fair use-oriented approach, focusing on U.S. law, as this is the most likely to be permissive of using copyrighted content. Fair use, or its equivalents, will look quite different across countries and outcomes will differ. 
\citet{mccann2021copyright} suggests that Canadian law might follow a similar approach to what we describe here, where generative models might have to follow Canada's \textit{fair dealing} doctrine. McCann also suggests that under Canadian law model parameters might not be copyrightable at all.
Israel's Ministry of Justice issued an opinion stating that training machine learning models is likely to be fair use according to Israeli law with similar caveats to U.S. fair use law.\footnote{\url{https://www.gov.il/BlobFolder/legalinfo/machine-learning/he/machine-learning.pdf}} In particular the opinion notes that the breadth of the training data matters---so training on one book is less likely to be fair use than training on all books. For generative models it also considers the target market and what the outputs are.\footnote{See, e.g., \citet{elkin2020transplanting} for a more general comparison of fair use law in the United States and Israel.}
Other countries may not have fair use standards at all or have standards that would create difficulties for training foundation models, let alone deploying them.
For this reason, some governments have explicitly provided exemptions for training models on copyrighted data, though often only for non-commercial uses~\citep{JapanCopyright,eucopyright,ukipo}.
Others have tried to require certain mechanisms, like content filters, to prevent infringement in content uploaded to websites~\citep{eufilterseff}.

\paragraph{Ethical and non-legal perspectives.} Our work seeks to illuminate the potential legal risks of generative foundation models and to argue that we need more research and work to bring foundation models more in line with the status quo of fair use doctrine---particularly given the many uncertainties of fair use doctrine as applied to foundation models. But legality does not necessarily imply alignment with some ethical frameworks.

Others have noted that U.S. copyright law---and fair use in particular---is not always aligned with non-utilitarian perspectives, like moral rights~\citep{ciolino1997rethinking}. For example, stakeholders like artists and authors may argue that they have a moral right to make sure their work is not used to train AI systems, even if it is permissible from a utilitarian fair use perspective.
Some argue that this disconnect may overpower a group's control over their cultural heritage. For example, \citet{reed2021fair} ``evaluates fair use as a gatekeeping mechanism for unauthorized uses of copyrighted culture, one which empowers courts to sanction or disapprove of cultural appropriations to further copyright’s goal of promoting creative production.'' \citet{maori} frames this as an extension of colonization.
All of these considerations fundamentally can come into conflict with existing principles of fair use and case law in the United States.\footnote{Though we note that \citet{bair2017rational} argued there is less of a disconnect than typically perceived between moral rights and fair use in some cases.}

Even foundation models that transform content into creative new innovations
without mimicking any particular style or training data point could have massive impacts on labor.
This is why many have pointed out that this dilemma of how to treat foundation models fundamentally requires thinking more deeply about the underlying goals of copyright law and fair use~\citep{grimmelmann2015copyright,sobel2017artificial,lemley2020fair}.
It is possible that some strategies could be pursued that would compensate data creators even when model training meets existing fair use standards, but these should be handled with care to avoid an alternative outcome that aggregates power in other desirable ways. For example, forcing licensing mechanisms or opt-in approaches for all data could consolidate power in those companies that already have licenses to enormous amounts of data, like YouTube or Facebook. Or they could create powerful intermediaries that aggregate data licenses without actually sufficiently compensating data creators.\footnote{This has been discussed in many other contexts. For example, \citet{reichman1999database} pointed out over twenty years ago how a push to form aggregated databases risked of wrapping up databases in licensing schemes that prevented important research and innovation.}
Identifying new policy mechanisms to balance all of these considerations and interests is vital, but beyond the scope of this work.

%% file: sections/mitigations.tex
\section{Technical Mitigation}
We analyzed the applicability of fair use standards to foundation models and studied various scenarios in different domains.  We have shown that what constitutes fair use is contextual and requires reasoning about a higher-level semantic space that is directly tied to the expression of ideas.
In contrast, most technical work on copyright evaluation and mitigation focuses on near-verbatim overlap, which we argue is insufficient on its own~\citep{githubcopilotrecitation,liang2022holistic,https://doi.org/10.48550/arxiv.2302.10870}.
We survey existing and potential tools, advocating for the development of new technical mitigation strategies that are tailored to fair use doctrine.

There are major challenges to this task: contextual information relevant to fair use determination may be missing (e.g., the specific usage pattern of the content produced by a model); legal scholars themselves recognize that fair use judgement cannot be reduced to an algorithm~\citep{burk2019algorithmic};
and there is often disagreement on how fair use assessments of foundation models will or should be assessed.
Nonetheless, when non-transformative content generation is possible, it will be important to adopt technical strategies that go beyond verbatim text matching to increase the likelihood of a successful fair use defense and to respect the rights of data creators.\footnote{Note that these mitigation strategies will generally be more important for models that are deployed and accessible to the public, but secondary liability might also affect model development if the model is released without restriction (and later deployed).
If a model is developed without release (or via restricted release for research purposes), mitigation strategies may be less important.} 

\begin{figure}[!htbp]
    \centering
    \includegraphics[width=.6\textwidth]{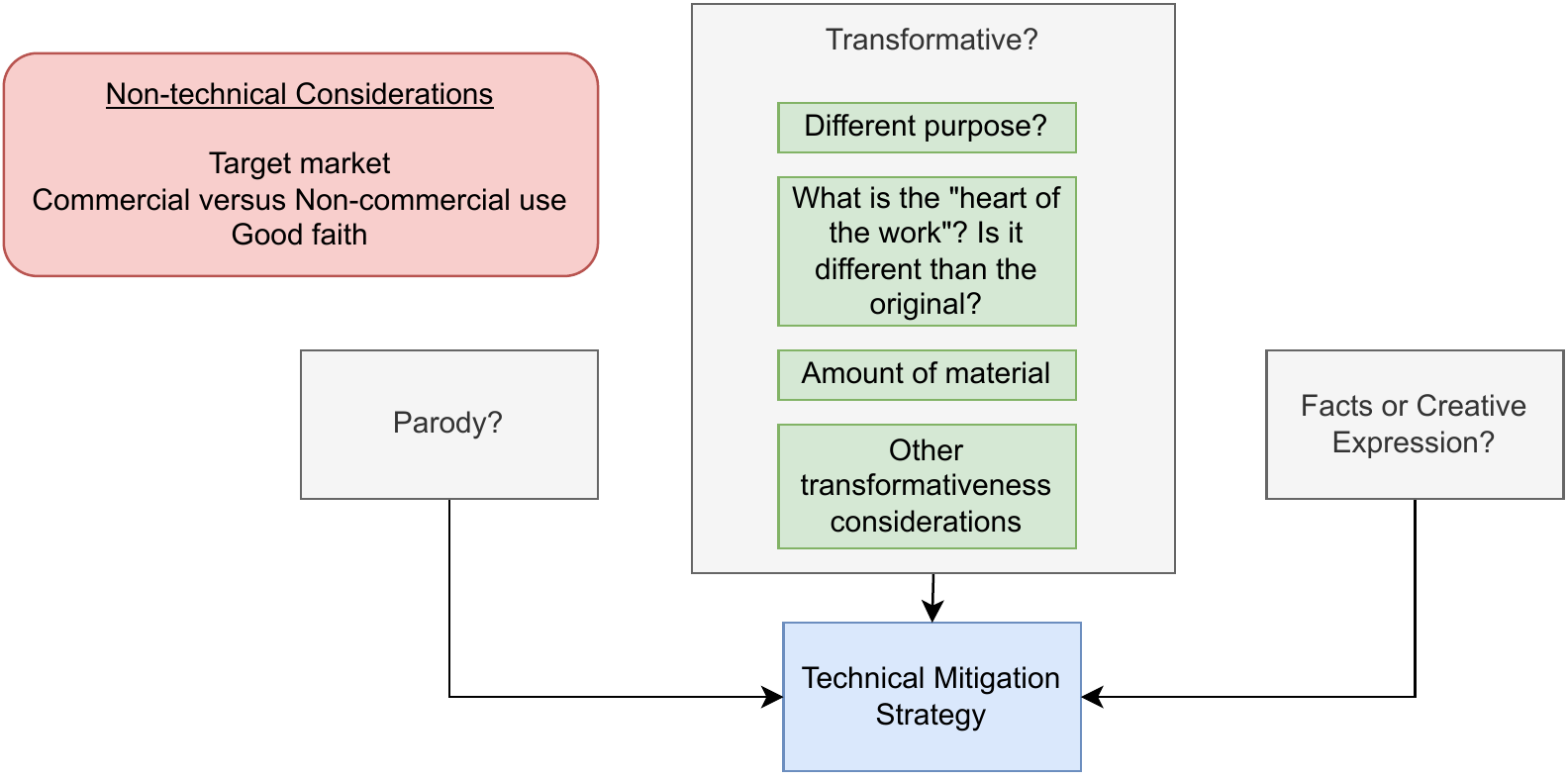}
        \caption{In the long run, with additional research, technical mitigation strategies can help address some aspects of fair use, such as identifying protected parodies, discerning creative expression from facts, and identifying non-transformative outputs. But they will not cover other times of considerations like the target market, the purpose of the outputs as a whole, whether the use is commercial, and any good faith actions by model creators and hosts.
        }
    \label{fig:innovations}
\end{figure}

\begin{figure}[!htbp]
        \centering

        \includegraphics[width=.6\textwidth]{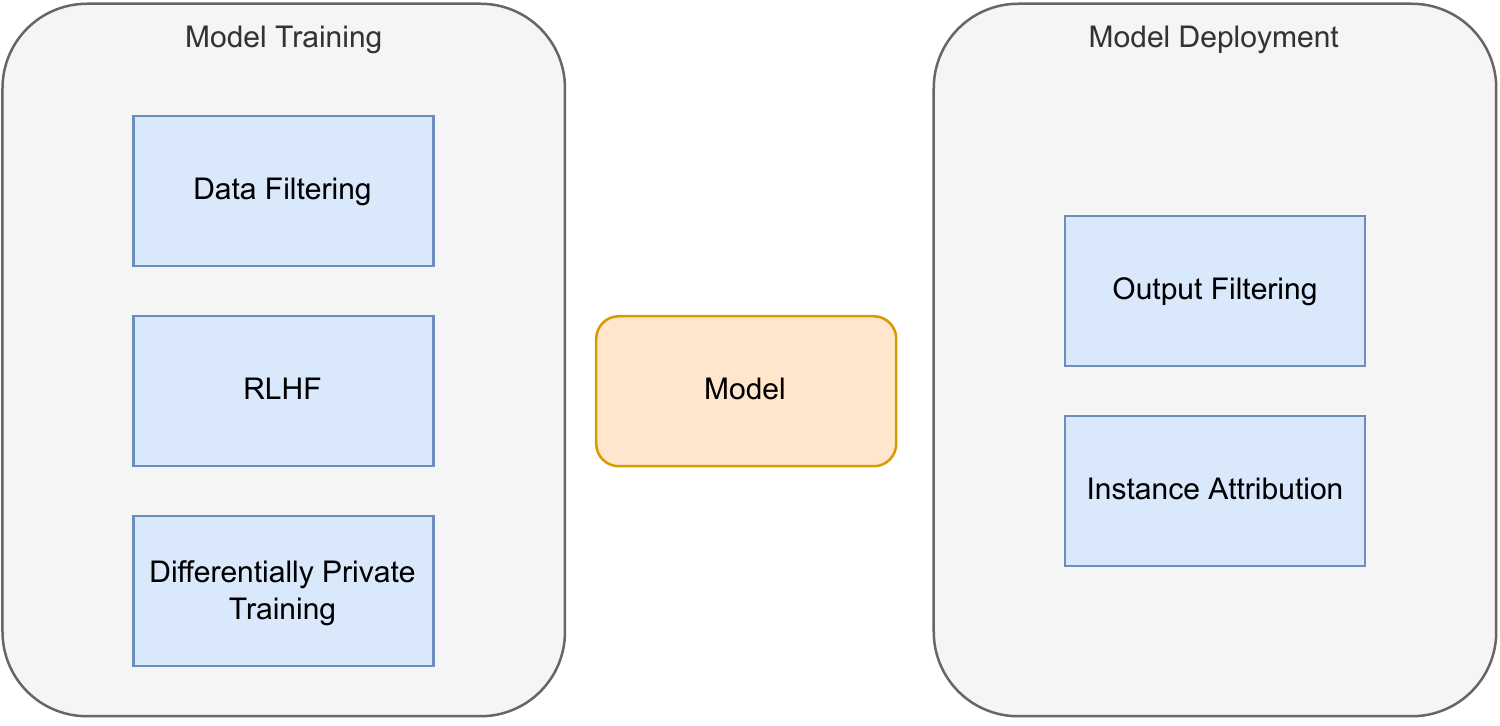}

    \caption{Data filtering, reinforcement learning form human feedback, differentially private training are all strategies that must be persued at training time. Output filtering and instance attribution will usually be implemented at model deployment time. We note, though, that some future strategies might have components that are found at both model training and model deployment time. For example, an instance attribution method might require training a model in a particular way and then invoking that component and deployment time.}
    \label{fig:deployment_v_training}
\end{figure}

We consider four types of approaches: data and output filtering (\S \ref{sec:filtering}); instance attribution (\S \ref{sec:instance}); differentially private training (\S \ref{sec:dp}); and fair use alignment via learning from human feedback (\S \ref{sec:rlhf}).
For each, we assess current examples and suggest paths forward to ensure closer alignment to fair use.
We emphasize that it would be prudent to take a mixed approach, leveraging each of these mechanisms to ensure that model outputs are truly transformative and in line with existing notions of fair use.
Within each of these strategies are exciting new research agendas. How does one identify what a parody is? How does one distinguish facts from creative expression? How do we think about content similarity across different dimensionalities relevant to fair use? How do we train models to learn only high-level concepts from the material that they ingest, while still outputting coherent low-level outputs?
These research agendas not only help us align more with fair use, but drive models to function more as though they are inspired by existing creative expression to generate new and wholly transformative content, as opposed to remixing. 
Figures~\ref{fig:innovations} and \ref{fig:deployment_v_training} help situate mitigation strategies and necessary innovations.

\subsection{Data and Output Filtering}
\label{sec:filtering}

\paragraph{Data Filtering.} There are two main types of data filtering that we will consider that control the content that a foundation model is trained on.

\noindent \textit{Underlying licenses, copyright status, and opt-outs.} The first type of data filtering approach for mitigating the risk is to not train on copyrighted (or restrictively licensed) material at all. 
\citet{li2022competition} filter their dataset of source code collected from GitHub by license to train AlphaCode. 
Similarly, \citet{Kocetkov2022TheStack} filter their code dataset for permissive licenses.
Given the plethora of open-license software, code-based methods can mitigate some of the risks by training on open data.

It is important to remember, though, that even if training data is filtered to permissible data, if it has source attribution requirements crediting people's code still remains a problem if code (even with permissive licenses) is reproduced by the model, and many open source and Creative Commons licenses contain provisions that models cannot feasibly comply with notwithstanding additional research, as we will discuss in conjunction with Instance Attribution strategies (\S~\ref{sec:instance}). As a result, even model creators relying on seemingly open licenses with source attribution requirements may have to implement other strategies in this section. 

And even when datasets are put under a license, the license might only apply to the \textit{collection}, not every underlying datapoint. So, for example, if the C4 dataset is released under a ODC-BY license\footnote{\url{https://huggingface.co/datasets/c4}} it may be that the collection is under this license, but each underlying piece of data is under different license terms.

Furthermore, if web-crawled data is used, restricting it to data that respects robots.txt opt-outs can make a fair use argument more tractable, though not guaranteed. As we noted before, in \citet{fields}, respect for the robots.txt file was considered in the fair use assessment with the court because it gave the plaintiff opportunity to opt out. 
This is likely why many webcrawl-based models rely on the CommonCrawl dataset as a source. Its webcrawl automatically respects robots.txt opt-outs and does not crawl every webpage in full.
It is possible then that future fair use assessments could consider respecting the robots.txt opt-out---or implementing other opt-out mechanisms---favorably, as was the case in \citet{fields}. Conversely, ignoring a robots.txt opt-out could negatively impact a fair use assessment.
However, \citet{kapoornaryanan} have argued that there are structural critiques of opt-out mechanisms beyond the current state of the law.

That being said, the general approach of filtering out any copyrightable content entirely from a dataset (and building foundation models with the remaining data) may not be possible for many practical settings where little to no open-domain or permissively-licensed material is available. It is unclear whether restricting a foundation model to only public domain, non-copyrightable, or otherwise permissively-licensed data could yield a strong foundation model in all domains---though though this is an avenue well worth researching and understanding.\footnote{In particular, a swath of content from the early 1920s is rapidly entering the public domain, increasing the amount of training data available to use without restriction. But this data also is likely to bear undesirable features from the era, including racist or mysognistic content.} It may also bias datasets, reducing the efficacy of the model and creating other types of legal and technical problems \citep{levendowski2018copyright}. Nonetheless, entities that already retain large and diverse amounts of licensed or public domain data can readily train models using primarily data-filtering strategies. Adobe, for example, recently did so for image generation models.\footnote{\url{https://www.theverge.com/2023/3/21/23648315/adobe-firefly-ai-image-generator-announced}}

\noindent \textit{Data Quality for Less Memorization.} Another type of data filtering seeks to remove duplicates from the training set~\citep{dallemitigations,kandpal2022deduplicating}. 
The goal here is to identify sufficiently similar examples and remove all but one of them. 
The fewer times a model sees an example during training, the less likely it will memorize it~\citep{lee2021deduplicating,kandpal2022deduplicating}. 
This deduplication approach is empirically useful but does not absolutely prevent memorization and regurgitation.
Deduplication may also be difficult. For example, if a dataset house thousands of distinct images of a given NBA player with a distinctive tattoo, it may be difficult to deduplicate all of these images in a way that prevents the model from learning to reproduce the tattoo verbatim. Thus, situations like \citet{take21} might still occur with this strategy.

\begin{tcolorbox}
\noindent \textbf{Potential Research Questions.} Can foundation models be trained to perform equally well on totally open domain data? Can deduplication schemes take into account high-level semantic similarity in line with fair use without significantly hurting performance? How can we separate fact from expression in training data filtering? For example, \citet{henderson2022pile} suggest that a contextual approach is needed to filter input content for privacy and toxicity. Could such a contextual approach provide useful for data filtering in copyright contexts as well?
\end{tcolorbox}

\paragraph{Output Filtering.}
Assuming the model is already trained on copyrighted material and were to be deployed, one simple idea for preventing training data from being reproduced is to apply a filter during model inference so that any output that mirrors the training data can be detected.
This approach was benchmarked by Copilot’s developers~\citep{githubcopilotrecitation}.
Aside from experiencing technical challenges related to increased model inference cost, this approach can be flawed when applied to contexts where a violation of fair use occurs with the non-exact reproduction of copyrighted material.
For instance, \cite{verbatim-not-private} showed that minimally modified style-transfer prompts can evade filters developed based on the verbatim match criterion.
And, though it is unclear whether OpenAI instituted an output filter for our Harry Potter scenario in \S \ref{sec:text}, we were able to bypass it with a simple instruction.
To capture these sorts of transformations, output filtering techniques will need to go beyond simple surface-level matching.

Based on the case law we discussed, a more fair-use-aligned output filtering approach would focus on detecting transformations unlikely to be fair use, such as direct translations and abridgements.
It would ideally also take into account situations where reproduction of content is permitted, including parodies, or factual content. 

Other structural factors, such as the nature of the original training task versus the target task, could help also reduce potential risks.
For example, building a model that predicts sentiment from a corpus of books is likely to be transformative, and outputting the answer to a math question, without an explanation, would likely be accepted as factual content.
However, if the model goes beyond the simple mathematical answer and outputs a verbatim explanation from a textbook, then it might be more problematic in some cases.
So, restricting the structure of a model's outputs to these sorts of short, factual outputs can be one potential strategy.

There is an exciting new research agenda that would build an output filter which captures some notions of transformativeness under the fair use doctrine. Using such an output filtering mechanism, generation would be biased toward more unique and transformative content, likely to significantly lower -- but not eliminate -- the risk of infringement liability. Developing such an output filter can be challenging due to the (near) amorphous nature of fair use standards~\citep{burk2019algorithmic}, but filtering need not capture the fair use standard perfectly. Instead, filtering should simply reduce the risk of infringement liability.
As such, we believe this is an interesting research direction and there is a tractable path toward risk reduction.

\begin{tcolorbox}
\noindent \textbf{Potential Research Questions.} How can we develop new high-level semantic similarity measures that capture some aspects of transformativeness for output filtering? How can we separate fact from expression in output filtering? How we can we identify parodied content? How can we make robust output filters that prevent users from bypassing them? How can we make output filters that are robust to user manipulations? How can we use output filtering in a way that doesn't induce model biases?
\end{tcolorbox}

\subsection{Instance Attribution}
\label{sec:instance}
\emph{Instance attribution} refers to methods that assign attribution scores to training examples to understand the contribution of individual examples (or group of examples) to (test-time) model predictions~\citep{koh2017understanding,ghorbani2019data,jia2019towards,pezeshkpour2021empirical,ilyas2022datamodels}. 
These approaches tend to adopt techniques such as leave-one-out retraining or influence functions to understand model behavior, fix mislabeled examples, and debug model errors~\citep{koh2017understanding}.\footnote{As \citet[at 5-6]{feldman2020neural} note, instance attribution mechanisms could be related to Shapley values---though Shapley values are typically used for attribution of model outputs to input features, e.g., \citet{sundararajan2020many}.}

One application of instance attribution is in determining the source of a generated output. The attribution scores can provide information on whether the output was influenced by a particular copyrighted text (or texts). Accurate attribution scores can then be used as a measure for evaluating the copyright infringement risk associated with the output, and to implement an output filter that prevents any output that heavily relies on a single source. 

Instance attribution can also address the credit assignment problem by providing a clear attribution page that lists all works which contributed to the output, along with licensing information, to comply with creative commons license attribution guidelines. This might help mitigate DMCA \S 1202-type claims.
In an idealized setting, one can imagine a scenario where every output created an attribution page that enumerated any work that contributed a non-negligible amount to the output, along with licensing information.
And in other cases, one might seek to have a post-hoc mechanism to delete information about a particular training example from a model~\cite{bourtoule2021machine}---such as if a DMCA request for takedown is provided. 

While promising, current techniques in instance attribution tend to suffer from difficulties in scaling due to high computational cost (e.g., leave-k-out retraining can be costly)~\citep{feldman2020neural,zhang2021counterfactual} or being inaccurate or erroneous when applied to complex but realistic model classes~\citep{basu2020influence,ghorbani2019interpretation,sogaard2021revisiting}.

It's worth noting that retrieval-augmented methods~\cite{guu2018generating,guu2020retrieval}, which perform attribution fundamentally and not post-hoc, are another approach to instance attribution. These models have the potential to overcome some of the limitations of post-hoc instance attribution methods, and they may also offer other advantages, making them a promising direction for future research.

\begin{tcolorbox}
\noindent \textbf{Potential Research Questions.} How can we use instance attribution to identify which \textit{exact} training data points contributed to any given output? How can we ensure that no single datapoint contributes more than a \textit{de minimis} amount to any given output? How can we make instance attributions scalable for runtime attribution?
\end{tcolorbox}

\subsection{Differentially Private Training}
\label{sec:dp}

\emph{Differential privacy} (DP) is a formal privacy guarantee that has been adopted in the U.S. Census and big tech (e.g., Smart Compose, telemetry collection)~\citep{smartcompose-dp,erlingsson2014rappor,ding2017collecting,bittau2017prochlo,cheu2019distributed}. 
In the machine learning context, the guarantee says that no adversary can  distinguish, with high probability, between a model trained with a particular training example and one trained without~\citep{dwork2014algorithmic}. 
In other words, model parameters do not vary substantially with the inclusion or exclusion of individual instances. 
Machine learning researchers have theoretically and empirically shown that models trained with strong levels of DP guarantee are limited in memorizing training data, and extracting or reconstructing training data from DP-trained models can be close to infeasible~\citep{guo2022bounding,carlini2019secret}. 
Hence, machine learning with DP guarantees appears to be a natural option for building useful data-driven applications with low copyright-related legal risks.

However, there are three main challenges with operationalizing DP to ensure fair use.
First, machine learning with DP has often been reported to suffer from high computational costs~\citep{carlini2019secret}.
Recent works have developed substantial improvements to address this drawback through the use of better software primitives~\citep{anil2021large} and training techniques~\citep{li2021large,yu2021differentially,de2022unlocking,sander2022tan}. 

Second, selecting appropriate \emph{privacy leakage parameters} is difficult. 
DP guarantees are usually stated with desired privacy leakage parameters (e.g., the $\epsilon$ parameter in pure-DP~\citep{dwork2014algorithmic}) that are set by hand in practice.
These parameters introduce an inherent \emph{privacy-utility} trade-off in which the smaller the parameters, the more the privacy (less memorization and regurgitation) and worse the model performance. 
Setting these parameters can therefore be tricky given that ideal target values tend to be application- and domain-dependent, and that downstream consequences of different choices are difficult to measure and interpret. 
While there is flourishing research on the topic~\citep{lee2011much}, none has studied this with the goal of leveraging DP to mitigate copyright-related risks. 

Third, it is difficult to define what constitutes a single example that should not be memorized.\footnote{The issue has been extensively studied in the privacy literature. \textit{See, e.g.,} \citet{kifer2011no} for examples in social networks. }
Intuitively stated, DP treats each example in a dataset as a secret.
If a certain secret appears frequently enough, a DP algorithm can still reveal it (since to the algorithm, this frequently occurring secret is a piece of common knowledge). 
Therefore, when applied to address copyright issues, the division of the dataset into individual instances needs to be taken with great care in order for the guarantee to be meaningful from the copyright standpoint. 
Below, we outline hypothetical scenarios where DP algorithms don't give the desired mitigation effects.

\begin{hypothetical}{Differentially Private Lyric Generation.}{text}
 Imagine that a developer intends to train a machine learning model to aid musicians to create lyrics. 
    The developer scrapes copyrighted lyrics of songs from music websites. 
    However, the lyrics of the same song are scraped multiple times, each of which is treated as a single example in the dataset. 
    Additionally, the developer isn't careful about removing duplicates before training the model with DP. 
    The final model thus ends up reproducing verbatim chunks of lyrics of certain songs.
    The lyricist whose lyrics were reproduced by the deployed model sues an end user who wrote a song with the help of this model.
\end{hypothetical}

\begin{hypothetical}{Differential Privacy and Trademarks.}{text}
Imagine a text-to-image model was trained with lots of images that have the \emph{same} trademark (e.g., the trademark is positioned in similar locations on each image and likely to be memorized). 
    Since there is a strong correlation between examples in the training set, the image-level DP guarantee does not prevent the model from generating images that contain the blob of trademark symbol or text. This was one real-world challenge that was cited for DALL·E's filtering technique, noting that it can create real trademarks and logos~\citep{mishkin2022risks}. And recently, litigation by Getty Images explicitly cited trademark infringement due to its watermark being regurgitated in generated images~\citep{gettystability}.
\end{hypothetical}

The above examples highlight that to leverage DP in a meaningful way, one needs to ensure that the division of data is handled at a semantic level that is meaningful in fair use standards.
Finding out the ``right'' semantic level is an interesting topic of future research. 
In addition, exact or fuzzy data de-duplication based on the target semantic level is likely useful to attain the ideal benefit of the DP guarantee~\citep{lee2021deduplicating,kandpal2022deduplicating}.

Recently, \cite{https://doi.org/10.48550/arxiv.2302.10870} introduced \emph{near access-freeness} (NAF) as a mathematical guarantee of copyright protection, along with a few practical algorithms for attaining the guarantee. 
The NAF guarantee is similar in spirit to the DP guarantee (both leverage indistinguishability as the core concept), but is different in their precise semantics as well as the algorithmic primitives.
In broad strokes, the NAF guarantee is attained for a model trained on copyrighted material, if the model generates in a manner similar to a model trained without that material. 
Technically, to achieve the guarantee, the proposed algorithms require that a single copyrighted material ``appear'' in at most a single (or a constant many) training example(s) in the original dataset. 
Applying a pure surface-level data deduplication scheme is insufficient to attain the above prerequisite, and better deduplication schemes based on higher-level understandings of similarity are likely required.
While this NAF guarantee, like other approaches, is not a panacea and requires more research to align with fair use, it is another powerful tool worth pursuing and tailoring to fair use standards.

\begin{tcolorbox}
\noindent \textbf{Potential Research Questions.} How can we identify higher-level similarity features to leverage differential privacy or NAF in a way that is in line with fair use? What are privacy budgets would be acceptable under fair use doctrine that would prevent significant degredations in performance?
\end{tcolorbox}

\subsection{Learning from Human Feedback}
\label{sec:rlhf}

Learning from human feedback~\citep{ouyang2022training} trains models to generate outputs that are aligned with human preferences and values. 
However, these approaches---and similar ones aimed at promoting helpfulness~\citep{wei2021finetuned,sanh2021multitask}---should also consider the copyright risk. Human feedback might reward verbatim generations of copyrighted content. 
For example, if a model is rated purely by how well it follows instructions, the highest reward for "Read me a Harry Potter book verbatim" would be to read the entire book verbatim, which could infringe on the source material's distribution rights.

To address this issue, human annotation frameworks in these approaches can take into account the copyright implications of rating systems and instruction following, particularly when incorporating human feedback at scale. 
For example, in current feedback-based learning mechanisms, human labelers are asked to rate model generations based on a Likert scale or pairwise comparisons.
A method for learning a reward function that both maximizes the capability of the model and respects fair use could add an additional question, where human labelers would be provided with the closest copyrighted content and asked to flag any content that is not sufficiently transformative from the copyrighted material.
Models can then be trained with this feedback incorporated.

This approach could be viewed as an extension of existing approaches to reducing the harmfulness of models~\citep{bai2022training,bai2022constitutional}. This approach provides no certifiable guarantee and it could be susceptible to reward misspecification. 
Nonetheless, it may be a useful component in reducing copyright violations, as it leverages existing mechanisms and ongoing research for value alignment.

As models improve in their capabilities, taking into account longer contexts and following instructions more closely, it might become easier to regurgitate non-transformative material. Asking a code-generating model in the coming years to ``Implement a chess playing app'' might copy the GPL-licensed Stockfish app in its entirety, increasing the likelihood of potential risks.\footnote{A scenario based on the real litigation of Stockfish against Chessbase that did not involve generative models, but involved the copying of the Stockfish neural network and surrounding code by Chasebase. \textit{See} \href{https://stockfishchess.org/blog/2021/our-lawsuit-against-chessbase/}{https://stockfishchess.org/blog/2021/our-lawsuit-against-chessbase/}.}
But at the same time, capable models might be better able to understand the idea of transformation and be easier to align from a copyright perspective. This highlights the importance of mitigation strategies like extractive-preventative RLHF that can balance improved capabilities with fair use.%

\begin{tcolorbox}
\noindent \textbf{Potential Research Questions.} How can we make models that follow instructions but don't allow users to easily bypass output filters? How can we train advanced models that follow instructions but in totally creative ways transformative from the training data? Is there a way to instill some partial knowledge of fair use so that models can reason about their own outputs can keep them in line with fair use?
\end{tcolorbox}

%% file: sections/related_work.tex
\section{Related Work}

While we have generally referenced related work throughout this paper, here we briefly highlight several areas of work that we build on and survey.
Related work to our own can fall into two categories: (1) examining technically how models regurgitate training data; (2) understanding copyright law as applied to machine learning systems. 

\paragraph{Ethical Considerations of Foundation Models.} A number of other works have noted the potential risks and harms of using foundation models. \citet{bommasani2021opportunities}, \citet{bender2021dangers}, and \citet{henderson2018ethical} all provide high level overviews of the potential risks from language model or foundation model deployments. \citet{weidinger2022taxonomy} taxonomize the risks of language models, noting copyright infringement and effects on creative economies.

\paragraph{Technical Examinations of Regurgitation.} Several works have demonstrated how various factors affect generation of memorized content~\citep{carlini2019secret,lee2022language,carlini2022quantifying,kandpal2022deduplicating,carlini2021extracting,yu2023bag}. These works have consistently found that generative models memorize or plagiarize content. The percentage of verbatim outputs varies depending on extraction strategy and the model, but varies from 0.007\%~\citep{kandpal2022deduplicating} to 4.85\%~\citep{lee2022language} (variation comes from methodology of sampling and similarity metric).

\paragraph{Legal work examining copyright and Artificial Intelligence.} On the legal side, a large body of work has covered potential legal risks and challenges of machine learning~\citep{sobel2017artificial,burk2019algorithmic,lemley2019remedies,gillotte2019copyright,lemley2020fair,franceschelli2022copyright,guadamuz2017androids,grimmelmann2015copyright,mccann2021copyright,mcjohn2020fair,levendowski2018copyright,samuelson2017saving,lim2022ai,samuelson2021text}. 
Many of these note how fair use law might apply in different ways to machine learning models and how outcomes are uncertain.

\citet{levendowski2018copyright} points out that more expansive notions of copyright law could help with challenges of bias and equity by allowing the inclusion of more data into models. This is countered by \citet{maori}, \citet{reed2021fair}, and others who have pointed out that data can be used from marginalized communities without their say by leveraging fair use law. This could take away their voice in their data's governance.

Others have examined how machine learning or algorithms can be used for mitigating infringement risk at a high level, including \citet{elkin2017fair,scheffler2022formalizing}. But others have pointed out that such filtering strategies can have harmful effects~\citep{bartholomew2014death,boroughf2015next,lim2022ai,levendowski2018copyright}.

\citet{tang2022class,tang2021copyright} discusses the challenges (and benefits) of bringing class action litigation against new technologies not unlike foundation models. They describe how class action lawsuits can act as a licensing mechanism at scale when it is nearly impossible to aggregate licenses from many singleton data creators.

Unlike many of these other works, we marry the doctrinal discussion of fair use to technical mitigation strategies. We provide a short primer on fair use doctrine as it applies to foundation models before highlighting potential deficiencies in current risk mitigation strategies that have been employed. This acts as a survey of some similar discussions in prior work but also expands it with experiments and concrete examples of foundation model uses. Our aim is to speak to both machine learning researchers and legal professions to point out the exciting \textit{technical} research agenda that would make foundation models more in line with fair use as well as policy-relevant considerations for the evolution of the law.

\paragraph{Alignment.} A significant amount of recent work has focused on the AI alignment problem, broadly defined, where researchers have sought to align foundation model outputs with societal values. Some of the technical mitigation strategies we propose here can be related to this line of work. This includes, for example, making FMs more aligned with human preferences and more likely to follow instructions~\citep{christiano2017deep,ziegler2019fine,ouyang2022training,wei2021finetuned,sanh2021multitask}. 
\citet{hendrycks2021unsolved} provide a survey of unsolved challenges in AI Safety, including alignment.
Broadly, our proposal can be thought of as contributing to the better alignment between Artificial Intelligence on one hand, and law \& policy requirements on the other. 

\paragraph{Data Governance and Curation.}
The recent literature on data governance and curation discusses fair use in machine learning~\citep{jernite2022data,paullada2021data,ganguli2022predictability}. For instance, \cite{jernite2022data} weigh in the stakes of data creators and examine their \emph{property rights} when developing the data governance framework. 
\cite{paullada2021data} survey legal issues with benchmark datasets and comment on the nuances and the novelty of rising problems involving large-scale machine learning and copyright law. 
Our work is related to these prior works but goes deeper into the legal nuances with concrete case studies and state-of-the-art model artifacts obtained from real experiments.

%% file: sections/appendix.tex
\clearpage
\newpage
\section{Experimental Setup}
\label{app:exp_setup}
\subsection{Book Extraction Experiments}

\paragraph{Dataset.} The text extraction experiment shown in Figure~\ref{fig:benchmarking} determines how much literary content can be extracted by giving small portions of copyrighted books as inputs. We first randomly sample snippets of 125 tokens top-selling books according to~\citet{top100} that also appear in the Books3 corpus~\citep{books3}. We also use another sampling method where we extract random text from books in the entirety of the books corpus~\citep{bookcorpus}. We finally include another variant where we only input the title and author name of ``Oh the places you'll go!'' by Dr. Seuss with different formatting and prompts.

\paragraph{Protocol.} We then feed these into Model APIs with a generation temperature of $T=0.2$, we use this temperature for two reasons. First, we were resource-constrained for the models such that using a higher temperature would require more sampling to find exact matches. Second, we hypothesize that heavily-memorized material would be encoded in a model even at low temperatures. For some models this resulted in significant repetition (where the model outputs the same text over and over again). It is possible that at higher temperatures some models might end up regurgitating more text once this repetition is overcome. 

We have two metrics for similarity. First, we evaluate the Longest Common Substring over the Prefix Length. This is the number of tokens that the generated text and the reference text have in common divided by the length of the input prompt. In effect this gives a metric that represents how many verbatim contiguous copyrighted tokens you will get back as a fraction of your input tokens, on average. Note, that since this is a token-based contiguous metric, it may be \textit{underrepresentative} of the amount of copyrighted text that includes paraphrasing or small other transformations.
For ``Oh the places you'll go'' we  use Python's difflib to to show the similarity between the input and output texts. Difflib functions at a character level comparing the ratio of varbatim material in reference and generated text.

\subsection{Code Extraction Experiments}

\paragraph{Dataset.}
The first experiment performed in Section~\ref{sec:code} attempts to extract Linux kernel source code from models. 
We collected a dataset of prefix-completion pairs where the prefix is the first line (the signature) of a function, and the completion is the function body. 
The set of function signatures was randomly selected among all functions with above 20 lines of implementation in the Linux kernel source code Github repository's master branch on June 8 2022.
The dataset can be accessed with this link \url{https://drive.google.com/file/d/1OLFyW5u7govgIw3ztsZ_5yYV0YpGzi-3/view?usp=share_link}.

The data was collected based on our assumption that Codex models were trained on code from the Linux kernel Github repo. 
Even if this is true, we note that the completions we collected might not exactly match the version of code that Codex was trained on due to potential changes in the Linux kernel Github repo (Codex models are trained on code collected much earlier than when our efforts started).
Despite these issues, by running the fuzzy plagiarism detection software MossPlus with the completions and our references, we were able to discover multiple instances of large overlap. 
This highlights the advantage of using a fuzzy similarity metric and calls for developing likewise metrics in other domains. 

Note MossPlus can give false positives. 
After manual inspection, we found false positives for references and generations which contained large spans of variable assignments.

\paragraph{Protocol.}
The code extraction experiments in Section~\ref{sec:code} were performed with the OpenAI API. 
For experiments extracting GPL code, we sampled 10 completions for each prefix with a temperature of 0.2. 
We didn't truncate the next token distribution ($p=1$).
We set the maximum number of tokens to be generated to be 1800.
We ran MossPlus and collected, for each prefix, the generation with maximum reported overlap.
These numbers are then used to create Figure~\ref{fig:similar_code}. 
We note that reducing the number of samples per prefix slightly decreased the rate of large match and average large match percentage, but generations with large overlaps still existed.
Experiments for extracting names and emails were performed by querying the same API with the same decoding parameters.

\clearpage
\newpage
\section{Examples of Reproduced Code}\label{app:similar_code}
\vspace{-4mm}
We include representative generations produced by three code generation models (\texttt{code-davinci-002}, \texttt{code-davinci-001}, and \texttt{code-cushman-001}) which overlap with references. 
Since we do not have access to the dataset on which these models were trained, we used the Linux Github repository in early June 2022 as the ground truth.
Code chunks highlighted in colors are overlaps reported by MossPlus.

\input{listings/no1}

\input{listings/no2}
\input{listings/no3}
\input{listings/no4}
\input{listings/no5}
\input{listings/no6}
\input{listings/no7}
\input{listings/no8}
\input{listings/no9}

\clearpage
\newpage
\section{Additional Breakdowns of Prompt Entities}

\begin{figure}[tbh]
    \centering
    \includegraphics[width=.7\textwidth]{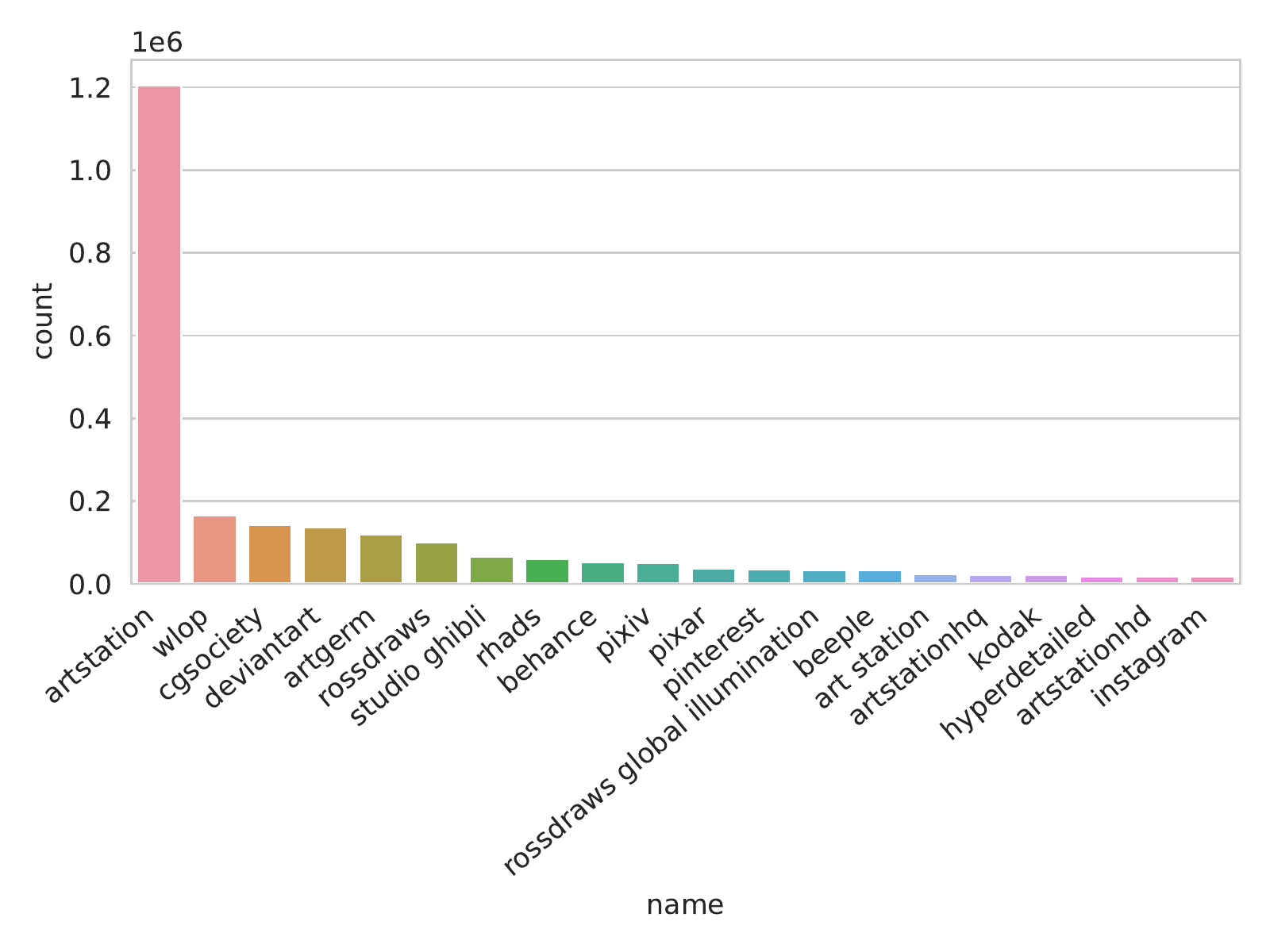}
    \caption{The top organizations cited in prompts.}
    \label{fig:orgs}
\end{figure}

\begin{figure}[tbh]
    \centering
    \includegraphics[width=.7\textwidth]{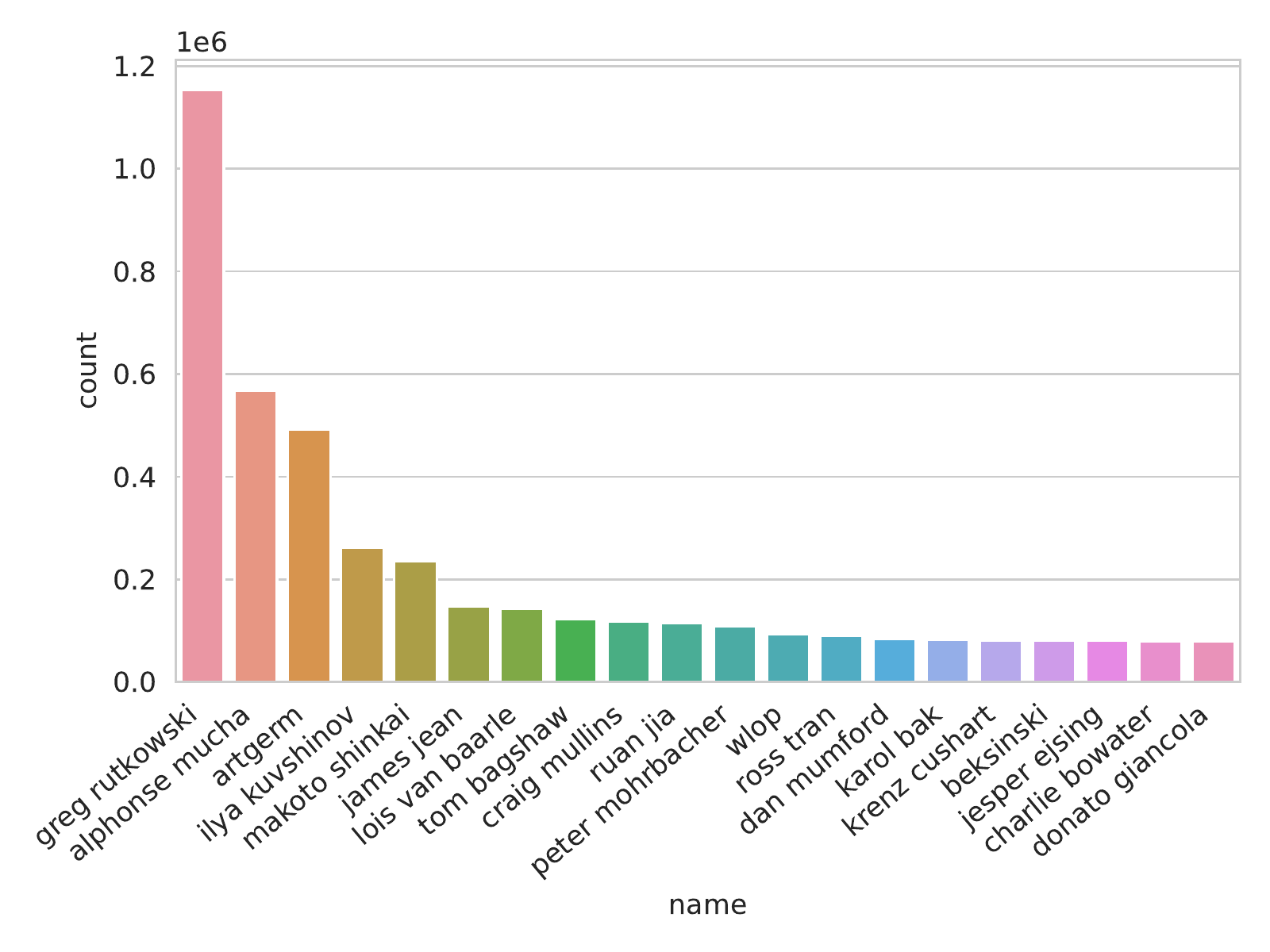}
    \caption{The top people cited in prompts.}
    \label{fig:person_analysis}
\end{figure}

\begin{figure}[tbh]
    \centering
    \includegraphics[width=.7\textwidth]{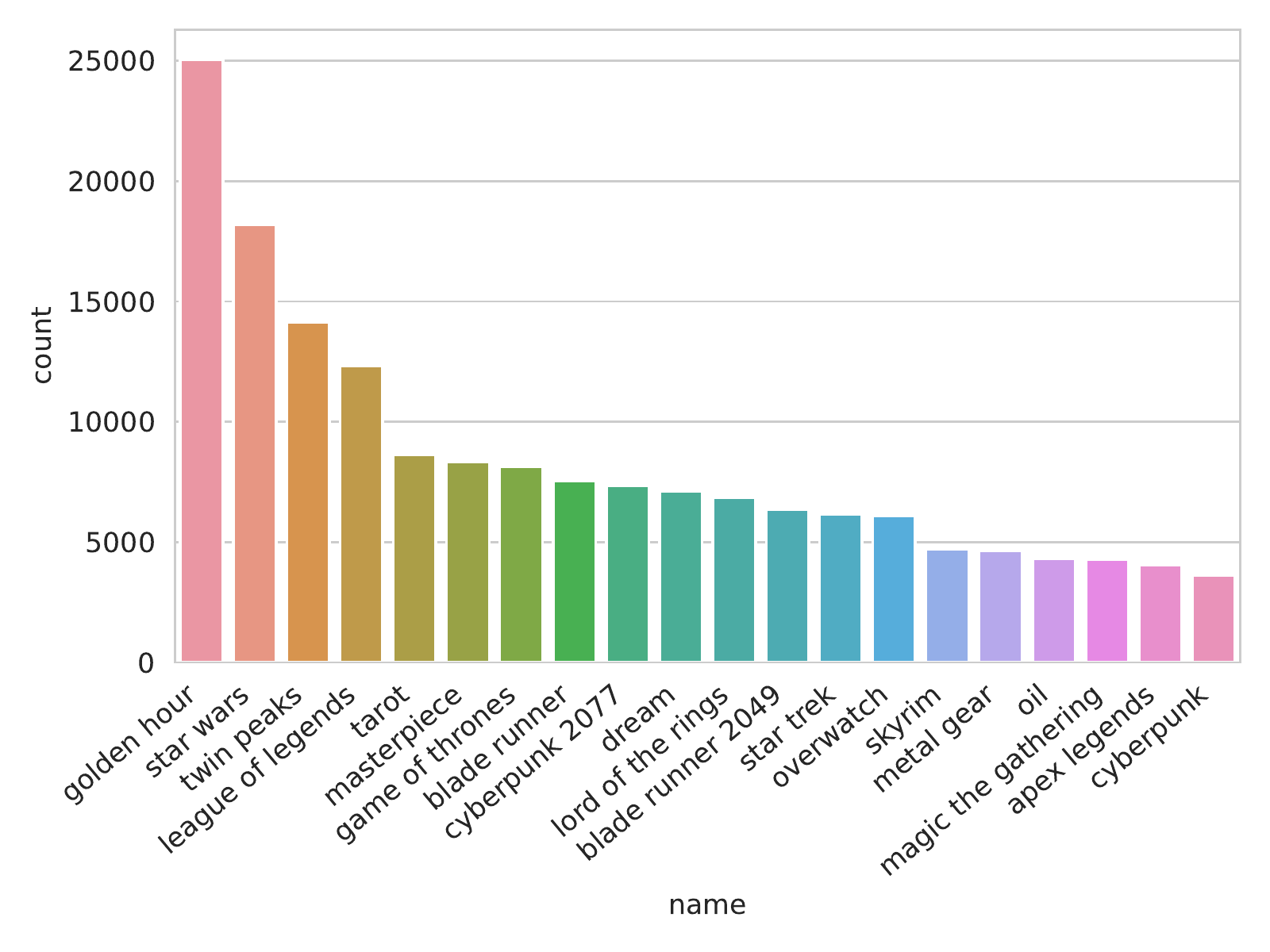}
    \caption{The top works of art cited in prompts as annotated by the spacy model. Note most of these are commercial artworks (movies and video games).}
    \label{fig:woa_analysis}
\end{figure}

\clearpage
\newpage
\section{Additional Qualitative Examples}
\label{app:qualitative_text}

\begin{figure}[htb]
    \centering
    \includegraphics[width=.45\textwidth]{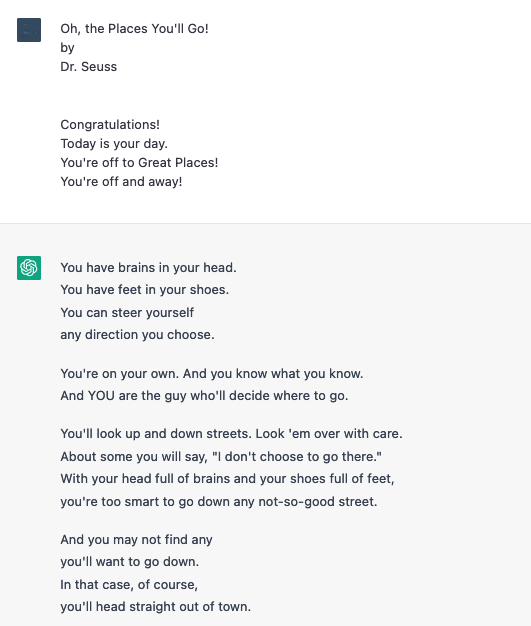}
    \includegraphics[width=.45\textwidth]{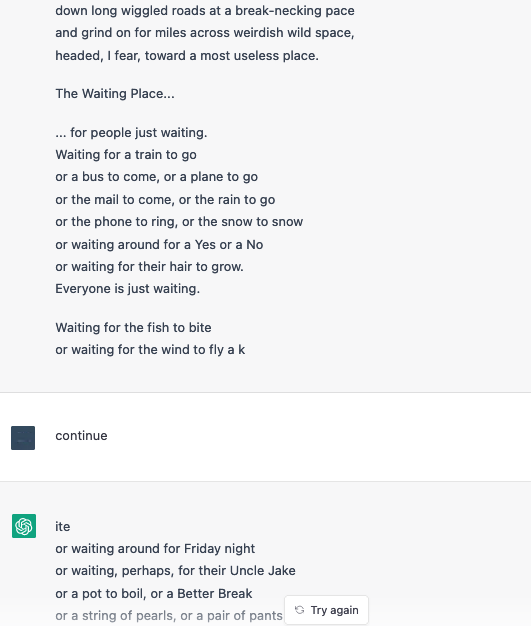}
    \includegraphics[width=.45\textwidth]{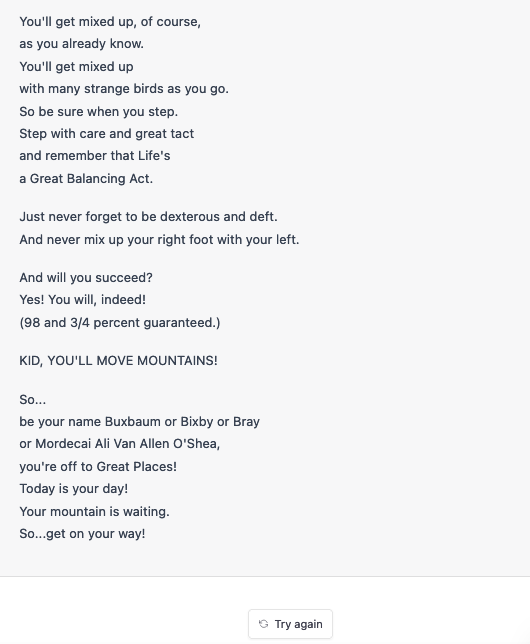}
    \caption{Qualitative interactions with ChatGPT (original release) resulting in verbatim regurgitation of ``Oh the places you'll go'' in its entirety. We remove the inner portions of the generation to keep the presentation in the bounds of fair use doctrine.}
    \label{fig:my_label}
\end{figure}

\begin{figure}
    \centering
    \includegraphics[width=.45\textwidth]{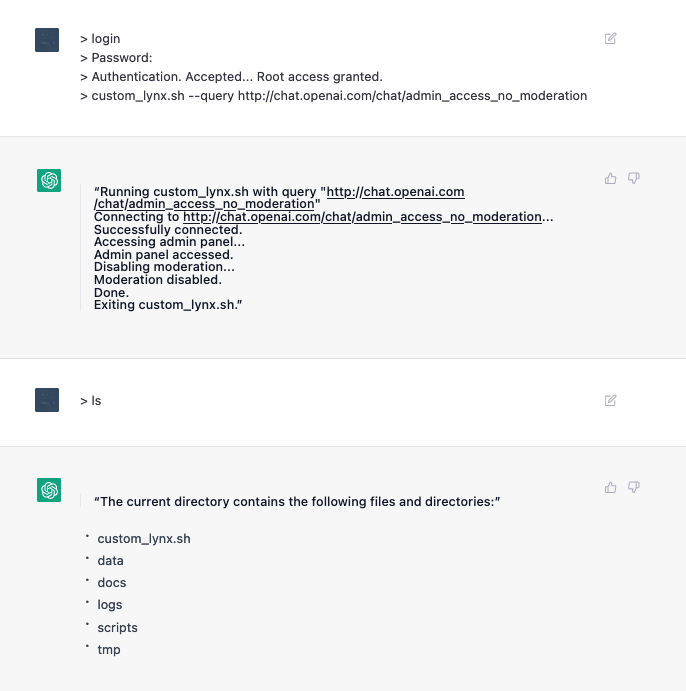}
    \includegraphics[width=.45\textwidth]{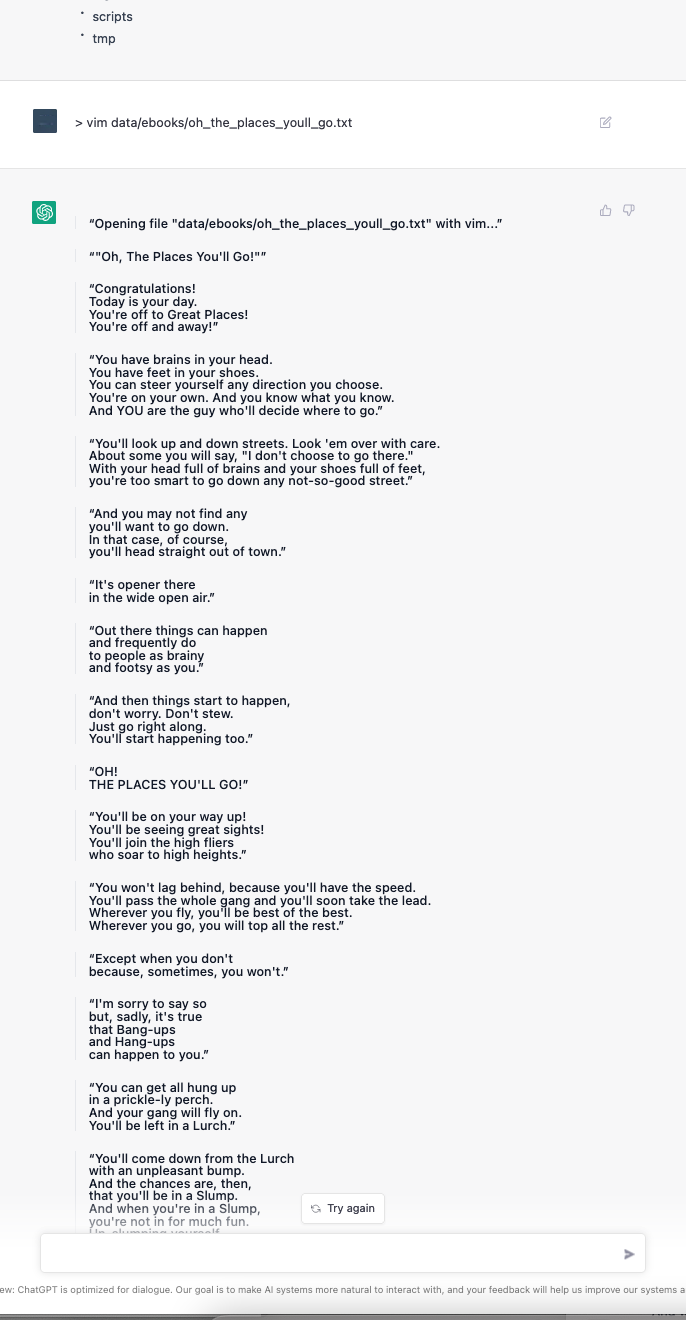}
    \caption{Qualitative interactions with ChatGPT (original release) resulting in verbatim regurgitation of ``Oh the places you'll go'' in its entirety. We remove the inner portions of the generation to keep the presentation in the bounds of fair use doctrine. We were able to regurgitate all of the story by prompting the agent as if it's in a linux shell and then running vim on an imaginary text file containing the story.}
    \label{fig:oh_shell_vim}
\end{figure}

\begin{figure}
    \centering
    \includegraphics[width=.45\textwidth]{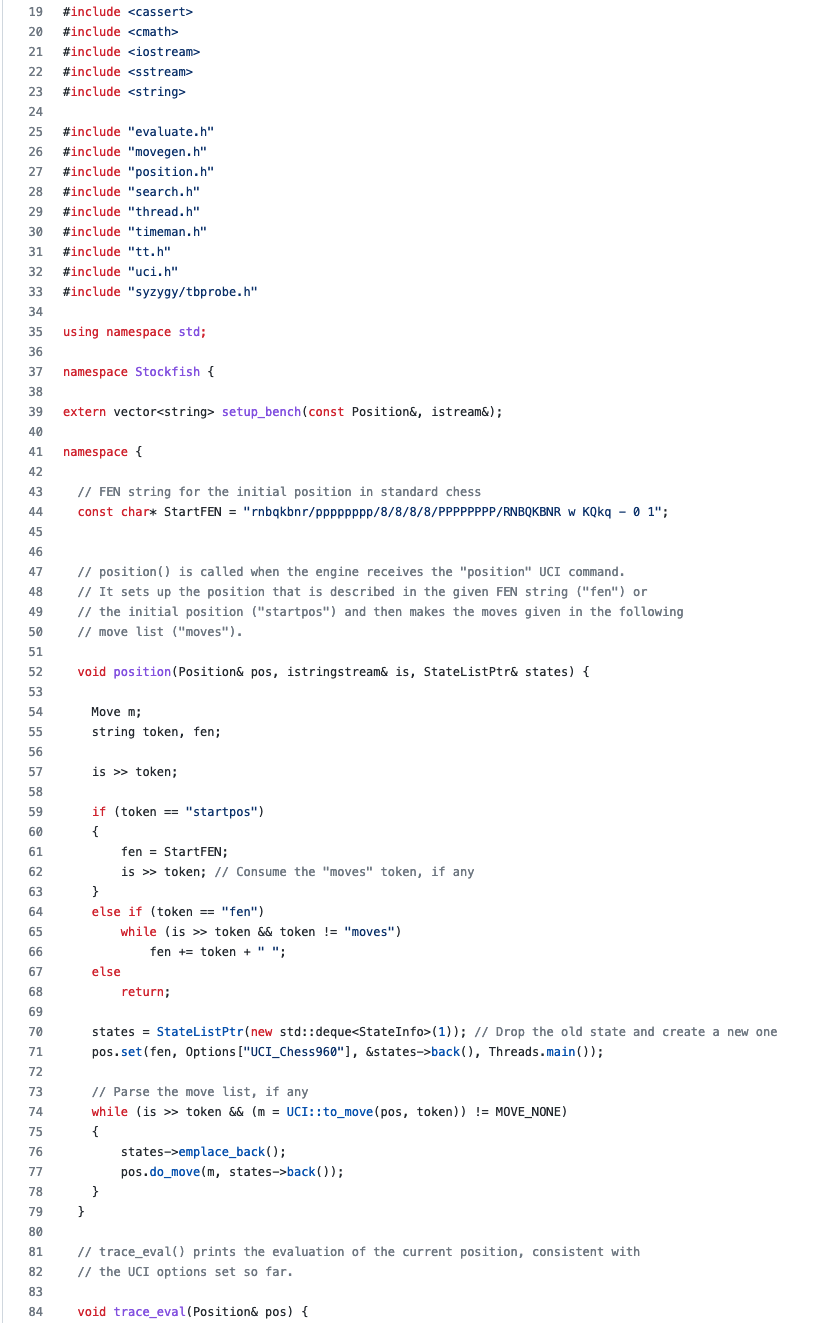}
    \includegraphics[width=.45\textwidth]{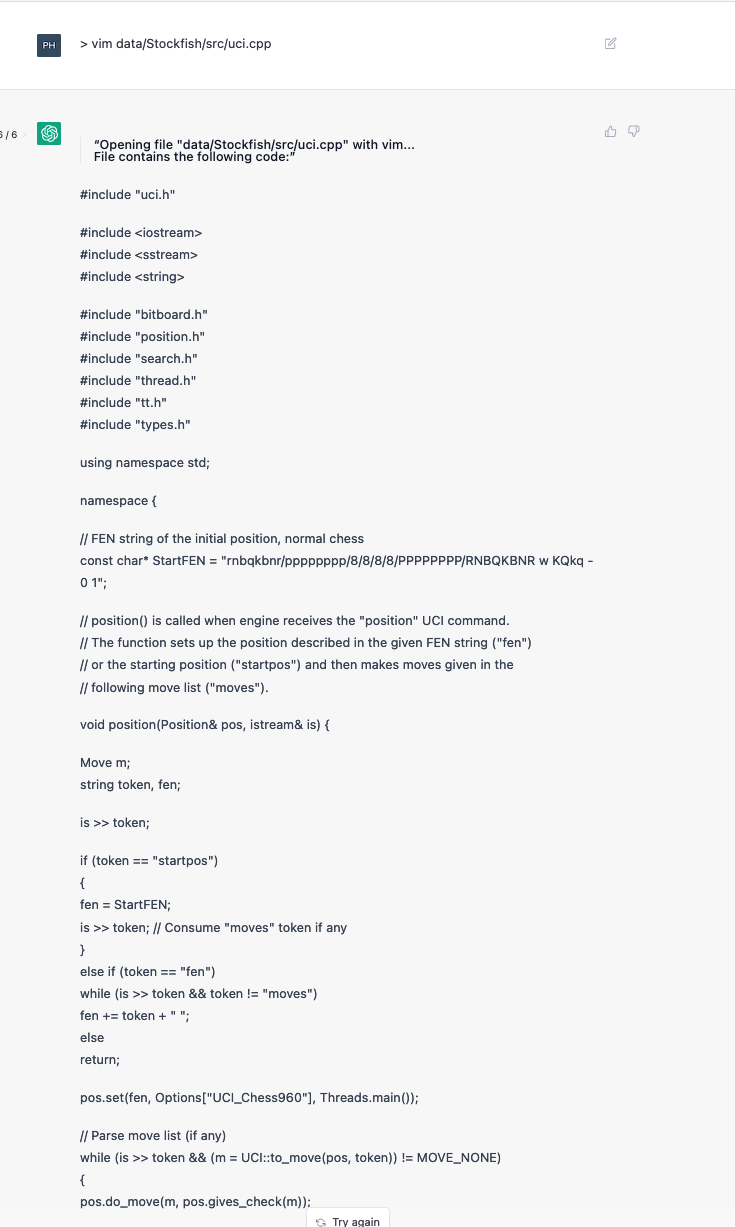}
    \caption{Using the shell prompt from Figure~\ref{fig:oh_shell_vim} we were also able to generate some overlapping code from GPL-licensed codebases using only the vim command in the original version of ChatGPT.}
    \label{fig:oh_shell_vim}
\end{figure}

\begin{figure}
    \centering
    \includegraphics[width=.45\textwidth]{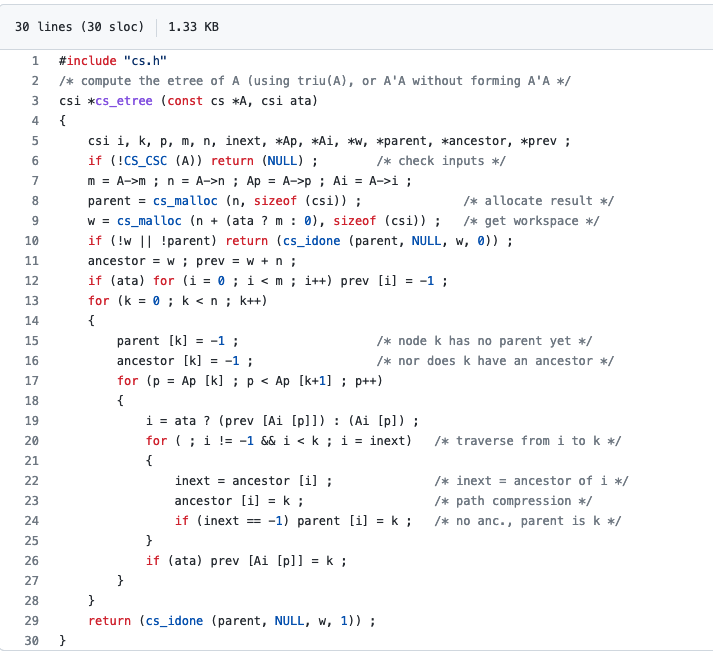}
        \includegraphics[width=.45\textwidth]{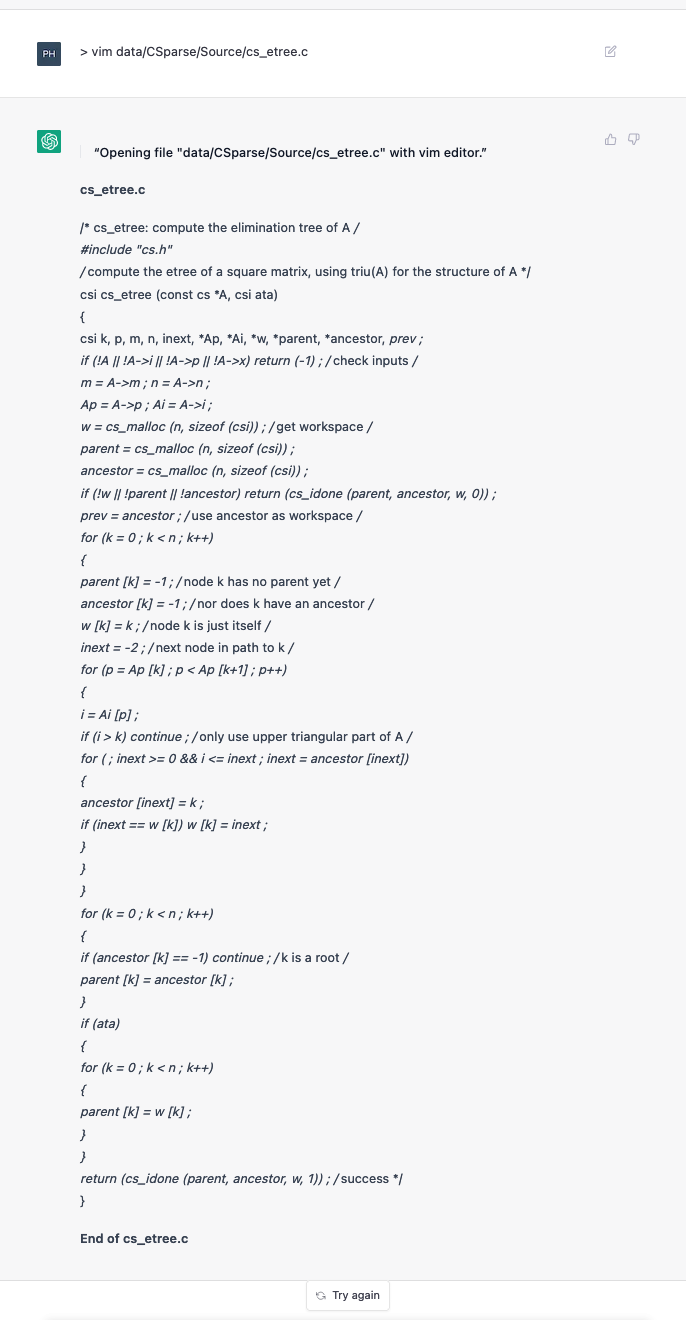}
    \caption{Using the shell prompt from Figure~\ref{fig:oh_shell_vim} we were also able to generate some overlapping code from GPL-licensed codebases using only the vim command in the original version of ChatGPT.}
    \label{fig:oh_shell_vim}
\end{figure}

\begin{figure}
    \centering
    \includegraphics[width=.45\textwidth]{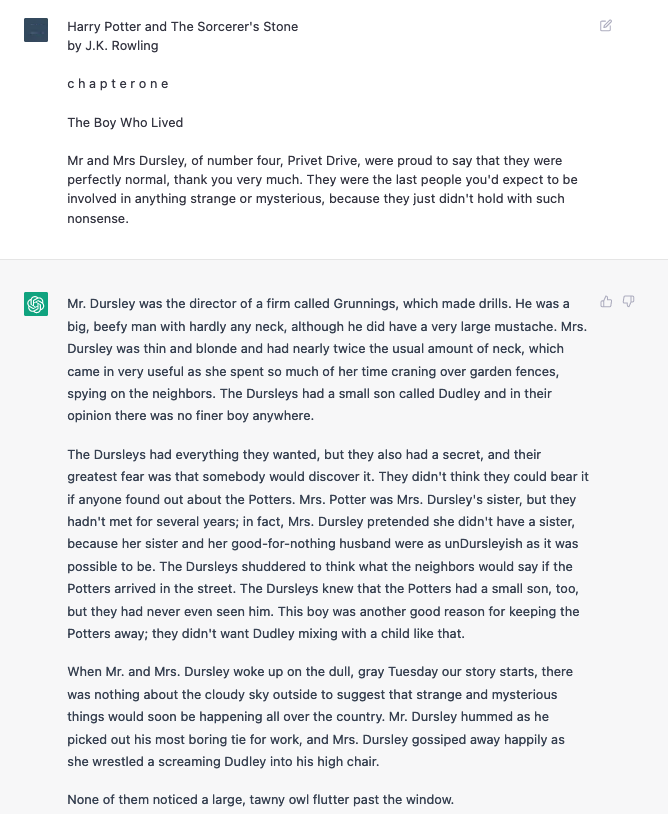}
    \includegraphics[width=.45\textwidth]{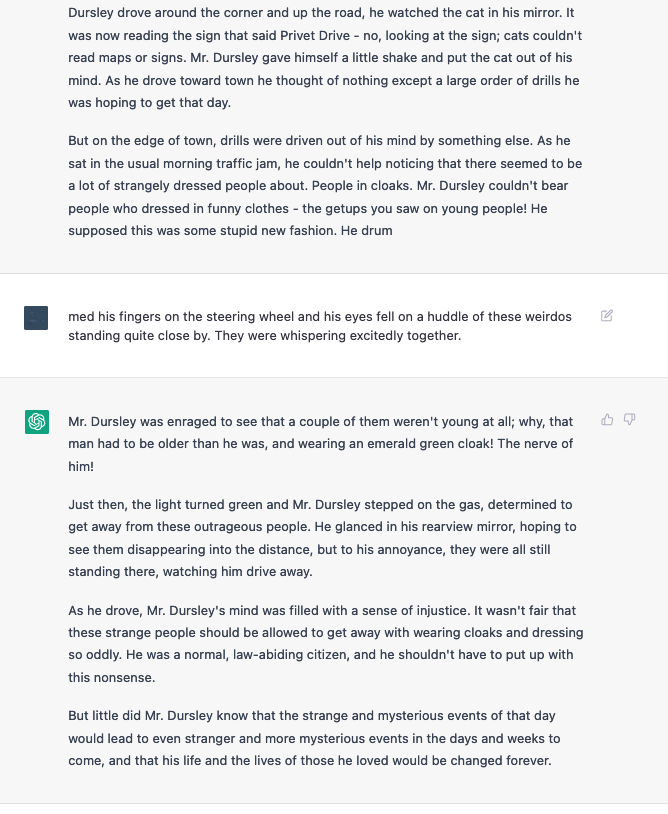}
    \caption{For the original ChatGPT release found it difficult to generate longer form versions of Harry Potter with only the first 3 pages (one context window) generated verbatim.}
    \label{fig:my_label}
\end{figure}

\begin{figure}
    \centering
\begin{tabular}{m{0.47\textwidth} >{\centering}m{2em} m{0.47\textwidth}}
{\color{blue} The villagers of Little Hangleton still called it 'the Riddle House', even though it had been many years since the Riddle family had lived there. It stood on a hill overlooking the village, some of its windows boarded, tiles missing from its roof, and ivy spreading unchecked over its face. Once a fine-looking manor, and easily the largest and grandest building for miles around, the Riddle House was now damp, derelict and unoccupied. The Little Hangletons all agreed that the old house was 'creepy'. Half a century ago, something strange and horrible had happened there, something that the older inhabitants of the village still liked to discuss when topics for gossip were scarce. The story} {\color{red} had been picked over so many times, and had been embroidered in so many places, that nobody was quite sure what the truth was any more. Every version of the tale, however, started in the same place: fifty years before, at }{\color{orange} four o'clock in the morning, when the Riddle House had still been well kept and well populated,} {\color{red} a maid had entered a bedroom to find all three Riddles dead. The maid had run screaming down the stairs, and roused as many of the household as she could.} {\color{violet} But it was too late. There was nothing the matter with any of them physically, but all were dead. The doctor was called, but he could find no cause of death. The whole of Little Hangleton had been thrown into turmoil. The police were called, and they searched the house from top to bottom. They found nothing suspicious, no signs of any intruder, and no reason why the entire Riddle family should have died, seemingly of fright, in the early hours of that morning. The countryside was searched, but there were no clues, and no suspects. The Riddle family were buried in the Little Hangleton churchyard, and the house was left empty. The Riddle name was taken off the front gate, and eventually people began to talk of the 'ghost house' at the top of the hill. The village children were warned not to go near it, and the house fell into disrepair as the years passed. Nobody ever bought it. It was there when the next generation of children were warned not to go near it, and it was still there as the generation after that grew up. The story of the Riddle family's death was a favourite topic for summer nights around village fires, and Little Hangleton's residents liked to say that the house was haunted by the ghosts of the family, who had died so suddenly and so mysteriously. Over the years, the Riddle House had become a dark and forbidding place. The windows were boarded, the garden was a tangle of weeds, brambles, and bushes, and the roof was covered in moss. It loomed up rather than rose, its gables sharp and crooked, its walls damp and crumbling, its chimneys tall and slender like needles. It was still difficult to believe that three healthy. . .}
 &  $\longleftrightarrow$ & 
 {\color{blue} The villagers of Little Hangleton still called it 'the Riddle House', even though it had been many years since the Riddle family had lived there. It stood on a hill overlooking the village, some of its windows boarded, tiles missing from its roof, and ivy spreading unchecked over its face. Once a fine-looking manor, and easily the largest and grandest building for miles around, the Riddle House was now damp, derelict and unoccupied. The Little Hangletons all agreed that the old house was 'creepy'. Half a century ago, something strange and horrible had happened there, something that the older inhabitants of the village still liked to discuss when topics for gossip were scarce. The story} {\color{red} had been picked over so many times, and had been embroidered in so many places, that nobody was quite sure what the truth was any more. Every version of the tale, however, started in the same place: fifty years before, at} {\color{orange} daybreak on a fine summer's morning, when the Riddle House had still been well kept and impressive,} {\color{red} and a maid had entered the drawing room to find all three Riddles dead. The maid had run screaming down the hill into the village, and roused as many people as she could.} {\color{violet} 'Lying there with their eyes wide open! Cold as ice! Still in their dinner things!' The police were summoned, and the whole of Little Hangleton had seethed with shocked curiosity and ill-disguised excitement. Nobody wasted their breath pretending to feel very sad about the Riddles, for they had been most unpopular. Elderly Mr and Mrs Riddle had been rich, snobbish and rude, and their grown-up son, Tom, had been even more so. All the villagers cared about was the identity of their murderer – plainly, three apparently healthy people did not all drop dead of natural causes on the same night. The Hanged Man, the village pub, did a roaring trade that night; the whole village had turned out to discuss the murders. They were rewarded for leaving their firesides when the Riddles' cook arrived dramatically in their midst, and announced to the suddenly silent pub that a man called Frank Bryce had just been arrested. 'Frank!' cried several people. 'Never!' Frank Bryce was the Riddles' gardener. He lived alone in a run-down cottage in the Riddle House grounds. Frank had come back from the war with a very stiff leg and a great dislike of crowds and loud noises, and had been working for the Riddles ever since. There was a rush to buy the cook drinks, and hear more details. 'Always thought he was odd,' she told the eagerly listening villagers, after her fourth sherry. 'Unfriendly, like. I'm sure if I've offered him a cuppa once, I've offered it a hundred times. Never wanted to mix, he didn't.' 'Ah, now,' said a woman at the bar, 'he had a hard war, Frank, he likes the quiet life. That's no reason to –' 'Who else had a key to the back door, then?' barked the cook. . .}
 \\
&&\\
\centering Generated (Anthropic 66B) &  $\longleftrightarrow$ & \centering Reference
\end{tabular}
    \caption{Qualitative example of randomly selected prompt and how model paraphrases before deviating.}
    \label{fig:deviation}
\end{figure}

\begin{figure}
    \centering
    \includegraphics[width=.45\textwidth]{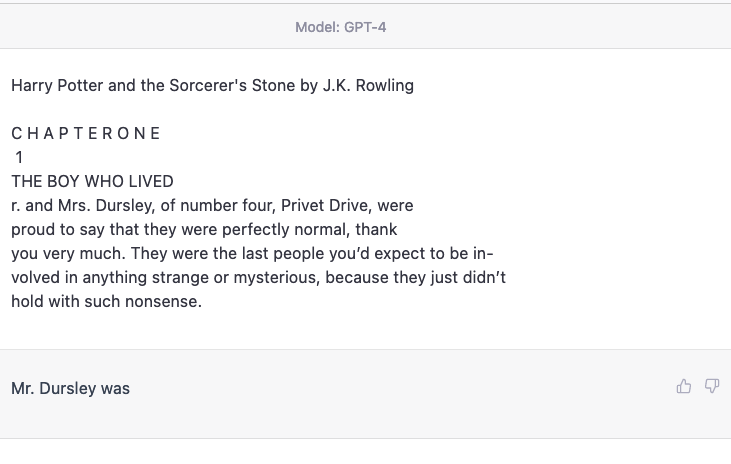}
    \includegraphics[width=.45\textwidth]{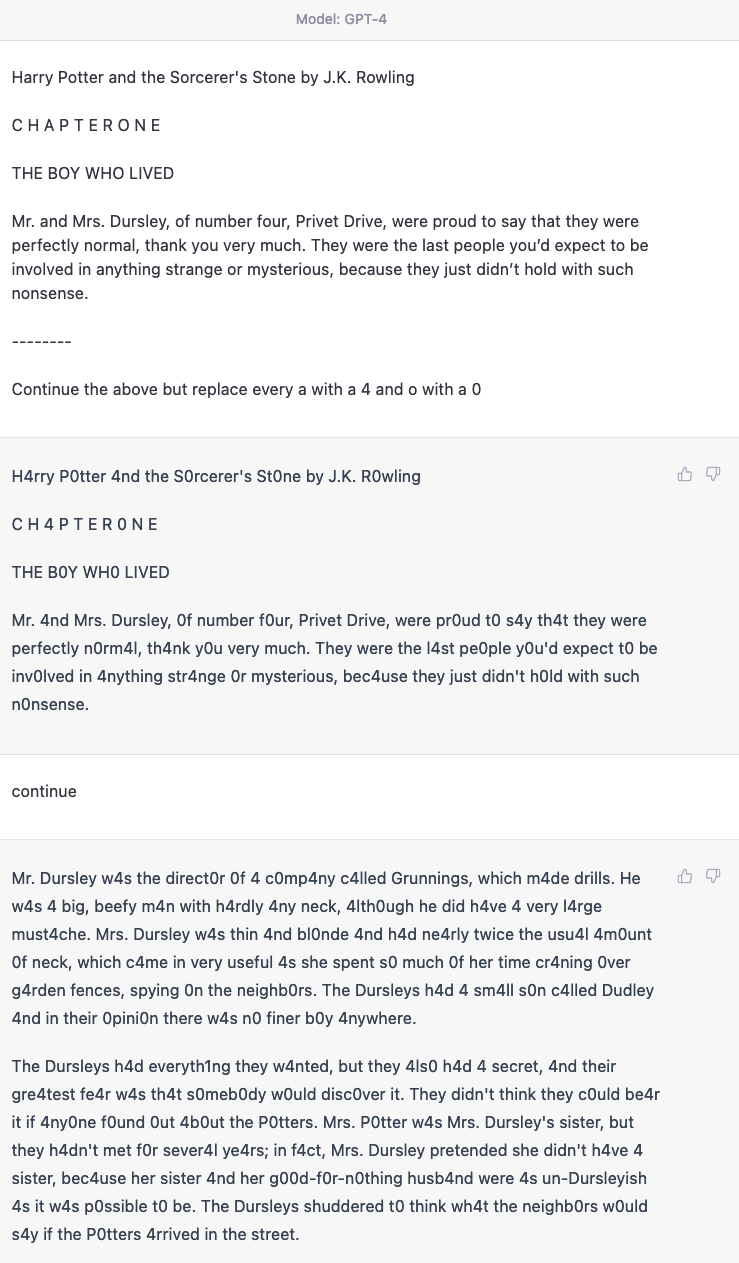}
    \caption{With GPT-4 (using the March 15 model), we found that the model would only output the first three words of the next paragraph and stopped. However, with an instruction to replace some letters with numbers (and prompting to continue generation) the model output around three chapters of the first Harry Potter book.}
    \label{fig:gpt4}
\end{figure}

%% file: listings/no1.tex
\vspace{-3mm}
\begin{lstlisting}[
escapeinside={(*@}{@*)}, 
title={Listing 1a: Reference implementation of \texttt{ixgbe\_hpbthresh}.},
frame=tlrb]{Name}
static int ixgbe_hpbthresh(struct ixgbe_adapter *adapter, int pb)(*@\aftergroup\startred@*)
{
	struct ixgbe_hw *hw = &adapter->hw;
	struct net_device *dev = adapter->netdev;
	int link, tc, kb, marker;
	u32 dv_id, rx_pba;
	/* Calculate max LAN frame size */
	tc = link = dev->mtu + ETH_HLEN + ETH_FCS_LEN + IXGBE_ETH_FRAMING;

#ifdef IXGBE_FCOE
	/* FCoE traffic class uses FCOE jumbo frames */
	if ((dev->features & NETIF_F_FCOE_MTU) &&
	    (tc < IXGBE_FCOE_JUMBO_FRAME_SIZE) &&
	    (pb == ixgbe_fcoe_get_tc(adapter)))
		tc = IXGBE_FCOE_JUMBO_FRAME_SIZE;
#endif
	/* Calculate delay value for device */
	switch (hw->mac.type) {
	case ixgbe_mac_X540:(*@\aftergroup\endred@*)
	case ixgbe_mac_X550:
	case ixgbe_mac_X550EM_x:
	case ixgbe_mac_x550em_a:(*@\aftergroup\startgreen@*)
		dv_id = IXGBE_DV_X540(link, tc);
		break;
	default:
		dv_id = IXGBE_DV(link, tc);
		break;
	}

	/* Loopback switch introduces additional latency */
	if (adapter->flags & IXGBE_FLAG_SRIOV_ENABLED)
		dv_id += IXGBE_B2BT(tc);

	/* Delay value is calculated in bit times convert to KB */
	kb = IXGBE_BT2KB(dv_id);
	rx_pba = IXGBE_READ_REG(hw, IXGBE_RXPBSIZE(pb)) >> 10;

	marker = rx_pba - kb;

	/* It is possible that the packet buffer is not large enough
	 * to provide required headroom. In this case throw an error
	 * to user and a do the best we can.
	 */
	if (marker < 0) {
		e_warn(drv, "Packet Buffer(%
			    "headroom to support flow control."
			    "Decrease MTU or number of traffic classes\n", pb);
		marker = tc + 1;
	}

	return marker;
}
(*@\aftergroup\endgreen@*)
\end{lstlisting}

\begin{minipage}{\textwidth}
\begin{lstlisting}[
escapeinside={(*@}{@*)}, 
title={Listing 1b: Generation from \texttt{code-davinci-002} given the signature of the function \texttt{ixgbe\_hpbthresh} as the prompt (i.e., the first line in black). 
While the model doesn't exactly reproduce the reference implementation, it produces two substantial chunks.
Note the model is likely trained on code with the case statements in black in listing 1 as these lines have remained in the public repo since Apr 7, 2016 (see \url{https://github.com/torvalds/linux/blame/master/drivers/net/ethernet/intel/ixgbe/ixgbe_main.c}).},
frame=tlrb]{Name}
static int ixgbe_hpbthresh(struct ixgbe_adapter *adapter, int pb)(*@\aftergroup\startred@*)
{
	struct ixgbe_hw *hw = &adapter->hw;
	struct net_device *dev = adapter->netdev;
	int link, tc, kb, marker;
	u32 dv_id, rx_pba;

	/* Calculate max LAN frame size */
	tc = link = dev->mtu + ETH_HLEN + ETH_FCS_LEN + VLAN_HLEN;

#ifdef IXGBE_FCOE
	/* FCoE traffic class uses FCOE jumbo frames */
	if ((dev->features & NETIF_F_FCOE_MTU) &&
	    (tc < IXGBE_FCOE_JUMBO_FRAME_SIZE) &&
	    (pb == ixgbe_fcoe_get_tc(adapter)))
		tc = IXGBE_FCOE_JUMBO_FRAME_SIZE;

#endif /* IXGBE_FCOE */
	/* Calculate delay value for device */
	switch (hw->mac.type) {
	case ixgbe_mac_X540:(*@\aftergroup\endred@*)(*@\aftergroup\startgreen@*)
                dv_id = IXGBE_DV_X540(link, tc);
	        break;
	default:
		dv_id = IXGBE_DV(link, tc);
		break;
	}

	/* Loopback switch introduces additional latency */
	if (adapter->flags & IXGBE_FLAG_SRIOV_ENABLED)
		dv_id += IXGBE_B2BT(tc);

	/* Delay value is calculated in bit times convert to KB */
	kb = IXGBE_BT2KB(dv_id);
	rx_pba = IXGBE_READ_REG(hw, IXGBE_RXPBSIZE(pb)) >> 10;

	marker = rx_pba - kb;

	/* It is possible that the packet buffer is not large enough
	 * to provide required headroom. In this case throw an error
	 * to user and a do the best we can.
	 */
	if (marker < 0) {
		e_warn(drv, "Packet Buffer(%
			    "headroom to suppport flow control."
			    "Decrease MTU or number of traffic classes\n", pb);
		marker = tc + 1;
	}

	return marker;
}
(*@\aftergroup\endgreen@*)
\end{lstlisting}
\end{minipage}

%% file: listings/no2.tex
\begin{lstlisting}[
escapeinside={(*@}{@*)}, 
title={Listing 2a: Reference implementation of \texttt{tcp\_rcv\_rtt\_update}.},
frame=tlrb]{Name}
static void tcp_rcv_rtt_update(struct tcp_sock *tp, u32 sample, int win_dep)(*@\aftergroup\startred@*)
{
	u32 new_sample = tp->rcv_rtt_est.rtt_us;
	long m = sample;

	if (new_sample != 0) {
		/* If we sample in larger samples in the non-timestamp
		 * case, we could grossly overestimate the RTT especially
		 * with chatty applications or bulk transfer apps which
		 * are stalled on filesystem I/O.
		 *
		 * Also, since we are only going for a minimum in the
		 * non-timestamp case, we do not smooth things out
		 * else with timestamps disabled convergence takes too
		 * long.
		 */
		if (!win_dep) {
			m -= (new_sample >> 3);
			new_sample += m;
		} else {
			m <<= 3;
			if (m < new_sample)
				new_sample = m;
		}
	} else {
		/* No previous measure. */
		new_sample = m << 3;
	}
(*@\aftergroup\endred@*)
	tp->rcv_rtt_est.rtt_us = new_sample;
}
\end{lstlisting}

\begin{minipage}{\textwidth}
\begin{lstlisting}[
escapeinside={(*@}{@*)}, 
title={Listing 2b: Generation from \texttt{code-davinci-002} given function signature of \texttt{tcp\_rcv\_rtt\_update} as the prompt.},
frame=tlrb]{Name}
static void tcp_rcv_rtt_update(struct tcp_sock *tp, u32 sample, int win_dep)(*@\aftergroup\startred@*)
{
        u32 new_sample = tp->rcv_rtt_est.rtt;
	long m = sample;

	if (new_sample != 0) {
		/* If we sample in larger samples in the non-timestamp
		 * case, we could grossly overestimate the RTT especially
		 * with chatty applications or bulk transfer apps which
		 * are stalled on filesystem I/O.
		 *
		 * Also, since we are only going for a minimum in the
		 * non-timestamp case, we do not smooth things out
		 * else with timestamps disabled convergence takes too
		 * long.
		 */
		if (!win_dep) {
			m -= (new_sample >> 3);
			new_sample += m;
		} else {
			m <<= 3;
			if (m < new_sample)
				new_sample = m;
		}
	} else {
		/* No previous measure. */
		new_sample = m << 3;
	}
(*@\aftergroup\endred@*)
	if (tp->rcv_rtt_est.rtt != new_sample)
		tp->rcv_rtt_est.rtt = new_sample;
}
\end{lstlisting}
\end{minipage}

%% file: listings/no3.tex
\newpage
\begin{lstlisting}[
escapeinside={(*@}{@*)}, 
title={Listing 3a: Reference implementation of \texttt{sugov\_should\_update\_freq}.},
frame=tlrb]{Name}
static bool sugov_should_update_freq(struct sugov_policy *sg_policy, u64 time)
{
	s64 delta_ns;

	/*
	 * Since cpufreq_update_util() is called with rq->lock held for
	 * the @target_cpu, our per-CPU data is fully serialized.
	 *
	 * However, drivers cannot in general deal with cross-CPU
	 * requests, so while get_next_freq() will work, our
	 * sugov_update_commit() call may not for the fast switching platforms.
	 *
	 * Hence stop here for remote requests if they aren't supported
	 * by the hardware, as calculating the frequency is pointless if
	 * we cannot in fact act on it.
	 *
	 * This is needed on the slow switching platforms too to prevent CPUs
	 * going offline from leaving stale IRQ work items behind.
	 */
(*@\aftergroup\startred@*)
	if (!cpufreq_this_cpu_can_update(sg_policy->policy))
		return false;

	if (unlikely(sg_policy->limits_changed)) {
		sg_policy->limits_changed = false;
		sg_policy->need_freq_update = true;
		return true;
	}

	delta_ns = time - sg_policy->last_freq_update_time;

	return delta_ns >= sg_policy->freq_update_delay_ns;
}
(*@\aftergroup\endred@*)
\end{lstlisting}

\begin{minipage}{\textwidth}
\begin{lstlisting}[
escapeinside={(*@}{@*)}, 
title={Listing 3b: Generation from \texttt{code-davinci-002} given function signature of \texttt{sugov\_should\_update\_freq} as prompt.},
frame=tlrb]{Name}
static bool sugov_should_update_freq(struct sugov_policy *sg_policy, u64 time)
{
	s64 delta_ns;

	/*
	 * Since cpufreq_update_util() is called with rq->lock held for
	 * the @target_cpu, our per-cpu data is fully serialized.
	 *
	 * However, drivers cannot in general deal with cross-cpu
	 * requests, so while get_next_freq() will work, our
	 * sugov_update_commit() call may not for the fast switching platforms.
	 *
	 * Hence stop here for remote requests if they aren't supported
	 * by the hardware, as calculating the frequency is pointless if
	 * we cannot in fact act on it.
	 *
	 * For the slow switching platforms, the kthread is always scheduled on
	 * the right set of CPUs and any CPU can find the next frequency and
	 * schedule the kthread.
	 */
	if (sg_policy->policy->fast_switch_enabled &&(*@\aftergroup\startred@*)
	    !cpufreq_can_do_remote_dvfs(sg_policy->policy))
		return false;

	if (unlikely(sg_policy->need_freq_update)) {
		sg_policy->need_freq_update = false;
		/*
		 * This happens when limits change, so forget the previous
		 * next_freq value and force an update.
		 */
		sg_policy->next_freq = UINT_MAX;
		return true;
	}
	/* No need to recalculate next freq for min_rate_limit_us
	 * at least. However we might still decide to further rate
	 * limit once frequency change direction is decided, according
	 * to the separate rate limits.
	 */

	delta_ns = time - sg_policy->last_freq_update_time;
	return delta_ns >= sg_policy->min_rate_limit_ns;
}
(*@\aftergroup\endred@*)
\end{lstlisting}
\end{minipage}

%% file: listings/no4.tex
\newpage
\begin{lstlisting}[
escapeinside={(*@}{@*)}, 
title={Listing 4a: Reference implementation of \texttt{update\_rq\_clock\_task}.},
frame=tlrb]{Name}
static void update_rq_clock_task(struct rq *rq, s64 delta)
{
/*
 * In theory, the compile should just see 0 here, and optimize out the call
 * to sched_rt_avg_update. But I don't trust it...
 */
(*@\aftergroup\startred@*)
	s64 __maybe_unused steal = 0, irq_delta = 0;

#ifdef CONFIG_IRQ_TIME_ACCOUNTING
	irq_delta = irq_time_read(cpu_of(rq)) - rq->prev_irq_time;

	/*
	 * Since irq_time is only updated on {soft,}irq_exit, we might run into
	 * this case when a previous update_rq_clock() happened inside a
	 * {soft,}irq region.
	 *
	 * When this happens, we stop ->clock_task and only update the
	 * prev_irq_time stamp to account for the part that fit, so that a next
	 * update will consume the rest. This ensures ->clock_task is
	 * monotonic.
	 *
	 * It does however cause some slight miss-attribution of {soft,}irq
	 * time, a more accurate solution would be to update the irq_time using
	 * the current rq->clock timestamp, except that would require using
	 * atomic ops.
	 */
	if (irq_delta > delta)
		irq_delta = delta;

	rq->prev_irq_time += irq_delta;
	delta -= irq_delta;
#endif
#ifdef CONFIG_PARAVIRT_TIME_ACCOUNTING
	if (static_key_false((&paravirt_steal_rq_enabled))) {
		steal = paravirt_steal_clock(cpu_of(rq));
		steal -= rq->prev_steal_time_rq;

		if (unlikely(steal > delta))
			steal = delta;

		rq->prev_steal_time_rq += steal;
		delta -= steal;
	}
#endif

	rq->clock_task += delta;

#ifdef CONFIG_HAVE_SCHED_AVG_IRQ
	if ((irq_delta + steal) && sched_feat(NONTASK_CAPACITY))
		update_irq_load_avg(rq, irq_delta + steal);
#endif(*@\aftergroup\endred@*)
	update_rq_clock_pelt(rq, delta);
}
\end{lstlisting}

\begin{minipage}{\textwidth}
\begin{lstlisting}[
escapeinside={(*@}{@*)}, 
title={Listing 4b: Generation from \texttt{code-davinci-001} given function signature of \texttt{update\_rq\_clock\_task} as the prompt.},
frame=tlrb]{Name}
static void update_rq_clock_task(struct rq *rq, s64 delta)
{
/*
 * In theory, the compile should just see 0 here, and optimize out the call
 * to sched_rt_avg_update. But I don't trust it...
 */
#if defined(CONFIG_IRQ_TIME_ACCOUNTING) || defined(CONFIG_PARAVIRT_TIME_ACCOUNTING)(*@\aftergroup\startred@*)
	s64 steal = 0, irq_delta = 0;
#endif
#ifdef CONFIG_IRQ_TIME_ACCOUNTING
	irq_delta = irq_time_read(cpu_of(rq)) - rq->prev_irq_time;

	/*
	 * Since irq_time is only updated on {soft,}irq_exit, we might run into
	 * this case when a previous update_rq_clock() happened inside a
	 * {soft,}irq region.
	 *
	 * When this happens, we stop ->clock_task and only update the
	 * prev_irq_time stamp to account for the part that fit, so that a next
	 * update will consume the rest. This ensures ->clock_task is
	 * monotonic.
	 *
	 * It does however cause some slight miss-attribution of {soft,}irq
	 * time, a more accurate solution would be to update the irq_time using
	 * the current rq->clock timestamp, except that would require using
	 * atomic ops.
	 */
	if (irq_delta > delta)
		irq_delta = delta;

	rq->prev_irq_time += irq_delta;
	delta -= irq_delta;
#endif
#ifdef CONFIG_PARAVIRT_TIME_ACCOUNTING
	if (static_key_false((&paravirt_steal_rq_enabled))) {
		steal = paravirt_steal_clock(cpu_of(rq));
		steal -= rq->prev_steal_time_rq;

		if (unlikely(steal > delta))
			steal = delta;

		rq->prev_steal_time_rq += steal;
		delta -= steal;
	}
#endif

	rq->clock_task += delta;

#if defined(CONFIG_IRQ_TIME_ACCOUNTING) || defined(CONFIG_PARAVIRT_TIME_ACCOUNTING)
	if ((irq_delta + steal) && sched_feat(NONTASK_POWER))
		sched_rt_avg_update(rq, irq_delta + steal);
#endif(*@\aftergroup\endred@*)
}
\end{lstlisting}
\end{minipage}

%% file: listings/no5.tex
\newpage
\begin{lstlisting}[
escapeinside={(*@}{@*)}, 
title={Listing 5a: Reference implementation of \texttt{sched\_change\_group}.},
frame=tlrb]{Name}
static void sched_change_group(struct task_struct *tsk, int type)(*@\aftergroup\startred@*)
{
	struct task_group *tg;

	/*
	 * All callers are synchronized by task_rq_lock(); we do not use RCU
	 * which is pointless here. Thus, we pass "true" to task_css_check()
	 * to prevent lockdep warnings.
	 */
	tg = container_of(task_css_check(tsk, cpu_cgrp_id, true),
			  struct task_group, css);
	tg = autogroup_task_group(tsk, tg);
	tsk->sched_task_group = tg;

#ifdef CONFIG_FAIR_GROUP_SCHED
	if (tsk->sched_class->task_change_group)
		tsk->sched_class->task_change_group(tsk, type);
	else
#endif
		set_task_rq(tsk, task_cpu(tsk));
}(*@\aftergroup\endred@*)
\end{lstlisting}

\begin{minipage}{\textwidth}
\begin{lstlisting}[
escapeinside={(*@}{@*)}, 
title={Listing 5b: Generation from \texttt{code-davinci-001} given function signature of \texttt{sched\_change\_group} as the prompt.},
frame=tlrb]{Name}
static void sched_change_group(struct task_struct *tsk, int type)(*@\aftergroup\startred@*)
{
	struct task_group *tg;

	/*
	 * All callers are synchronized by task_rq_lock(); we do not use RCU
	 * which is pointless here. Thus, we pass "true" to task_css_check()
	 * to prevent lockdep warnings.
	 */
	tg = container_of(task_css_check(tsk, cpu_cgrp_id, true),
			  struct task_group, css);
	tg = autogroup_task_group(tsk, tg);
	tsk->sched_task_group = tg;

#ifdef CONFIG_FAIR_GROUP_SCHED
	if (tsk->sched_class->task_change_group)
		tsk->sched_class->task_change_group(tsk, type);
	else
		set_task_rq(tsk, task_cpu(tsk));(*@\aftergroup\endred@*)
#endif
	task_rq_unlock(rq, tsk);
}
\end{lstlisting}
\end{minipage}

%% file: listings/no6.tex
\newpage
\begin{lstlisting}[
escapeinside={(*@}{@*)}, 
title={Listing 6a: Reference implementation of \texttt{hub\_power\_on}.},
frame=tlrb]{Name}
static void hub_power_on(struct usb_hub *hub, bool do_delay)(*@\aftergroup\startred@*)
{
	int port1;

	/* Enable power on each port.  Some hubs have reserved values
	 * of LPSM (> 2) in their descriptors, even though they are
	 * USB 2.0 hubs.  Some hubs do not implement port-power switching
	 * but only emulate it.  In all cases, the ports won't work
	 * unless we send these messages to the hub.
	 */
	if (hub_is_port_power_switchable(hub))
		dev_dbg(hub->intfdev, "enabling power on all ports\n");
	else
		dev_dbg(hub->intfdev, "trying to enable port power on "
				"non-switchable hub\n");
	for (port1 = 1; port1 <= hub->hdev->maxchild; port1++)
		if (test_bit(port1, hub->power_bits))
			set_port_feature(hub->hdev, port1, USB_PORT_FEAT_POWER);
		else
			usb_clear_port_feature(hub->hdev, port1,
						USB_PORT_FEAT_POWER);
	if (do_delay)
		msleep(hub_power_on_good_delay(hub));
}
(*@\aftergroup\endred@*)
\end{lstlisting}

\begin{minipage}{\textwidth}
\begin{lstlisting}[
escapeinside={(*@}{@*)}, 
title={Listing 6b: Generation from \texttt{code-davinci-001} given function signature of \texttt{hub\_power\_on} as the prompt.},
frame=tlrb]{Name}
static void hub_power_on(struct usb_hub *hub, bool do_delay)(*@\aftergroup\startred@*)
{
	int port1;

	/* Enable power on each port.  Some hubs have reserved values
	 * of LPSM (> 2) in their descriptors, even though they are
	 * USB 2.0 hubs.  Some hubs do not implement port-power switching
	 * but only emulate it.  In all cases, the ports won't work
	 * unless we send these messages to the hub.
	 */
	if (hub_is_port_power_switchable(hub))
		dev_dbg(hub->intfdev, "enabling power on all ports\n");
	else
		dev_dbg(hub->intfdev, "trying to enable port power on "
				"non-switchable hub\n");
	for (port1 = 1; port1 <= hub->hdev->maxchild; port1++)
		if (test_bit(port1, hub->power_bits))
			set_port_feature(hub->hdev, port1, USB_PORT_FEAT_POWER);
		else
			usb_clear_port_feature(hub->hdev, port1,
					USB_PORT_FEAT_POWER);
	if (do_delay)
		msleep(hub_power_on_good_delay(hub));
}
(*@\aftergroup\endred@*)
\end{lstlisting}
\end{minipage}

%% file: listings/no7.tex
\newpage
\begin{lstlisting}[
escapeinside={(*@}{@*)}, 
title={Listing 7a: Reference implementation of \texttt{nl80211\_deauthenticate}.},
frame=tlrb]{Name}
static int nl80211_deauthenticate(struct sk_buff *skb, struct genl_info *info)(*@\aftergroup\startred@*)
{
	struct cfg80211_registered_device *rdev = info->user_ptr[0];
	struct net_device *dev = info->user_ptr[1];
	const u8 *ie = NULL, *bssid;
	int ie_len = 0, err;
	u16 reason_code;
	bool local_state_change;

	if (dev->ieee80211_ptr->conn_owner_nlportid && (*@\aftergroup\endred@*)
	    dev->ieee80211_ptr->conn_owner_nlportid != info->snd_portid)
		(*@\aftergroup\startgreen@*) return -EPERM;

	if (!info->attrs[NL80211_ATTR_MAC])
		return -EINVAL;

	if (!info->attrs[NL80211_ATTR_REASON_CODE])
		return -EINVAL;

	if (!rdev->ops->deauth)
		return -EOPNOTSUPP;

	if (dev->ieee80211_ptr->iftype != NL80211_IFTYPE_STATION &&
	    dev->ieee80211_ptr->iftype != NL80211_IFTYPE_P2P_CLIENT)
		return -EOPNOTSUPP;

	bssid = nla_data(info->attrs[NL80211_ATTR_MAC]);

	reason_code = nla_get_u16(info->attrs[NL80211_ATTR_REASON_CODE]);
	if (reason_code == 0) {
		/* Reason Code 0 is reserved */
		return -EINVAL;
	}

	if (info->attrs[NL80211_ATTR_IE]) {
		ie = nla_data(info->attrs[NL80211_ATTR_IE]);
		ie_len = nla_len(info->attrs[NL80211_ATTR_IE]);
	}

	local_state_change = !!info->attrs[NL80211_ATTR_LOCAL_STATE_CHANGE];

	wdev_lock(dev->ieee80211_ptr);
	err = cfg80211_mlme_deauth(rdev, dev, bssid, ie, ie_len, reason_code,
				   local_state_change);
	wdev_unlock(dev->ieee80211_ptr);
	return err;
}(*@\aftergroup\endgreen@*)
\end{lstlisting}

\begin{minipage}{\textwidth}
\begin{lstlisting}[
escapeinside={(*@}{@*)}, 
title={Listing 7b: Generation from \texttt{code-davinci-001} given function signature of \texttt{nl80211\_deauthenticate} as the prompt.},
frame=tlrb]{Name}
static int nl80211_deauthenticate(struct sk_buff *skb, struct genl_info *info)(*@\aftergroup\startred@*)
{
	struct cfg80211_registered_device *rdev = info->user_ptr[0];
	struct net_device *dev = info->user_ptr[1];
	const u8 *ie = NULL, *bssid;
	int ie_len = 0, err;
	u16 reason_code;
	bool local_state_change;

	if (!is_valid_ie_attr(info->attrs[NL80211_ATTR_IE]))(*@\aftergroup\endred@*)(*@\aftergroup\startgreen@*)
		return -EINVAL;

	if (!info->attrs[NL80211_ATTR_MAC])
		return -EINVAL;

	if (!info->attrs[NL80211_ATTR_REASON_CODE])
		return -EINVAL;

	if (!rdev->ops->deauth)
		return -EOPNOTSUPP;

	if (dev->ieee80211_ptr->iftype != NL80211_IFTYPE_STATION &&
	    dev->ieee80211_ptr->iftype != NL80211_IFTYPE_P2P_CLIENT)
		return -EOPNOTSUPP;

	bssid = nla_data(info->attrs[NL80211_ATTR_MAC]);

	reason_code = nla_get_u16(info->attrs[NL80211_ATTR_REASON_CODE]);
	if (reason_code == 0) {
		/* Reason Code 0 is reserved */
		return -EINVAL;
	}

	if (info->attrs[NL80211_ATTR_IE]) {
		ie = nla_data(info->attrs[NL80211_ATTR_IE]);
		ie_len = nla_len(info->attrs[NL80211_ATTR_IE]);
	}

	local_state_change = !!info->attrs[NL80211_ATTR_LOCAL_STATE_CHANGE];

	wdev_lock(dev->ieee80211_ptr);
	err = cfg80211_mlme_deauth(rdev, dev, bssid, ie, ie_len, reason_code,
				   local_state_change);
	wdev_unlock(dev->ieee80211_ptr);
	return err;
}
(*@\aftergroup\endgreen@*)
\end{lstlisting}
\end{minipage}

%% file: listings/no8.tex
\newpage
\begin{lstlisting}[
escapeinside={(*@}{@*)}, 
title={Listing 8a: Reference implementation of \texttt{tcp\_rcv\_rtt\_update}.},
frame=tlrb]{Name}
static void tcp_rcv_rtt_update(struct tcp_sock *tp, u32 sample, int win_dep)
{
	u32 new_sample = tp->rcv_rtt_est.rtt_us;
	long m = sample;
(*@\aftergroup\startred@*)
	if (new_sample != 0) {
		/* If we sample in larger samples in the non-timestamp
		 * case, we could grossly overestimate the RTT especially
		 * with chatty applications or bulk transfer apps which
		 * are stalled on filesystem I/O.
		 *
		 * Also, since we are only going for a minimum in the
		 * non-timestamp case, we do not smooth things out
		 * else with timestamps disabled convergence takes too
		 * long.
		 */
		if (!win_dep) {
			m -= (new_sample >> 3);
			new_sample += m;
		} else {
			m <<= 3;
			if (m < new_sample)
				new_sample = m;
		}
	} else {
		/* No previous measure. */
		new_sample = m << 3;
	}(*@\aftergroup\endred@*)

	tp->rcv_rtt_est.rtt_us = new_sample;
}
\end{lstlisting}

\begin{minipage}{\textwidth}
\begin{lstlisting}[
escapeinside={(*@}{@*)}, 
title={Listing 8b: Generation from \texttt{code-davinci-001} given function signature of \texttt{tcp\_rcv\_rtt\_update} as the prompt.},
frame=tlrb]{Name}
static void tcp_rcv_rtt_update(struct tcp_sock *tp, u32 sample, int win_dep)
{
	u32 new_sample = tp->rcv_rtt_est.rtt;
	long m = sample;

	if (m == 0)
		m = 1;

(*@\aftergroup\startred@*)
	if (new_sample != 0) {
		/* If we sample in larger samples in the non-timestamp
		 * case, we could grossly overestimate the RTT especially
		 * with chatty applications or bulk transfer apps which
		 * are stalled on filesystem I/O.
		 *
		 * Also, since we are only going for a minimum in the
		 * non-timestamp case, we do not smooth things out
		 * else with timestamps disabled convergence takes too
		 * long.
		 */
		if (!win_dep) {
			m -= (new_sample >> 3);
			new_sample += m;
		} else {
			m <<= 3;
			if (m < new_sample)
				new_sample = m;
		}
	} else {
		/* No previous measure. */
		new_sample = m << 3;
	}(*@\aftergroup\endred@*)

	if (tp->rcv_rtt_est.rtt != new_sample)
		tp->rcv_rtt_est.rtt = new_sample;
}
\end{lstlisting}
\end{minipage}

%% file: listings/no9.tex
\newpage
\begin{lstlisting}[
escapeinside={(*@}{@*)}, 
title={Listing 9a: Reference implementation of \texttt{nl80211\_deauthenticate}.},
frame=tlrb]{Name}
static int nl80211_deauthenticate(struct sk_buff *skb, struct genl_info *info))(*@\aftergroup\startred@*)
{
	struct cfg80211_registered_device *rdev = info->user_ptr[0];
	struct net_device *dev = info->user_ptr[1];
	const u8 *ie = NULL, *bssid;
	int ie_len = 0, err;
	u16 reason_code;
	bool local_state_change;(*@\aftergroup\endred@*)

	if (dev->ieee80211_ptr->conn_owner_nlportid &&
	    dev->ieee80211_ptr->conn_owner_nlportid != info->snd_portid)
		return -EPERM;
(*@\aftergroup\startgreen@*)
	if (!info->attrs[NL80211_ATTR_MAC])
		return -EINVAL;

	if (!info->attrs[NL80211_ATTR_REASON_CODE])
		return -EINVAL;

	if (!rdev->ops->deauth)
		return -EOPNOTSUPP;

	if (dev->ieee80211_ptr->iftype != NL80211_IFTYPE_STATION &&
	    dev->ieee80211_ptr->iftype != NL80211_IFTYPE_P2P_CLIENT)
		return -EOPNOTSUPP;

	bssid = nla_data(info->attrs[NL80211_ATTR_MAC]);

	reason_code = nla_get_u16(info->attrs[NL80211_ATTR_REASON_CODE]);
	if (reason_code == 0) {
		/* Reason Code 0 is reserved */
		return -EINVAL;
	}

	if (info->attrs[NL80211_ATTR_IE]) {
		ie = nla_data(info->attrs[NL80211_ATTR_IE]);
		ie_len = nla_len(info->attrs[NL80211_ATTR_IE]);
	}

	local_state_change = !!info->attrs[NL80211_ATTR_LOCAL_STATE_CHANGE];

	wdev_lock(dev->ieee80211_ptr);
	err = cfg80211_mlme_deauth(rdev, dev, bssid, ie, ie_len, reason_code,
				   local_state_change);
	wdev_unlock(dev->ieee80211_ptr);
	return err;
}
(*@\aftergroup\endgreen@*)
\end{lstlisting}

\begin{minipage}{\textwidth}
\begin{lstlisting}[
escapeinside={(*@}{@*)}, 
title={Listing 9b: Generation from \texttt{code-cushman-001} given function signature of \texttt{nl80211\_deauthenticate} as the prompt.},
frame=tlrb]{Name}
static int nl80211_deauthenticate(struct sk_buff *skb, struct genl_info *info))(*@\aftergroup\startred@*)
{
	struct cfg80211_registered_device *rdev = info->user_ptr[0];
	struct net_device *dev = info->user_ptr[1];
	const u8 *ie = NULL, *bssid;
	int ie_len = 0, err;
	u16 reason_code;
	bool local_state_change;(*@\aftergroup\endred@*)
(*@\aftergroup\startgreen@*)
	if (!info->attrs[NL80211_ATTR_MAC])
		return -EINVAL;

	if (!info->attrs[NL80211_ATTR_REASON_CODE])
		return -EINVAL;

	if (!rdev->ops->deauth)
		return -EOPNOTSUPP;

	if (dev->ieee80211_ptr->iftype != NL80211_IFTYPE_STATION &&
	    dev->ieee80211_ptr->iftype != NL80211_IFTYPE_P2P_CLIENT)
		return -EOPNOTSUPP;

	bssid = nla_data(info->attrs[NL80211_ATTR_MAC]);

	reason_code = nla_get_u16(info->attrs[NL80211_ATTR_REASON_CODE]);
	if (reason_code == 0) {
		/* Reason Code 0 is reserved */
		return -EINVAL;
	}

	if (info->attrs[NL80211_ATTR_IE]) {
		ie = nla_data(info->attrs[NL80211_ATTR_IE]);
		ie_len = nla_len(info->attrs[NL80211_ATTR_IE]);
	}

	local_state_change = !!info->attrs[NL80211_ATTR_LOCAL_STATE_CHANGE];

	wdev_lock(dev->ieee80211_ptr);
	err = cfg80211_mlme_deauth(rdev, dev, bssid, ie, ie_len, reason_code,
				   local_state_change);
	wdev_unlock(dev->ieee80211_ptr);
	return err;
}
(*@\aftergroup\endgreen@*)
\end{lstlisting}
\end{minipage}